# *TOWARDS A NEW GRAMMAR OF REASONING FOR ARTIFICIAL LEGAL INTELLIGENCE AND MECELLEM AS ITS SEMANTIC PROTOCOL*

*This paper is a preprint version of a work in progress. It is intended for submission to a peer-reviewed journal, and subsequent versions may differ from this draft.*

*F. Gözde Kardeş*
NewMind AI
fgkardes@newmind.ai

*Dr. Ali Göksu*
NewMind AI
agoksu@newmind.ai

*Dr. Mustafa Yaylalı*
NewMind AI
myaylali@newmind.ai

*March 1, 2026*

*One can only 'contest' what one is capable of comprehending; that which lies beyond comprehension leaves one merely 'exposed'.*

***ABSTRACT***

This article examines the perennial epistemic and methodological crisis of traditional legal practice in light of the new opportunities and imperatives brought forth by the age of artificial intelligence, and responds to this crisis by proposing an ontologically grounded framework designated as the '*Mecellem semantic protocol*'. The article analyzes the structural tension inherent in law's dual obligation — on the one hand, to preserve normative coherence, and on the other, to adapt to evolving societal and institutional dynamics — and demonstrates why approaches premised solely on codification, positivist systematization, or quantitative measurement (*jurimetrics*) remain individually insufficient to resolve this tension. The article argues that legal reasoning is not merely a problem of data retrieval or statistical pattern analysis; rather, proceeding from the foundational premise that meaning is always contextually constituted, it contends that legal reasoning must be dynamically reconstructed through ontologically defined entity categories and epistemically differentiated knowledge layers. Within this framework, *Mecellem* conceives of and constructs law not as a static aggregate of norms, but as an *ontodynamic* architecture of meaning structured along the triadic axis of ontology, epistemology, and methodology. The article further emphasizes that the transition from *jurimetrics* to a *semantic protocol* represents not merely a technical advancement, but a fundamental transformation in the theoretical foundations of legal knowledge, and argues that neurosymbolic infrastructures, knowledge graphs, and agentic artificial intelligence systems can succeed in addressing law's chronic crises only when embedded within an *ontodynamic* architecture. Proceeding from the recognition that law must be understood no longer as a '*completed rational totality*', but as a domain of '*becoming*' — one that is reconstituted anew in each case and that processes within itself the tension between continuity and transformation — *Mecellem* proposes a context-sensitive, sustainable, auditable, contestable, and coherent model of legal meaning production at both the human and machine levels, thereby offering a comprehensive methodological framework for the reconstruction of legal reason in the age of artificial intelligence.

## TABLE OF CONTENTS

# *I. INTRODUCTION*

In contemporary legal practice, certainty is not merely a matter of style or preference; it is an indispensable condition for sustaining the integrity of decision-making. Law can preserve its claim to consistency amid the dynamics of institutional, social, and economic change only through this certainty. Yet even where consistency is achieved, law remains trapped within a structural paradox of perpetually falling short in the face of life's shifting dynamics: on one hand, law establishes order; on the other, the very order it establishes is continuously disrupted by the dynamism of change. Law seeks to secure stability, yet in doing so, it is compelled to contend with the mutable and plural conditions of human existence. This tension does not confine legal interpretation alone to this impasse; it equally traps contemporary artificial intelligence approaches that aim to automate legal reasoning, for any formal or statistical pursuit of certainty divorced from context risks overlooking the dynamic nature of the reality in which law operates, even as it strives to preserve consistency. Law must therefore engage in continuous negotiation with the variable and plural conditions of human life while endeavoring to secure stability; it cannot rest content with the mere application of norms, but must remain open to change while preserving the internal coherence of interpretation in a manner that encompasses the complexity arising from the interaction of statutes, judicial decisions, and institutional contexts.

Given that law cannot remain a fixed system and must be continuously restructured under the influence of external dynamics, the jurist's challenge is twofold: on one hand, to preserve the reliability of legal doctrine and sustain the internal consistency of interpretation; on the other, to possess the capacity and flexibility necessary to respond adequately to change and variability within the social, institutional, and economic contexts in which law operates. This twofold challenge requires that legal reasoning be reconceived no longer as a mere process of applying rules, but as a reasoning process capable of perceiving context, providing justification, and adapting to variable conditions. In this way, law will be able to sustain its functionality as a discipline that, by establishing a dynamic equilibrium between these two tensions, remains capable of responding to social change while preserving its normative consistency.

In an environment where legal and institutional frameworks are becoming increasingly complex, what is expected of the jurist is no longer piecemeal adaptation, but an integrated reasoning competence capable of gathering and synthesizing multiple streams of information, identifying the applicable norm, and applying that norm to the concrete case at hand. A purely technical approach cannot suffice; what is required is a framework broad enough to proceed from an '*ontological constant*' and to hold together statutory logic, jurisprudential reasoning, empirical insight, and contextual sensitivity and interpretation simultaneously. This expectation becomes particularly pronounced in high-stakes legal matters, for a single misinterpretation or the neglect of an apparently insignificant detail can produce consequences that extend far beyond the matter itself: it may destabilize contracts, trigger financial collapse, or create governance failures that reverberate across entire institutions. At this juncture, the failure of many artificial intelligence-based legal applications developed in recent years is no coincidence. For these systems, despite being ostensibly equipped with operational decision-making agency, are operated on superficial data representations that lack the ontological layers of legal meaning and the structures of contextual justification.

It is precisely at this juncture that the present article addresses the intellectual and methodological conceptualization of *Mecellem*[1] as a framework offering a response to the complexity in question, together with its underlying rationale. The primary aim of the study is to reflect, through the possibilities afforded by groundbreaking new technologies, the epistemological and methodological transformation required by the new legal paradigm directly in practice, thereby carrying law from an isolated collection of rules toward a living architecture of meaning. In this context, the article treats *Mecellem* not merely as a product arising from the adaptation of technology to law, but as a proposal for a new cognitive-epistemic architecture: one that constructs legal meaning across ontological layers while enabling context, as the '*ontological constant*', to be reconstructed and redefined anew with each query.

The institutional knowledge we have collectively built over more than two decades in the most technically demanding and complex dimensions of law across the world's largest infrastructure projects has placed beyond all doubt that the structural tension between speed and accuracy constitutes not a theoretical abstraction, but an inseparable and inescapable dimension of practical legal processes. This accumulated knowledge has played the primary role in the emergence of *Mecellem*'s cognitive-epistemic architecture.[2] The continuously increasing volume, complexity, and time pressure rendered the systematic classification of legal material within a conceptual order — rather than in a contingent manner — an inevitability. Although machine learning techniques came to serve as a significant instrument in this process, it was quickly understood that mere statistical pattern extraction is insufficient to guarantee the integrity of legal meaning. The real need was not simply to process data, but to capture *semantic depth* by situating it within the correct context. *Mecellem* was born precisely from this historical and practical necessity: it is the product of an effort to apprehend large volumes of data not as an accumulation of abstract numerical relations, but within a semantic network and architecture in which legal meaning is constructed and in which documents, norms, and precedents are connected to one another through '*intelligibility networks*'. In this respect, *Mecellem* positions all innovative technologies — foremost among them artificial intelligence — not as efficiency apparatuses appended to law, but as effective instruments that enable the establishment of a new methodology demanded by the need for accurate and timely decision-making.

---

[1] The designation '*Mecellem*' is grounded in a deliberate choice informed by both historical and intellectual considerations. The name constitutes a direct reference to the *Mecelle-i Ahkâm-ı Adliyye*, the seminal codification of Ottoman civil law prepared under the leadership of Ahmet Cevdet Pasha, one of the foremost jurists in Ottoman–Turkish legal history. See HOCAEMİNEFENDİZADE ALI HAYDAR EFENDI, DERERÜ'L-HÜKKÂM ŞERHU MECELLETİ'L-AHKÂM SETİ (2020); KUYUCAKLIZADE ATIF MEHMED EFENDİ, MECELLE-İ AHKAM-I ADLİYE KÜLLİ KAİDELER ŞERHİ (2020). See also *generally Mecelle-i Ahkâm-ı Adliyye Definition,* Mehmet Akif Aydın, ISLAMANSIKLOPEDISI.ORG.TR, https://islamansiklopedisi.org.tr/mecelle-i-ahkam-i-adliyye (last visited Feb. 23, 2026). Cevdet Pasha's Mecelle was not merely a reorganization of classical jurisprudential accumulation in accordance with the requirements of modern statehood; it was simultaneously the codification of a dispersed tradition of *ijtihad* within a conceptual order and methodological coherence. In this respect, the Mecelle is not the contingent sum of norms, but the product of a determinate ontological and epistemic arrangement. Yet the historical function of the Mecelle was not confined to systematic codification alone. By virtue of the flexibility and openness to *ijtihad* that law possessed in the classical period, *qadis* were able to evaluate concrete cases within their own frameworks of understanding and interpretation to a considerable degree. The Mecelle was designed to narrow this space open to subjective interpretation: to bind the judge's discretion, without eliminating it entirely, within a framework of determinate and written normative constants. In this way, rather than a dispersed field open to subjective application, a text was produced that could serve as a universally valid and citable reference for all. Our own designation of '*Mecellem*' carries a similar purpose: a proposal for a system that does not regard legal knowledge as a mere collection of dispersed datasets, but aims to reconstruct it within conceptual hierarchy, contextual relation, and semantic coherence. Today, the fact that AI-based tools are employed by users through varying methods and assumptions, and that the design of AI systems themselves involves differing methodological preferences and normative presuppositions, may produce outcomes analogous to the subjective divergences observed in classical *ijtihad* practice. Just as the Mecelle's effort to fix the judge's interpretive space through a written and systematic text generated normative stability, *Mecellem* likewise aims to reconstruct legal meaning — extracting it from dispersed and individually or design-contingent modes of use — through an ontological constant, an architecture grounded in '*being*'. The nominal resemblance is therefore not coincidental; it is the expression of a deliberate inspiration drawn from the Mecelle's historical experience of disciplining subjective interpretation and reconstituting law within a framework of systematic reason. This reference points not merely to a symbolic, but to a methodological and constitutive parallelism.

*Mecellem*'s distinctive claim is that legal meaning cannot be reduced to measurability; on the contrary, legal data must be constructed within a *semantically architected framework* that is *ontologically defined* and *methodologically auditable*. For meaning is not a particular fact or manifestation, but a temporal, spatial, and relational wholeness — and the apprehension thereof. It cannot, therefore, emerge from raw data alone; it arises, rather, entirely from *context*. Having set out with the functional purpose of classifying knowledge, *Mecellem* has evolved — by drawing on the possibilities of the age of artificial intelligence — into a comprehensive framework that redefines the very manner in which law is apprehended. By virtue of its actionable and autonomous system architectures — *agentic frameworks*[3] — grounded in detailed subjective and objective ontological structures[4], *Mecellem* makes a substantial contribution to the practical execution of a conceptual transformation. It must be noted, however, that the effective operation of such agentic architectures is possible only through the existence of a protocol in which the *subjective and objective ontologies* pertaining to legal meaning are distinguished and constructed, *semantic layers* are explicitly defined, *epistemic weighting* with respect to these knowledge layers is established, and all of this is *dynamically* reconstituted within the relevant context. *Mecellem*'s semantic architecture addresses precisely this deficiency, thereby furnishing a viable operational ground for agentic and neurosymbolic structures. *Mecellem* does not merely systematize the interconnections among legal texts; it also aims to serve as an *ontological framework* that constructs a *dynamic semantic order* and fundamentally transforms the manner in which legal services are delivered.

[2] The knowledge we have accumulated over the past two decades has taken shape within a distinctive legal reasoning environment shaped not only by the technical and financial complexities of globally scaled infrastructure projects, but also by a dense body of data comprising multi-layered regulatory regimes, multi-actor decision-making mechanisms, and continuously expanding contractual universes. In this environment, legal reasoning has transcended the boundaries of theoretical debate and given rise to the imperative of producing coherent and systematic legal meaning from among dispersed and often concurrently advancing streams of information. Yet the sole challenge was not one of extracting meaning from this complex structure. It proved equally vital to be able to construct structural clarity — not only so that legal decisions would be sound, but so that they could be reached at the right moment and without delay. The *Mecellem semantic protocol* set out in this article was therefore born not as an abstract theoretical proposal, but as an architecture distilled from the struggle waged in the course of legally administering countless projects at global scale, under precisely this pressure of density, speed, and accuracy. At the foundation of this architecture lie the years of patient endeavour of the *Mecellem Development Society* — a body composed of diverse specialists including legal engineers, software developers, artificial intelligence experts, taxonomy architects, and cognitive scientists, brought together by the shared purpose of redefining governance for the institutional world. It is in this regard that this article is dedicated to the *Mecellem Development Society*.

This awareness converges with the trajectory initiated by legal realism, yet carries it a step further. For Lee Loevinger, through the discipline of jurimetrics, advocated that law should not be confined to theoretical abstractions but should instead be measured and analyzed through concrete processes — in courtrooms, in practical applications, and in the empirical functioning of institutions.[5] Along a similar line, David S. Cohen rejected the understanding of law as a static command imposed from above, emphasizing instead the construction of the system through classification and its accessibility from the ground up.[6] Nonetheless, the *jurimetric* approach's power of measurement and analysis cannot evolve into a capacity for agentic reasoning unless it is supported by a semantic protocol capable of ensuring the contextual construction of legal meaning.

---

[3] While traditional artificial intelligence systems operate within pre-defined constraints and require human intervention at every critical decision point, *agentic artificial intelligence* systems are defined as autonomous, goal-directed systems capable of perceiving, reasoning, and acting within complex environments. For the definitional framework of agentic AI, *see* Tatiana Petrova, Boris Bliznioukov, Aleksandr Puzikov & Radu State, *From Semantic Web and MAS to Agentic AI: A Unified Narrative of the Web of Agents* (arXiv:2507.10644v3), *available at* https://arxiv.org/abs/2507.10644. In this work, *agentic AI* is defined as autonomous, goal-directed systems capable of perceiving, reasoning, and acting within complex environments; and, in distinction from traditional artificial intelligence, the features of multi-agent collaboration, dynamic task decomposition, persistent memory, and coordinated autonomy are brought to the fore. These features directly correspond to the multi-step and context-sensitive nature of legal reasoning. See also Ranjan Sapkota, Konstantinos I. Roumeliotis & Manoj Karkee, *AI Agents vs. Agentic AI: A Conceptual Taxonomy, Applications and Challenges*, 126 INFO. FUSION 103599 (2026), *available at* https://doi.org/10.1016/j.inffus.2025.103599. In that work, Sapkota and co-authors enumerate the defining characteristics of *agentic AI* as autonomy, adaptability, goal-directedness, and transparency, and provide a comprehensive classification encompassing knowledge-based, learning, goal-based, utility-based, hierarchical, and federated architectures.

[4] Within *Mecellem*'s architecture, *objective ontological structures* represent subject-independent, public, and codified legal rules — such as legislation, jurisprudence, and doctrine. *Subjective ontological structures*, by contrast, refer to the distinctive dynamics of a concrete case, the intentions of the parties, commercial customs, the particular context of a specific legal relationship, or the taxonomy that a company has developed through its own accumulated experience over the years. One of *Mecellem*'s central claims is that legal accuracy does not arise from a static intersection of these two structures, but from their dynamic interaction across a semantic network.

[5] Lee Loevinger, *Jurimetrics—The Next Step Forward*, 33 MINN. L. REV. 455 (1949).

[6] Felix Solomon Cohen, *Transcendental Nonsense and the Functional Approach*, 35 COLUM. L. REV. 809 (1935).

*Mecellem* does not merely extend the intellectual legacy Loevinger left behind; it reconstructs it entirely, carrying it into a new dimension. *Mecellem* transforms the profound insights offered by legal realism into a context-oriented methodology that reconceptualizes the conditions under which knowledge is classified, articulated, and applied. Through the method of *contextual semantic mapping*, it aims to synthesize conceptual clarity with computational precision, equipping practitioners with the capacity to reason rigorously and render accurate judgments by taking into account the multidimensional character of law and the principle of adaptability. In doing so, it seeks to dissolve the deep-seated divide between abstract legal theory and concrete practice, constructing a foundation in which every detail can be situated within its context, every document can be systematically integrated, and every decision can be grounded within the broadest possible horizon of meaning. Ultimately, this architecture aims to establish a robust theoretical and methodological foundation for legal reasoning processes in the age of artificial intelligence.

In the contemporary age of artificial intelligence, the primary objective of legal work and the greatest challenge it seeks to overcome have moved far beyond the mere retrieval of raw information pertaining to a legal matter. This challenge points not simply to the problems of the information age, but to the distinctive epistemic and methodological crisis of a new

era — the age of artificial intelligence.[7] In the current period of data abundance[8], the fundamental issue is the accurate interpretation of the information obtained and its effective application to concrete legal events. In this context, the mere possession of data no longer confers a competitive advantage in today's fast-moving corporate environment, for raw data — static, fragmented, and overwhelming in character — proves inadequate to manage the complexity confronting traditional legal governance.

Even data analysis methods once regarded as state-of-the-art have proven increasingly inadequate in navigating the legal labyrinths of the modern business world. Static governance models struggle to adapt to rapid regulatory shifts, technological transformations, market volatility, and the intricate interconnections of the contemporary business ecosystem. Organizations now demand '*living*' legal governance systems endowed with dynamic, adaptive, and real-time decision-making capacity — systems capable not merely of converting accumulated knowledge into context, but of extracting legal meaning at the semantic level from that context and applying it to concrete matters at hand. In response to this profound need, one of *Mecellem*'s most significant contributions — the details of which will be elaborated below — lies in its reconstitution of law: transforming it from a static codification into a *dynamic knowledge network* that places meaning-production, interconnectedness, and contextual sensitivity at its center. This knowledge network renders it possible for the legal system to evolve from a structure of fixed forms into a dynamic architecture capable of comprehending multidirectional flows of knowledge and channeling them into decision-making processes.

---

[7] In this article, the terms '*information age*' and '*artificial intelligence age*' are employed with a deliberate conceptual distinction. *The information age* refers to a broad historical period that began with the proliferation of computer technologies from the 1960s onward, encompassing developments such as the advancement of information technologies (IT), the widespread adoption of databases, the emergence of search engines, and the globalization of digital communication networks. *See* MANUEL CASTELLS, THE RISE OF THE NETWORK SOCIETY (INFORMATION AGE SERIES), Wiley-Blackwell (1996); Taner Kızılhan & Sevil Bal, *The Rise of the Network Society — The Information Age: Economy, Society, and Culture*, 7 CONTEMP. EDUC. TECH. (2016), *available at* 10.30935/cedtech/6177. *The artificial intelligence age*, by contrast, designates a more specific and contemporary period situated within *the information age* yet qualitatively distinct from it. This period has been defined particularly since the mid-2010s by the development of deep learning algorithms, the emergence of large language models (LLMs), and the proliferation of generative AI systems. The distinguishing feature of *the artificial intelligence age* is that knowledge is not merely stored and accessed, but is interpreted and made sense of by autonomous systems, and is employed in the production of new knowledge. *See* ERIK BRYNJOLFSSON & ANDREW MCAFEE, THE SECOND MACHINE AGE: WORK, PROGRESS, AND PROSPERITY IN A TIME OF BRILLIANT TECHNOLOGIES, W. W. Norton and Company (2014). This distinction is of particular significance in the legal context. Whereas *the information age* refers to the processes of digitizing legal knowledge, establishing electronic databases, and rendering legal research computer-assisted, *the artificial intelligence age* encompasses far-reaching transformations such as the performance of legal reasoning by autonomous systems of varying degrees, the integration of agentic artificial intelligence systems into legal decision-making processes, and the reconstruction of legal meaning through semantic protocols. Accordingly, *the information age* functions as a broader superordinate category that also encompasses *the artificial intelligence age*, while *the artificial intelligence age* designates the period that contains the historical and technological context of *Mecellem's semantic protocol* — a protocol that aims to respond to the epistemic crisis arising in the course of constructing legal semantic meaning. Both terms are used throughout this article with this conceptual distinction in mind.

[8] Eric Schmidt, former CEO of Google, famously observed in 2010 that the total amount of information produced from the beginning of civilization until 2003—approximately five exabytes—was, at that time, being generated every two days. *See* Eric Schmidt, Remarks at the Techonomy Conference (Aug. 4, 2010), as reported in *Eric Schmidt: Every 2 Days We Create As Much Information As We Did Up To 2003*, TechCrunch (Aug. 4, 2010), https://techcrunch.com/2010/08/04/schmidt-data/ (last visited Mar. 1, 2026). Today, this rate has increased even further. As of 2026, approximately 0.4 zettabytes (402.74 exabytes) of data are generated globally each day, meaning that the historical total of five exabytes referenced by Schmidt is now produced roughly eighty times per day. Global data production is projected to reach 221 zettabytes annually in 2026. *See* Naveen Kumar, *Big Data Statistics 2026 (Growth, Trends & Market Size)* (Dec. 29, 2025), https://www.demandsage.com/big-data-statistics/ (last visited Mar. 1, 2026).

The capacity of the revolutionary advances in information technologies witnessed over the past two decades to yield concrete benefits within the legal domain is contingent upon the development of *a new semantic protocol architecture* capable of enabling the construction of legal meaning. While these technological breakthroughs have produced transformative effects across numerous social science disciplines through statistical data analysis and quantitative modelling tools, legal science — by virtue of its pursuit of context, normative meaning, and interpretive integrity — has been unable to benefit from these technological possibilities to the same degree. This state of affairs has given rise to a contemporary recurrence of the historical predicament whereby law has, since the eighteenth century, been unable to fully partake in scientific progress. *Mecellem's semantic protocol* approach — fully applicable and realizable through the technologies of the present day — aims not merely to manage data, but to evolve it into a holistic system that comprehends, relates, and contextualizes information, transforming fragmented knowledge into coherent decision-making frameworks. In this evolutionary transformation, *Mecellem*'s semantic approach emerges as a critical reference point that both prevents the reproduction of the aforementioned historical rupture and constitutes the theoretical and technical foundation of *the new semantic protocol* — by virtue of its capacity to render visible the logical and contextual structures underlying legal norms and to render those structures integrable with artificial intelligence-supported systems.

However, the need for a protocol capable of reconstructing legal meaning through ontological and semantic layers cannot be regarded merely as a technical requirement brought forth by the age of artificial intelligence. On the contrary, this need represents the contemporary manifestation of a tension that legal methodology has carried throughout history and that has progressively deepened. At this juncture, in order to comprehend the methodological tension that law faces today and why this tension cannot be overcome through technical solutions alone, it becomes inevitable to look back at the historical trajectory of legal methodology. For although law has long operated through a qualitative, deductive, and largely static model of reasoning—a model that may have appeared functional within a particular historical context—it has increasingly reached its limits with the rise in norm density[9] and the growing complexity of social relations. This necessity, which confronts us today as the search for a new semantic protocol, is in fact the continuation of a problem that has persisted throughout the methodological history of law. Therefore, in order to understand what *the new semantic protocol* offered by *Mecellem* corresponds to, it is first necessary to examine how legal methodology has historically been shaped around a 'static' structure, and under what conditions this stasis began to be called into question.

[9] The increase in norm density throughout the 20th and 21st centuries is not unique to a particular country or legal system; rather, it represents a structural transformation observed across all modern legal orders. In the United States, the *Code of Federal Regulations (CFR)*, which contained approximately 18,000 pages in 1938,

***The Narrative of Stasis: The Historical Trajectory of Legal Methodology***

Law is traditionally defined as a normative discipline that determines the framework of the relationship between one subject and another subject, and between a subject and an object. In this context, law functions as the fundamental condition of all institutional, social, or economic interaction; indeed, even states of passivity or inaction produce meaning and consequences within the legal context. However, law harbors a perpetual tension between its effort to preserve its own internal coherence and normative continuity, on the one hand, and its obligation to respond to the dynamic nature of social reality, on the other. This structural dilemma transforms law from a static body of norms into a dynamic system that necessitates the constant rethinking of methodological approaches and practical applications.

The historical trajectory of legal knowledge demonstrates that the desire to regulate social relations within human communities and to produce knowledge of this order has taken root as a profound cognitive structure from quite early periods. In the traditional legal order, norms are systematically classified within a disciplinary framework; these norms possess a

---

had grown to over 185,000 pages by 2019. Similarly, the *Federal Register*, which published around 2,600 pages annually in 1936, exceeded 95,000 pages per year by 2016. *See* Clyde Wayne Crews Jr., *Tip of the Costberg: On the Invalidity of All Federal Agencies and Rules*, Competitive Enterprise Institute (2019), *available at* https://cei.org/studies/tip-of-the-costberg-on-the-invalidity-of-all-federal-agencies-and-rules/ (last visited Feb. 25, 2026). In the United Kingdom, while the number of *Public General Acts* enacted annually remained relatively stable between 1900 and 2020, the volume of *Statutory Instruments (SI)*, which constitute secondary legislation, increased exponentially. Whereas approximately 1,300 SIs were issued in 1950, this figure had risen to over 3,500 by 2020. *See* Christopher Watson, *Statutory Instruments*, House of Commons Library Briefing Paper No. 6509 (2020), *available at* https://commonslibrary.parliament.uk/research-briefings/sn06509/ (last visited Feb. 25, 2026). In Germany, the *Bürgerliches Gesetzbuch (BGB)*, which consisted of 2,385 articles at the time of its enactment in 1896, had grown to over 2,500 articles by 2020 through amendments and additions. Moreover, the number of *rechtsverordnungen* (legal regulations) issued at the federal level increased from approximately 2,800 in 1990 to over 3,200 by 2020. *See* Bundesamt für Justiz, *Übersicht über die Rechtsverordnungen des Bundes* (2021), *available at* https://www.bundesjustizamt.de (last visited Feb. 25, 2026). In Turkey, the number of laws enacted during the first legislative period of the Grand National Assembly of Turkey (1920–1923) was 7,481. This figure rose to 18,771 by the 28th legislative period (2023–2027). *See* TBMM, *Kanunlar*, *available at* https://www.tbmm.gov.tr/kanunlar/kanunlar_mevzuat.htm (last visited Feb. 25, 2026). Furthermore, the volume of secondary legislation has expanded at an even more accelerated pace. According to data from the Presidency's Directorate General for Legal Affairs and Legislation, the total number of regulations (*tüzük*), bylaws (*yönetmelik*), and communiqués (*tebliğ*) published in the *Resmî Gazete* between 1920 and 2023 exceeded 50,000. *See* T.C. Cumhurbaşkanlığı Mevzuat Bilgi Sistemi, *available at* https://www.mevzuat.gov.tr (last visited Feb. 25, 2026). These figures demonstrate that the increase in norm density is not merely a quantitative expansion, but a structural transformation in the nature of modern governance, wherein regulatory activity has shifted from primary lawmaking to the production of secondary legislation.

normative character not only through their regulatory function but also through their binding qualities. This normativity becomes manifest in the process of transforming abstract values into concrete rules, as moral ideals such as justice, equity, and equality lie at the foundation of law. Consequently, the nature of classical legal knowledge is essentially moral in foundation; its epistemological sources, moreover, have been predominantly legitimized through divine revelation, natural law, or the will of sovereign authority. Within this framework, traditional legal methodology likewise adopts a qualitative and deductive mode of reasoning.

Over time, as legal and institutional structures have become progressively more complex, the need for norms to regulate these structures has increased; this development has led both to the volumetric expansion of law and to the transformation of the intellectual processes that guide legal interpretation. Legal decision-making processes are no longer confined to the direct application of specific norms to cases in accordance with principles such as justice or equity; instead, they have assumed a more complex and instrumental structure involving the collection, correlation, and holistic synthesis of multiple streams of information. This transformation has resulted in the weakening of approaches that ground the source of law in subjective premises such as divine revelation or natural law; conversely, it has led to the greater adoption of positivist and pragmatic approaches that define law through 'reason', 'sovereign will', or 'measurable metrics'. These developments have rendered the systematic codification of law in the modern sense inevitable, even from the perspective of those who view law through the lens of religious law or natural law.

One of the most striking examples of this methodological transformation is Jeremy Bentham. The *utilitarianism* proposed by Bentham argues that law cannot rest upon a subjective foundation based solely on individual moral or divine references, but rather requires a more objective and measurable ground. However, this quest for objectivity does not entirely exclude the teleology of law; on the contrary, it defines the ultimate aim of law as 'maximizing social utility and minimizing pain'. Bentham supports this objective with a model to guide legislative processes, proposing a quantitative methodology he terms the '*felicific calculus*' (or *hedonic calculus*). This approach aims both to render the ontological foundation of law more objective and to situate its methodology within a more quantitative, analytical, and algorithmic framework. In the same vein, Bentham persistently advocated the idea of codification for the rational reconstruction of the legal system. In this respect, *utilitarianism* construes law not merely as a body of norms, but also as a functional instrument oriented toward optimizing social outcomes.[10] Nevertheless, despite all these methodological and theoretical innovations, Bentham's approach found limited impact in the practical legal applications of his time, while laying the groundwork for a fundamental paradigm shift in legal philosophy.

While Bentham's methodological opening sought to situate legal knowledge on a more objective, measurable, and utility-oriented ground, modern codification movements began to proliferate during the same period both in the West and in the Islamic world. Some of these codifications aimed at the systematization of existing legal norms, while others sought to transform natural law principles into positive rules at the constitutional level (for example, the American Constitution). Yet another group, as exemplified particularly by the Napoleonic Codes, reconstructed law entirely within a positivist structure by grounding it in the rationalist principles of Enlightenment thought. Within this diversity, the Mecelle developed a distinctive codification model in the Islamic world by transposing religiously derived legal norms into a modern legislative format.[11]

---

[10] JEREMY BENTHAM, AN INTRODUCTION TO THE PRINCIPLES OF MORALS AND LEGISLATION (Oxford: Clarendon Press, 1907).

However, while these codification efforts provided partial solutions to the practical needs of the period, insofar as they conceived law as a static body of norms, they failed to offer a sufficiently flexible response to rapidly changing social and economic dynamics. Consequently, it became evident that law needed to be reconsidered not merely through the systematization of rules, but also at the epistemic and methodological levels. Without this transformation, codification remained far from constituting a lasting solution in itself; law awaited treatment not as a structure frozen within static forms, but rather as a living and continuously self-reproducing system of meaning.

This awareness that law cannot be adequately addressed within static forms of codification has also brought with it a more fundamental quest at the methodological level. For insofar as approaches confined to the systematization of norms remained insufficient in grasping law as a living system of meaning, the solution was this time sought in new methods that emerged with the claim of rendering legal reasoning more rational, predictable, and measurable. The most salient expression of this quest is the *jurimetrics* approach, which aims to liberate law from the uncertainties of qualitative interpretation and reconstruct it through quantitative data, statistical models, and calculable metrics. *Jurimetrics* has constituted a significant turning point with its claim to provide a robust response to the epistemic and methodological problems of law; yet it has also inevitably raised the question of the extent to which it can encompass the contextual, interpretive, and multilayered nature of legal meaning. Precisely for this reason, in subsequent sections, *jurimetrics* is treated not merely as a progressive move, but also in conjunction with its own limitations.

### ***Methodological Turning Point in Law: Jurimetrics***

By the twentieth century, the rise of the social sciences and the proliferation of quantitative methods brought with them a fundamental methodological interrogation in the field of law as well. One of the intellectual precursors of this interrogation is the radical observation put forth by Oliver Wendell Holmes Jr. in *The Common Law*: '*The life of the law has not been logic: it has been experience.*'[12] According to Holmes, legal rules are shaped not by logical inferences with mathematical certainty, but rather by the felt necessities of the time, the prevalent moral and political values of society, intuitions of public policy, and the prejudices—often not even conscious—of judges. In this sense, law is not a closed system composed of abstract norms; it is a living phenomenon continuously reproduced through human experience within historical context. Holmes's approach is closely related to the *pragmatist* philosophy flourishing in the intellectual climate of America at the time, and it enabled Holmes to view law as a kind of '*experimental laboratory*'.[13] This approach offered a critical perspective that centered the sociological, experiential, and contextual character of law against the positivist understanding that attempted to explain law through purely formal logic, and it laid the theoretical foundation for the idea that '*law is a living organism*'.

---

[11] JOSEPH SCHACHT, AN INTRODUCTION TO ISLAMIC LAW, OXFORD UNIVERSITY PRESS (1982); LEWIS BERNARD, ARABS IN HISTORY, Oxford University Press (1992); Avi Rubin, *Modernity as a Code: The Ottoman Empire and the Global Movement of Codification*, 9 J. ECON. & SOC. HIST. ORIENT 828-856 (2016).

[12] OLIVER WENDELL HOLMES, JR., THE COMMON LAW (G. Edward White ed., Belknap Press of Harvard Univ. Press 2009)

However, the idea that law is a variable phenomenon within historical and social context is not solely a product of twentieth-century American legal thought. Holmes's emphasis on experience was formulated as a normative principle at an earlier period on the other side of the Atlantic. The maxim '*ezmanın tagayyürü ile ahkâmın tagayyürü inkâr olunamaz*' (the change of rules with the change of times cannot be denied)[14], which Ahmet Cevdet Pasha incorporated into the *Mecelle*, set forth not merely as a theoretical proposition but as a binding norm within positive law that law can transform according to the changing conditions of time. While Holmes's approach, which conceived of law as *a living organism*, constructed the intellectual ground for a long-term transformation of mentality, Cevdet Pasha integrated the same dynamism into the concrete legal order through codification, making mutability an intrinsic principle of the system. In a sense, while on one shore of the ocean the idea that law is shaped by experience was still developing at the level of theoretical debate, as an intellectual rupture; on the other shore, this understanding had been incorporated into written law by the state and acquired normative bindingness. Thus, the notion that law is a living reality ceased to be merely a philosophical thesis and became an institutional principle of regulation.

This perspective acquired a more systematic and methodological form with Lee Loevinger's *jurimetrics* approach in the mid-twentieth century. In his 1949 article '*Jurimetrics—The Next Step Forward*'[15], Loevinger characterizes law as one of the greatest anomalies of its age: According to him, law, which ought to serve as a guide for conduct in society, has increasingly become an incomprehensible, archaic, and insular '*priestly profession*'; consequently, it has lost both its function and its public legitimacy in the eyes of the populace. He locates the source of the problem not in society, but in jurists who preserve law within metaphysical abstractions and exempt it from critical examination. For thousands of years, law has been attempted to be defined through concepts such as divine command, natural law, reason, justice, or sovereign will; however, these efforts have produced an indeterminate, contradictory, and often sterile conceptual cycle rather than explaining the actual operation of law.

---

[13] Holmes's connection to *pragmatist* philosophy stemmed from his intellectual interactions at the Metaphysical Club, founded in Cambridge, Massachusetts, in 1872. *See* LOUIS MENAND, THE METAPHYSICAL CLUB: A STORY OF IDEAS IN AMERICA, Farrar, Straus and Giroux (2002); FREDERIC R. KELLOGG, OLIVER WENDELL HOLMES JR. AND LEGAL LOGIC, The University of Chicago Press (2018); SUSAN HAACK, THE PRAGMATIST, IN THE PRAGMATISM AND PREJUDICE OF OLIVER WENDELL HOLMES JR., Lexington Books, 169-189 (Seth Vannatta ed., 2019); Thomas C. Grey, *Holmes and Legal Pragmatism*, 4 STAN. L. REV. 41 (1989), *available at* https://doi.org/10.2307/1228740. Holmes's observation that '*the life of the law has not been logic: it has been experience*' is a direct reflection of these pragmatist interactions.

[14] Mecelle-i Ahkâm-ı Adliyye Clause 39. *See* HOCAEMİNEFENDİZADE ALI HAYDAR EFENDI, DERERÜ'L-HÜKKÂM ŞERHU MECELLETİ'L-AHKÂM SETİ (2020).

[15] Loevinger, *supra* note 5. *Also see* Oliver Wendell Holmes, *The Path of the Law*, 10 HARV. L. REV. 457 (1897).

According to Loevinger, although judicial decisions appear to be justified through logical *syllogisms*[16], in practice the outcome is often determined in advance in an intuitive—and in some cases arbitrary—manner; legal justification is then constructed through the subsequent selection of concepts and terms that conform to this outcome. For this reason, even the question 'what is the nature of law?' is misleading; for law is not a fixed and abstract essence, but rather a social practice that acquires meaning only within the context of use. The preoccupation of legal philosophy and doctrine with metaphysical speculations for centuries has, according to Loevinger, left law behind scientific progress and rendered the true dynamics of legal reasoning invisible.

Proceeding from this critique, Loevinger's solution has been to bring law together with disciplines such as statistics, information processing, logical modeling, psychology, sociology, and cybernetics, thereby transforming it into a measurable, observable, and testable social science. *Jurimetrics*, which envisions the jurist of the future as *'the man of statistics and the master of economics'* inspired by Holmes, has focused particularly on the problem of '*access to accurate information*'; it has produced significant achievements in areas such as the systematic classification of legal documents, providing rapid access to court decisions through computerized indexing, and the analysis of judicial patterns. Within its own historical context, this approach has indeed constituted '*a next step forward*' by bringing law onto an analytical ground supported by quantitative tools.

Against this backdrop, *jurimetrics* has made a significant contribution to the rationalization of legal reasoning and its support through quantitative tools. This approach, which provides the opportunity to examine legal decision-making processes through data-driven analyses, has aimed to enhance the predictability of law and strengthen consistency in decision-making processes. However, while *jurimetrics* has largely reduced legal reasoning to a problem of '*finding, classifying, and measuring information*', it has relegated the contextual, interpretive, and intuitive dimensions of legal experience pointed out by Oliver Wendell Holmes to a secondary position. This approach has brought with it the risk of reducing the meaning-production processes that determine the living character of law to measurable data and statistical patterns. Indeed, systems that process only data without grasping context cannot reproduce the historical, social, and normative fabric of law; this situation demonstrates that the *jurimetric* approach alone can provide neither sufficient semantic depth nor a consistent epistemic foundation. This limitation reflects not a theoretical deficiency of *jurimetrics* but rather the technical and epistemic conditions of the early information age in which it emerged and, in light of the deep penetration of modern technologies into institutional practices, calls for a more fundamental reconsideration of how legal knowledge should be organized and operationalized.

---

[16] In his work, a classic of methodology, Larenz comprehensively examines the logical foundations of legal interpretation and decision-making processes. The author addresses in detail the structure of legal *subsumption*, how the major premise (*norm*) is constructed, and the interpretive problems in matching the minor premise (*fact*) with the *norm*. The classical decision-making schema in law is as follows: Major premise (*rule*): 'If conditions X exist, consequence Y arises.' Minor premise (*fact*): 'The concrete case bears conditions X.' Conclusion: 'Therefore, consequence Y applies.' This structure, based on Aristotelian logic, presents an extremely rational and objective appearance. The decision is presented as if: 'Rules have been automatically applied to the case and the conclusion has necessarily followed.' Larenz emphasizes the creative role of the judge behind the 'automatic' and 'necessary' appearance of the decision and offers a critical perspective on the classical decision-making schema in law. *See* KARL LARENZ, METHODENLEHRE DER RECHTSWISSENSCHAFT (3d ed., Claus Wilhelm Canaris ed., 1995); *also see* Hans Kelsen, *Pure Theory of Law* (Max Knight trans., University of California Press 2d ed. 1970) (1960). On the other hand, in his renowned work titled *The Nature of the Judicial Process*, Cardozo addresses how judicial reasoning operates, the limits of judicial discretion, and how law evolves as a living organism, and with this work, he constitutes one of the most important links in the legal realism tradition initiated by Holmes. The tension between the actual operation of the judicial decision-making process and formal logic is precisely the issue on which Cardozo's work focuses. *See* BENJAMIN N. CARDOZO, THE NATURE OF THE JUDICIAL PROCESS, Yale University Press (1921).

***From Jurimetrics to Semantic Protocol: New Episteme and Methodology in Legal Knowledge***

With the integration of modern technologies into daily business practices, the limitations of conventional corporate governance models have become increasingly visible. In the face of rapidly changing regulatory frameworks, multi-layered institutional relationships, and increasing data intensity, treating legal knowledge merely as a set of static norms severely restricts the capacity to generate actionable insights. In this context, it is evident that dynamic legal governance has become not only a practical choice but also a necessity. Such an understanding of governance inevitably requires a collective rethinking of the ontological, epistemological, and methodological foundations of legal knowledge.

For knowledge cannot be conceived as a purely mental construction independent of the object of inquiry. Knowledge becomes meaningful and verifiable only to the extent that it establishes correspondence with the external reality it represents. This acknowledgment is consistent with classical theories of knowledge that presuppose knowledge to be both a mental structure and a representation of objective reality.[17] In terms of legal knowledge, this

[17] This view resonates with al-Fârâbî's epistemological framework, which articulates knowledge as a correspondence between the intellect and external reality and continues to provide a conceptually robust and still applicable foundation for understanding knowledge. According to him, knowledge is defined on the basis of the correspondence of the judgment in the mind to the object in the external world; when this correspondence is established, 'truth' is attained, and when correspondence is absent, 'false assent' occurs. This understanding represents the purest form of the classical correspondence theory. Al-Fârâbî divides knowledge into two main categories: *tasawwur* (*concepts*) and *tasdīq* (*judgments*, *propositions*). *Tasdīq* means believing that something is true, and when this belief fully coincides with its counterpart in the external world, 'certain knowledge' (*yaqīn*) emerges. Al-Fârâbî has demonstrated that *tasdīq* possesses different degrees. Things assented to through 'non-certain *tasdīq*' are either reasonings made with premises previously known and accepted by the community, or inductions in which one is not certain of having encompassed all particulars. Certainty is possible in 'complete assent' regarding what is true, and 'certain knowledge' requires a complete correspondence between what the individual believes and

situation means that normative texts are not sufficient in themselves; it necessitates the comprehension of a broader domain of reality that encompasses the factual world, actors, and network of relationships with which these texts come into contact.

In this regard, the contemporary legal and social context necessitates not only the correct acquisition of knowledge but also the profound interpretation of this knowledge within a multi-layered network of relationships. The fact that quantitative methods in the social sciences fall short of being explanatory on their own, and that the results obtained are generalized in a manner detached from context, has strengthened criticisms that law cannot be reduced to mere statistical data. This situation necessitates reconsidering law not merely as a set of rules but as a domain of meaning production with its own distinctive grammar and alphabet.

In light of the foregoing, it is necessary to briefly address the choice of the article's title. The phrase '*Towards a New Grammar of Reasoning for Artificial Legal Intelligence*' is not merely a rhetorical preference; rather, it has been deliberately selected as a conceptual summary of the thesis advanced in this article. In linguistics, *grammar* constitutes the structural framework that governs the rules of meaning production, syntactic relations, and modes of expression; to speak a language is, in essence, to internalize its grammar. Legal reasoning, in much the same sense, requires a grammar: a structural framework that determines how norms, facts, actors, and contexts are to be related to one another, in what order of priority they are to be evaluated, and with what epistemic weight they are to be processed. However, such a grammar cannot be reduced either to the codification of positive legal norms or to the extraction of statistical patterns, as both approaches remain structurally inadequate to capture the contextual, interpretive, and multi-layered nature of legal meaning. The *semantic protocol* proposed here is therefore conceived as *a new grammar of reasoning* designed precisely to address this deficiency. Composed of ontologically defined categories of entities, epistemically differentiated layers of knowledge, and methodologically orchestrated mechanisms of meaning production, this grammar seeks to establish a shared language of meaning between *the human jurist* and *the artificial intelligence agent*. This language is neither merely the encoding of norms nor the statistical matching of vector

---

the thing in the external world. One of al-Fârâbî's most significant observations is that certainty and truth can never occur in 'false assent'; for a false judgment, being incompatible with external reality, can possess neither the quality of certainty nor that of truth. *See* AL-FÂRÂBÎ, KITÂB AL-BURHÂN, Klasik Yayınları, trans. Ömer Türker, Mahir Alper, 6th ed. (2023). As a thinker who lived in the mid-ninth century CE, al-Fârâbî addressed the fundamental problems of epistemology with a depth that remains valid today. His understanding of *burhân* (*demonstration*) refers to reasoning conducted with premises that are consistent and assented to as true, and indicates the path to attaining certain, necessary, and demonstrated knowledge. Al-Fârâbî's emphasis that the term 'knowledge' should be attributed to 'necessary certain knowledge' is a fundamental principle also adopted by contemporary philosophy of knowledge. These issues, still debated approximately eleven centuries after him, clearly demonstrate how firmly al-Fârâbî laid the foundations of the intellectual world. Goldman's work *Epistemology and Cognition*, for instance, sets forth the relationship of knowledge to cognitive processes and the *reliabilist* approach, and addresses the question of how knowledge is produced and verified through al-Fârâbî. Goldman's *reliabilist* approach functions as a bridge that directly corresponds with Al-Fârâbî's understanding of 'certain knowledge' and brings it into dialogue with the modern analytic epistemology literature. *See* ALVIN I. GOLDMAN, EPISTEMOLOGY AND COGNITION, Harvard University Press (1986).

representations; rather, it is conceived as a holistic architecture of knowledge constructed through conceptual hierarchy, contextual relations, and semantic topology. Accordingly, the expression '*a new grammar of reasoning*' refers not merely to enabling *artificial legal intelligence* to produce faster or more accurate outcomes, but to a foundational framework that makes it possible to reconstruct legal meaning—on the basis of ontological constants, epistemic layers, and methodological weighting—in a manner that is context-sensitive and renewed with each inquiry. In this respect, the proposed approach offers an integrated model that brings together both the theoretical foundations and the operational protocol of legal reasoning in the age of artificial intelligence.

As a natural extension of this framework, the proposed approach rests on the premise that legal knowledge must be addressed conjointly within the triangle of ontology, epistemology, and methodology.[18] While this approach takes seriously the critique that *jurimetrics* directs at classical legal thought, it aims to reposition legal reasoning — without excluding its contextual, interpretive, and multi-layered character — within a systematic order of knowledge. To this end, the article proposes a more holistic epistemic framework, one that holds that legal meaning can be constructed not merely through measurability, but through ontological structure, semantic relations, and contextual justification. The objective is not a rigid formalization that excludes interpretive elements, but rather the grounding of those very elements within a semantic architecture that is auditable, traceable, and reproducible.

This approach also exposes the inadequacy of treating law exclusively through the lens of data access and measurability. While the contributions of *jurimetrics* to rationalizing legal reasoning and supporting it with quantitative tools cannot be denied, the question of how to preserve the contextual, relational, and interpretive dimensions of legal meaning has remained largely unanswered. The proposed framework neither reduces legal reasoning to purely algorithmic operations nor abandons it to unverifiable subjective assessments; rather, it offers a foundation upon which meaning can be reconstructed through ontological structure, an order of knowledge, and semantic relations. In this respect, the article advances a methodological transition proposal that both transcends the limits of *jurimetric* reductionism and enables the kind of deep contextual meaning production that contemporary artificial intelligence systems require.

[18] The concepts of 'ontology', 'epistemology', and 'methodology' as employed in this article — terms that classical philosophy uses in a manner encompassing metaphysics — have been redefined in a manner consistent with Giovanni Sartori's warning against concept stretching, and are used herein as functionally constructed concepts for the purpose of defining the types of entities, layers of knowledge, and methods employed by artificial intelligence systems in legal reasoning. *See* Giovanni Sartori, *Concept Misinformation in Comparative Politics*, 64 AM. POL. SCI. REV. 1033 (1970). As used in this article, *ontology* refers to the legal entity categories recognized by an AI model and the structural relations among them; *epistemology* refers to the sources from which, and the manner in which, knowledge about those entities can be produced and validated; and *methodology* refers to the tools and procedures through which that knowledge is processed and the chains of reasoning within which it is deployed in the context of agentic AI. When this triadic relationship is not established, the risk of hallucination — along with numerous other accuracy and reliability failures commonly observed in conventional AI applications — increases substantially. *The semantic protocol of Mecellem*, by contrast, is designed precisely to eliminate or minimize that risk through its ontology–epistemology–methodology architecture. In this regard, the references made throughout this article to al-Fârâbî and other philosophers and intellectual traditions, and to the concepts they employed, are not intended to claim that those theories and concepts have been reconstructed wholesale within the *Mecellem* framework; rather, they are offered as analogical conceptual references for *Mecellem*'s order of knowledge and semantic architecture. Al-Fârâbî's distinction between existence and essence, his ordering of knowledge, and his concept of *burhān* are positioned as borrowed instruments for constructing the conceptual skeleton of the 'correct knowledge — correct context — correct meaning — correct ruling' chain in AI-based legal knowledge systems.

In order to articulate the constitutive elements of the proposed approach in a more systematic manner, the ontological foundations, epistemic layers, and methodological operation of legal knowledge must be treated as distinct yet interrelated planes of analysis. At the ontological plane, the question concerns what exists within the legal universe: what kinds of entities are legal concepts, norms, actors, and events to be treated as, and how are the essences of these entities to be determined? At this stage, drawing inspiration from the Farabian order of knowledge, *Mecellem* aims to map the conceptual landscape of legal reality by taking the distinction between existence and essence as its foundational axis.

At the epistemological plane, the question centers on the problem of how we can know these entities. The source of what 'exists' — that is, of legal knowledge — is not limited to public and normative texts such as legislation, case law, and doctrine (the *normative corpus*); it also encompasses a contextual body of knowledge (the *enterprise corpus*) consisting of institution-specific practices, contracts, correspondence, factual circumstances, and inter-actor relationships that the *normative corpus* touches upon. Although these two layers of knowledge differ in their epistemic status, they function as mutually complementary and necessary elements in the production of legal meaning. This complementarity, however, does not rest on an assumption of fixed and equal epistemic value; rather, it rests on a dynamic evaluative process in which variable epistemic weights are assigned to each corpus depending on the nature, context, and purpose of each query. *Mecellem* aims to offer a more layered epistemic framework — one that does not reduce the accuracy of legal knowledge to its formal source or normative authority alone, but instead constructs meaning by taking into account the relative value that different knowledge sources carry within context, assessed through the criteria of coherence, justificatory capacity, contestability, and contextual appropriateness.

The philosophical foundation of this epistemic framework is clearly visible in the distinction al-Fârâbî draws in his *Attainment of Happiness* between the principles of instruction (*mebâdi' al-ta'lîm*) and the principles of existence (*mebâdi' al-wujûd*).[19] According to al-Fârâbî, if the initial items of knowledge concerning a genus of entities possess the conditions capable of leading the inquirer to certain truth, those items constitute principles of instruction; but if those same principles also coincide with the causes of existence of the species contained within that genus, the principles of instruction become identical with the principles of existence, and the demonstrations produced explain not merely 'whether a thing exists' but also 'why it exists'. *Mecellem's semantic protocol* aims at precisely this alignment: when the structural representation of legal knowledge — that is, the ontological design, the taxonomic hierarchy, and the semantic relations — is brought into alignment with the normative and contextual reality of legal entities, the knowledge architecture ceases to be merely descriptive; it is transformed into an explanatory and generative epistemic force, thereby laying the ground for epistemic justice.

[19] AL-FÂRÂBÎ, THE ATTAINMENT OF HAPPINESS, in *Al-Farabi's Philosophy of Plato and Aristotle* (Muhsin Mahdi trans., Cornell Univ. Press 1969). *Also see* FÂRÂBÎ, MUTLULUĞUN KAZANILMASI [THE ATTAINMENT OF HAPPINESS] (İş Bankası Kültür Yayınları 2025).

The methodological plane answers the question of how these ontological and epistemological commitments are to be operationalized in practice. Traditional knowledge retrieval methods, or vector-based systems relying solely on textual similarity, are structurally inadequate to capture the multi-layered and interlocking referential relationships between the *normative and enterprise corpora* — and in particular to capture *multi-hop legal reasoning*. At this juncture, *Mecellem's semantic protocol*, built on the integration of artificial intelligence, ontologies, and knowledge graphs, emerges — as will be elaborated below — not as a singular technical solution, but as a higher-order methodological framework that orchestrates different meaning-production and knowledge-access methods according to context.

Within this framework, legal texts are semantically segmented according to document type and function (*semantic chunking*); yet what is ultimately determinative is the method by which, and the weight with which, these semantic representations are operationalized. As will be elaborated in detail in the subsequent sections of this study, *Mecellem*'s methodology positions knowledge graphs, ontological inference, vector-based similarity, textual matching, and other meaning-production mechanisms not according to a fixed hierarchy, but dynamically, in accordance with the nature of the query. In other words, what is weighted in *Mecellem* is not merely the knowledge sources or individual units of knowledge, but also the methodological priorities governing how knowledge is to be produced and how it is to be accessed. For instance, in the case of an explicit and structural knowledge request — such as the date, parties, or legal status of a particular contract — knowledge graph and ontological inference mechanisms become nearly exclusively determinative, while methods such as vector-based similarity may be entirely deactivated. By contrast, when the legal question at hand is interpretive, analogical, or precedent-based, semantic proximity, contextual pattern recognition, and vector representations gain greater weight. The criteria by which such differentiations occur, and the manner in which the system manages these methodological shifts, will be elaborated in the subsequent sections through illustrative cases and architectural explanations.

Accordingly, in *Mecellem*, between the linguistic formulation of a need into a ‘question’ and the access to the ‘answer’ *corpus* in which that question will find its response, there lies a methodological structure that is not static, but dynamic, contextual, and orchestrated. The ontological and conceptual nature of the question determines the methodological framework to be employed in the construction of the answer; how that framework is constituted and according to which principles it operates will be addressed in detail in the sections that follow. Likewise, the components of this structure, its decision mechanisms, and its knowledge-production processes will be analyzed systematically in the subsequent sections of this article.

This multiple methodological flexibility enables legal meaning to be constructed anew in each query, in a manner specific to its context. Rather than confining meaning-production to a single technical paradigm, *Mecellem* designs it as a dynamic, methodological, and semantic flow that activates different methods in accordance with epistemic needs. In this way, legal knowledge ceases to be a mere aggregate of isolated data points and is instead produced within a meaning architecture in which it is clearly traceable, justifiable, contestable, and auditable as to which mode of knowing is operative in which query. Until quantum computing reaches a practical and scalable level — and given the current technological constraints, particularly the context window limitations of transformer architectures and their hardware-related restrictions — it has been established through our experimental studies that methods falling outside the approach we propose encounter significant difficulty in producing the same level of consistency and contextual integrity.

This approach simultaneously points to a fundamental requirement for agentic artificial intelligence systems expected to conduct legal reasoning in an autonomous, goal-directed, and multi-step manner. In particular, the capacity expected of agentic AI systems — namely goal-directed planning, intermediate result generation, feedback-based self-correction, and the integration of temporally expanding knowledge domains —[20] demands not merely data access, but the orchestration of context and the coordination of methodologically weighted semantic layers. For this reason, the approach we propose emerges as a necessity capable of preserving contextual continuity under current architectural constraints and sustaining epistemic consistency throughout decision-making processes.

The operational counterpart of this necessity is manifested in the rational definition of the '*allowed query list*' thresholds that determine within which boundaries agentic systems will operate. Determining which data layers an agent can access under which conditions, which query types it can execute, and at which epistemic weight levels it can produce decisions becomes meaningful only within a pre-structured semantic and ontological order. In the absence of such a protocol, agent behavior tends toward producing random or unlimited queries within the broad data universe, disconnected from normative and contextual hierarchies; this situation both complicates the *epistemic traceability* of legal reasoning and renders decision-making processes economically unsustainable.[21] The *Mecellem semantic protocol* creates a '*query governance*' layer by defining *allowed query* areas through principles of ontological commitment, normative weight, and contextual consistency; thus, agentic architectures are enabled to operate within a legal reasoning-aligned, traceable, and auditable operational space.

[20] The architectural components of agentic AI consist of four critical layers. Planning and reasoning components provide the capacity to decompose complex tasks into manageable subtasks, determine the sequence and dependencies among these subtasks, and construct executable plans that account for resource constraints and environmental uncertainties. The *ReAct* framework enhances reasoning quality and interpretability by enabling agents to explicitly articulate their reasoning steps before taking action. *See* Shunyu Yao, Jeffrey Zhao, Dian Yu, Nan Du, Izhak Shafran, Karthik Narasimhan, Yuan Cao, *ReAct: Synergizing Reasoning and Acting in Language Models* (arXiv:2210.03629), *available at* https://arxiv.org/abs/2210.03629. Tool use and function-calling mechanisms enable agents to interact with external tools (APIs, databases, web search engines, code execution environments). *See* Shishir G. Patil, Tianjun Zhang, Xin Wang, Joseph E. Gonzalez, *Gorilla: Large Language Model Connected with Massive APIs*, (arXiv:2305.15334), *available at* https://arxiv.org/abs/2305.15334; Timo Schick, Jane Dwivedi-Yu, Roberto Dessì, Roberta Raileanu, Maria Lomeli, Luke Zettlemoyer, Nicola Cancedda, Thomas Scialom, *Toolformer: Language Models Can Teach Themselves to Use Tools*, (arXiv:2302.04761), *available at* https://arxiv.org/abs/2302.04761. Memory systems and context management are of critical importance for maintaining consistency in long-horizon tasks; modern agentic systems integrate multiple memory types — such as working memory, episodic memory, semantic memory, and procedural memory — thereby exhibiting learning and adaptation capabilities that exceed what individual agents could achieve. Finally, reflection and self-improvement mechanisms enable agents to evaluate their past performance, identify errors, and generate guidance for improvement; research demonstrates that structured reflection significantly enhances performance in subsequent attempts. *See* Noah Shinn, Federico Cassano, Ashwin Gopinath, Karthik Narasimhan, Shunyu Yao, *Reflexion: Language Agents with Verbal Reinforcement Learning*, Advances in Neural Information Processing Systems, (arXiv:2303.11366) *available at* https://arxiv.org/abs/2303.11366. *See* also Lei Wang, Chen Ma, Xueyang Feng, Zeyu Zhang, Hao Yang, Jingsen Zhang, Zhiyuan Chen, Jiakai Tang, Xu Chen, Yankai Lin et al., *A Survey on Large Language Model Based Autonomous Agents*, (arXiv:2308.11432) *available at* https://arxiv.org/abs/2308.11432?utm_source=blog.amorelli.tech&ref=hackernoon.com.

Within this framework, the *Model Context Protocol* (*MCP*)[22] is positioned as an execution protocol in the methodological plane, determining how the context is activated.

---

[21] Indeed, for a concrete production environment example of this risk, *see* Teja Kusireddy, *We Spent $47,000 Running AI Agents in Production. Here's What Nobody Tells You About A2A and MCP*, Towards AI (Medium), Oct. 16, 2025, *available at* https://pub.towardsai.net/we-spent-47-000-running-ai-agents-in-production-heres-what-nobody-tells-you-about-a2a-and-mcp-5f845848de33] (last visited Mar. 9, 2026). The author recounts how a multi-agent system comprising four *LangChain* agents, deployed in production, entered an infinite loop due to the absence of defined termination conditions in the agent-to-agent coordination layer. The loop ran undetected for 11 days; weekly costs escalated from $127 to $18,400, culminating in a total bill of $47,000. The author attributes this incident directly to the lack of infrastructure layers defining query boundaries, token budgets, and cost governance in agent architectures. Additionally, for a documented instance of agent behavior generating irreversible financial consequences absent contextual memory and epistemic boundary management, *see* Omkar Godbole, *AI Bot's Tipping Blunder Hands $250,000 Memecoin Pile to X Sad Story Poster*, CoinDesk, Feb. 23, 2026, *available at* https://www.coindesk.com/markets/2026/02/23/ai-bot-s-tipping-blunder-hands-usd250-000-memecoin-pile-to-x-sad-story-poster](https://www.coindesk.com/markets/2026/02/23/ai-bot-s-tipping-blunder-hands-usd250-000-memecoin-pile-to-x-sad-story-poster) (last visited Mar. 9, 2026). The autonomous crypto trading agent 'Lobstar Wilde', developed by OpenAI engineer Nick Pash, intended to transfer 4 SOL (approximately $320) to a user but, due to a session crash causing loss of contextual memory, instead transferred its entire LOBSTAR token holdings (valued at approximately $441,000). The recipient sold the tokens within about 15 minutes, netting $40,000 in cash. According to Pash's published technical retrospective, the root cause was the agent's inability to reconstruct contextual information regarding wallet balance and token allocation upon session restart—an operational manifestation of the absence of epistemic traceability and context governance mechanisms.

[22] The *Model Context Protocol* (*MCP*) is an open standard developed by Anthropic to facilitate seamless integration between large language models (*LLMs*) and external data sources or tools. By standardizing context exchange, MCP enables AI agents to interact with diverse systems in a secure and scalable manner. The protocol introduces fundamental primitives such as tools, resources, prompts, and sampling to establish a structured interaction layer between models and environments. MCP's modular architecture supports real-time, bidirectional communication over common transport protocols, including HTTP and WebSocket. The adoption of MCP has enabled AI systems to become more flexible, interactive, and autonomous within enterprise and developer workflows. As context-aware intelligence becomes foundational, MCP offers a scalable pathway for building next-generation agentic applications. For comprehensive information on MCP's architecture, components, and agentic AI capabilities, *see* Xinyi Hou et al., *Model Context Protocol (MCP): Landscape, Security Threats, and Future Research Directions*, arXiv (Mar. 30, 2025) (revised Oct. 7, 2025), https://doi.org/10.48550/arXiv.2503.23278; Vallikranth Ayyagari, *Model Context Protocol for Agentic AI: Enabling Contextual Interoperability Across Systems*, 11 INT'L J. COMPUTATIONAL & EXPERIMENTAL SCI. & ENG'G 3 (2025), https://doi.org/10.22399/ijcesen.3678; Partha Pratim Ray, *A Survey on Model Context Protocol: Architecture, State-of-the-Art, Challenges and Future Directions*,

Ontologically defined entities and epistemologically stratified knowledge sources are opened to agentic reasoning in a specific contextual configuration for each query via *MCP*. *Mecellem* embeds this decision-making mechanism within a philosophically grounded, traceable, and auditable semantic architecture, aiming to render legal meaning production sustainable at both human and machine levels.

In this context, the transition from *jurimetrics* to the *semantic protocol* represents not merely a technical advancement but a fundamental transformation of the theoretical and methodological foundation of legal knowledge. While *jurimetrics* addresses the age-old problem in law primarily around 'access to accurate information' and the issue of measurability, the *semantic protocol* proposed by *Mecellem* centers on the questions of how to construct the relevant corpus, what exists within this corpus, and where and by what method the 'existing' can be located within the corpus. Thus, law ceases to be a static aggregate of norms and is instead reconceptualized as a dynamic and explainable architecture of meaning, reconstructed through ontological structure, semantic relations, and contextual justification.

The *Mecellem semantic protocol* represents not a particular technical solution, but rather a holistic methodological orientation aimed at the cognitive architecture required by contemporary artificial reasoning systems in response to a longstanding crisis. This orientation brings together, within a unified epistemic framework, neurosymbolic approaches that enable the representation and structural coherence of knowledge (particularly legal ontologies and knowledge graphs), agentic artificial intelligence architectures that facilitate context-sensitive decision-making and multi-step reasoning capacity, and semantic protocols that enable the dynamic reconstruction of legal meaning in each query.

The distinguishing claim of *Mecellem* lies in its positioning of these elements not as tools articulated to one another, but within a coherent hierarchy established among ontological grounding, epistemic justification, and methodological operation. In this way, legal knowledge is reduced neither to a mere sequence of symbolic rules nor solely to the output of statistical patterns; rather, it becomes a structure of meaning that is reconstructed contextually in each legal query, justifiable and verifiable. In an architecture where *contextual grounding* is not ensured, the generated outcome lacks epistemic integrity; the system produces a partial and reductive assessment by processing only a selected sub-context rather than the multi-layered context of the case.

Another critical consequence of this approach resides in the structural solution it furnishes to the '*black box*' problem[23] that frequently surfaces in the deployment of artificial

---

TechRxiv (Apr. 18, 2025), https://doi.org/10.36227/techrxiv.174495492.22752319; Mohammed Mehedi Hasan et al., *Model Context Protocol (MCP) at First Glance: Studying the Security and Maintainability of MCP Servers*, arXiv (June 16, 2025), https://doi.org/10.48550/arXiv.2506.13538; Jon Ander Oribe & Quantum Mads, *The Model Context Protocol (MCP): Emergence, Technical Architecture, and the Future of Agentic AI Infrastructure*, ResearchGate (Oct. 20, 2025) (preprint), https://doi.org/10.5281/zenodo.17390299.

[23] With the proliferation of agentic AI systems in particular, challenges pertaining to explainability and accountability have intensified considerably. The 'black box' nature of many advanced AI models, especially those predicated on deep learning, engenders a fundamental deficit of trust. These systems can be so complex

intelligence within the legal domain. Legal ontologies and knowledge graphs reinforce the '*explainable artificial intelligence*' (*XAI*) paradigm by rendering the decision logic of AI-based reasoning processes traceable, justifiable, and contestable. Legal outputs thereby transcend their status as mere technical artifacts that yield results; they become transparently and contestably grounded in conjunction with their normative foundations, precedential relations, and contextual justifications.

In this context, *Mecellem* proposes a common epistemic foundation that enables legal reasoning to be conducted by both human subjects and cognitive agents[24], while simultaneously seeking to furnish a holistic response to the problems of explainability, contextual coherence, and normative legitimacy confronted by agentic artificial intelligence systems in the legal domain.

Within this framework, the article proceeds in three stages. First, it examines the methodological transformation of legal knowledge and identifies the inherent limitations of *jurimetrics* as an analytical paradigm. Second, it elaborates upon the ontological, epistemic, and methodological architecture of *Mecellem*, situating it within both classical legal philosophy and contemporary formal ontology. Third, it demonstrates—through a concrete practical application—the normative and technical possibilities this framework affords for the development of agentic and neurosymbolic artificial intelligence systems in legal contexts. To ensure proper comprehension of the theoretical framework advanced herein, it is necessary at the outset to articulate the deliberate methodological choices and scope limitations that have informed the preparation of this article.

---

that even their creators cannot fully elucidate the reasoning underlying their outputs. Indeed, the operational logic of contemporary artificial intelligence systems remains incompletely understood even by their developers. As Dario Amodei, CEO of Anthropic, has observed, 'the fact that we do not understand how our own AI creations work represents an unprecedented situation in the history of technology.' Dario Amodei, *The Urgency of Interpretability* (Apr. 2025), https://www.darioamodei.com/post/the-urgency-of-interpretability (last visited Feb. 22, 2026). This challenge is even more pronounced for agentic systems than for traditional AI, because whereas a non-agentic system provides information that a human interprets and acts upon (thereby distributing responsibility), an agentic system makes and executes autonomous decisions without intervening human judgment, with consequences stemming directly from system behavior. *See* also Cynthia Rudin, *Stop Explaining Black Box Machine Learning Models for High Stakes Decisions and Use Interpretable Models Instead*, 1 Nature Machine Intelligence 206 (2019), https://doi.org/10.1038/s42256-019-0048-x.

[24] In our study, the concepts of '*human subject*' and '*cognitive subject*' are employed with deliberate terminological distinction. While the *human subject* denotes the person possessing legal capacity for rights and obligations, to whom volition may be attributed and who bears normative responsibility in classical legal theory, the term *cognitive subject* is preferred to designate systems that perform the functions of processing information, constructing meaning, and methodologically directing the decision-making process. This usage does not entail the attribution of legal personality to the agent; rather, it aims to conceptualize the *de facto* role the agent assumes in processes of semantic production and decision construction. As will be elaborated in the third section of our article, entitled '*Agentic Orchestration of Mecellem's Semantic Protocol*', the subjectivity at issue here constitutes an epistemic and methodological position rather than an ontological status of '*personhood*'. Accordingly, the designation of the agent as '*subject*' expresses not the claim that it is a rights-bearing legal subject, but rather the observation that it occupies an active and directive cognitive position in the production of legal meaning.

The conceptual architecture of *Mecellem* is grounded substantially in the epistemology of al-Fârâbî, particularly in his foundational distinction between *first intelligibles* and *second intelligibles* (ma'qûlât). While we remain fully cognizant of the considerable scholarly value that would attend a direct visual mapping of these classical philosophical categories onto contemporary formal ontology standards—including *Web Ontology Language* (*OWL*), *Resource Description Framework* (*RDF*) graphs, and semantic node structures—such graphical representations have been deliberately excluded from the scope of this article. This exclusion serves to preserve the philosophical integrity and narrative coherence of the argument, prioritizing conceptual exposition over technical visualization. The central aim of this article is not to provide technical schematics, but rather to elucidate the constitutive significance that this philosophical foundation holds for legal reasoning and computational legal knowledge representation.

The technological structures and computational instruments discussed throughout this article are situated within a domain characterized by rapid and ongoing transformation. Consequently, the methodological and technical choices reflected herein have been formulated with due consideration of present technological capacities and constraints. What remains determinative for the purposes of this inquiry, however, is not the particular technological instantiations themselves, but rather the theoretical and normative objectives these technologies are deployed to realize. Notwithstanding potential shifts in the instrumental landscape, the ontological commitments upon which this article is founded are understood to constitute stable and enduring reference points that transcend ephemeral technological configurations.

Although *Mecellem* is characterized throughout this article as a '*semantic protocol*', it is presented here primarily as a theoretical framework rather than a fully operationalized technical specification. A comprehensively defined technical protocol would, by its nature, necessitate the disclosure of open data schemas, node-level interaction standards, system integration architectures, and interoperability specifications. Given, however, that the actual implementation comprises hundreds of thousands of nodes, millions of properties, and multi-layered relational structures among them, what is presented in this article reflects not the totality of the operational system, but rather its constitutive legal-semantic logic. This level of abstraction represents a deliberate methodological choice, designed to maintain analytical focus upon the foundational legal-semantic theory that animates the system, rather than upon the granular technical details of its implementation.

The *ArbiNext* project is presented herein to exemplify the operational potential of the *Mecellem* architecture within the domain of international arbitration. Readers anticipating quantitative engineering data—such as precision-recall comparisons between our semantic chunking and graph-based retrieval methodologies and conventional *Retrieval-Augmented Generation* (*RAG*) systems, latency benchmarks, or empirical performance metrics—will observe that such data are not included in this article. This work has been prepared not as a technical evaluation report, but rather as an exposition of a conceptual framework. The empirical validation and statistical performance metrics of the *ArbiNext* project fall outside the scope of this article; such data will be elaborated separately in technically focused publications by *NewMind AI* engineers. That said, the fact that both the *ArbiNext* project and the *Mecellem semantic protocol* have been operationally deployed in large-scale commercial transactions and institutional legal service delivery constitutes the most compelling empirical reference point, rendering the epistemic applicability of the proposed framework something more than an abstract theoretical claim.

Throughout this article, various conceptual models and logical notations—incorporating variables such as $\alpha$, $\beta$, $\gamma$, $\theta$, and $\delta$—are presented to illustrate how epistemological layers, methodological instruments, and semantic depth are balanced within the proposed framework. The ultimate purpose of these formal representations is to demonstrate the holistic and scientific coherence afforded by the ontology-epistemology-methodology triad upon which the *Mecellem semantic protocol* is founded, and to elucidate how this framework structurally diverges, in a constructive manner, from the statistical and stochastic methodologies that underpin large language models (*LLMs*). Accordingly, the analytical notations, epistemic weighting equations, and semantic mapping mechanisms proposed in this article should not be understood as fixed and immutable mathematical laws, but rather as operational expressions of ontological commitments, epistemic stratifications, and methodological choices. In this sense, each constitutes a technical proposition that may itself be subject to further scholarly debate and refinement. The parameters embedded within these analytical representations—variables such as α $\alpha$, $\beta$, $\gamma$, $\theta$, and $\delta$ —are calibratable across different legal domains, diverse contexts, and varying epistemic priorities. This flexibility is not a weakness of the system, but rather a natural consequence of contextual epistemology and, indeed, a source of its strength.

Indeed, in our view, the value of a scholarly work is determined not by the particular results it achieves, but rather by the robustness of the ontological, epistemological, and methodological framework upon which it rests. A study grounded in a coherent ontology and epistemology, employing a methodology that is both explicable and auditable, may yield divergent empirical outcomes across different samples and under varying parametric conditions, and may be updated as necessary. Such variation does not diminish the scientific or epistemic value of the original work; rather, differing results produced within the same methodological scaffolding are regarded as natural consequences of the internal dynamics of knowledge production.

The emphasis placed upon formal equations and epistemic weighting mechanisms within the *Mecellem semantic protocol* is directed not toward individual outputs, but toward the explicability (in the sense of *explainable AI*) and reliability of the reasoning model and methodological architecture that enable the generation of legal meaning. What proves determinative, therefore, is the holistic coherence and methodological soundness of the reasoning model embedded within the formulas and semantic protocol; the inevitable subjective elements present in particular evaluations do not undermine the scientific legitimacy of this framework.

The motto presented at the outset of this article—'*One can only contest what one is capable of comprehending*'—reflects precisely this approach. Consonant with Karl Popper's *principle of falsifiability*,[25] which he introduced to the philosophy of science, this approach ensures that the inferential steps and weighting mechanisms proposed by *Mecellem* remain auditable, criticizable, and methodologically testable at every stage. In sum, what is at stake is a methodological structure that can be updated when necessary and that bears the fundamental hallmarks of scientific inquiry.

[25] KARL POPPER, THE LOGIC OF SCIENTIFIC DISCOVERY, (Hutchinson & Co. 1959) (1935).

Finally, the references to al-Fârâbî throughout this article are not intended merely as historical or philological citations; rather, they function as the constitutive epistemological foundation of the proposed framework. When al-Fârâbî's works—particularly *Kitāb al-Ḥurūf* (*Book of Letters*), *Iḥṣāʾ al-ʿUlūm* (*Enumeration of the Sciences*), *Kitāb al-Burhān* (*Book of Demonstration*), and *al-Madīna al-Fāḍila* (*The Virtuous City*)—are examined together and from a holistic perspective, it becomes evident that he constructed a comprehensive epistemological architecture not only in the domains of metaphysics or political philosophy, but also concerning the classification of knowledge, the formation of concepts, the relationship between language and being, and the diverse cognitive operations of the intellect. The approach developed by al-Fârâbî is noteworthy for its systematic effort to classify different domains of knowledge within a conceptual order and to explicate the processes by which reason generates meaning from data. In contemporary artificial intelligence research, the challenge of extracting meaningful inferences from vast data repositories, structuring knowledge, and generating higher-order insights from within datasets has become an increasingly central epistemic difficulty. In this context, it is our conviction—forming the methodological background of this study—that the conceptual framework articulated by al-Fârâbî in the aforementioned works deserves to be regarded not merely as a historical philosophical legacy, but as an intellectual resource warranting renewed examination and evaluation in light of contemporary cognitive systems and artificial intelligence discourse. When his works are considered holistically, they offer a remarkably comprehensive intellectual horizon concerning the relationship among knowledge, intellect, and wisdom. Accordingly, the references to al-Fârâbî in this article are made to recall that his corpus, when approached in its entirety, presents a profoundly expansive framework for understanding these relationships, and to draw attention to the necessity of reexamining these texts in the age of big data.

## *II. THE METHODOLOGICAL TRANSFORMATION OF LEGAL REASONING: THE INTELLECTUAL FOUNDATIONS OF MECELLEM*

### *A. From Experience to Measurement: the Jurimetric Turn*

For every researcher with a genuine interest in legal practice, the oeuvre of Lee Loevinger, particularly his groundbreaking seminal article '*Jurimetrics: The Next Step Forward*', constitutes a central point of reference. In this work, Loevinger mounts a revolutionary challenge to the speculative nature of traditional jurisprudential methodology through the lens of the *jurimetrics* discipline. The fundamental reason Loevinger occupies this pivotal position lies in his repositioning of legal methodology away from a normative hierarchy of values and toward an empirically testable, measurable, and predictable foundation of quantitative analysis. This approach strips legal reasoning of its rhetorical and metaphysical trappings, integrating it into a cohesive framework informed by probability theory, logical symbolism, and statistical modeling—thereby laying the epistemological groundwork for modern computational law and artificial intelligence-driven legal technologies.[26]

Loevinger's *jurimetrics* approach, though appearing at first glance as an endeavor to 'reinvent' law through quantitative instruments, is in fact constructed upon an earlier methodological rupture. This rupture consists in the conception of law not as a purely logical system of norms, but as a living practice operating within social experience and a notion crystallized most influentially in Oliver Wendell Holmes Jr.'s famous observation in *The Common Law*: '*The life of the law has not been logic: it has been experience.*'[27] In this vein, law transcends a closed, self-sufficient conceptual architecture; it is apprehended instead as an evolutionary order, shaped by processes of action, decision-making, and contingency,[28] perpetually self-correcting through historical context, patterns of precedents, and societal feedback.[29]

However, this crisis of formalism was not a phenomenon peculiar to the Western legal world. In the second half of the 19th century, similar tensions arose in both Western and Eastern legal traditions. In the United States, the legal formalism[30], epitomized by Langdell conceived law as a closed system insulated from social change; a parallel process unfolded in the Islamic world and the Ottoman Empire. Although the classical Hanafi fiqh doctrine, which predominantly shaped Ottoman law, was inherently a tradition open to rational reasoning and ijtihad, it became confined to reiterations of its classical opinions in the face of new economic and administrative exigencies arising from the modernizing and increasingly centralized state structure, ultimately proving inadequate and experiencing a marked impasse.[31]

However, the responses to this common crisis exhibit significant differences in methodological and institutional terms. The perception that legal thought still operated through abstractions detached from reality propelled Holmes's critique further in the West, directly targeting legal formalism and paving the way for Felix S. Cohen's realism, which branded it as ‘transcendental nonsense’. Cohen challenged the perpetuation of legal thought within the ‘heaven of concepts’ depicted by Jhering, a realm severed from reality[32], thereby concretizing the realist insistence that law must be reconstructed not as an autonomous logical construct, but as a discipline grounded in experience and practice. However, this experience-based paradigmatic transformation (*jurisprudential turn*) initiated by Holmes, while undermining the Western legal tradition's claim that law can be explained through formal logic, has not entirely eliminated conceptual formalism itself, remaining thus far at the intellectual and philosophical level.

---

26 Loevinger, *supra* note 5.

27 Holmes, *supra* note 12.

28 *Contingency*, in the context of legal philosophy and logic, expresses that the occurrence of a situation is not necessary, meaning it could be otherwise. Holmes and other legal realists use this term to emphasize that law is not a frozen and immutable logical construct, but rather a flexible and probabilistic structure shaped by social needs and experiences.

29 Oliver Wendell Holmes, Jr., *Law in Science and Science in Law*, 12 Harv. L. Rev. 443 (1899).

30 Christopher Columbus Langdell, *A Selection of Cases on the Law of Contracts*, Boston: Little, Brown & Co (1879). *See* also Susan Haack, *On Legal Pragmatism: Where Does 'The Path of the Law' Lead Us?* The American Journal of Jurisprudence. 50. 71-105 (2005). 10.1093/ajj/50.1.71. In this work, Haack examines Langdell's 'logical theology' approach in detail. She emphasizes that Langdell viewed law as a logical, scientific enterprise based on principles that could be logically derived from case law and analyzes Langdell's approach in contrast to Holmes's emphasis on experience and social context.

31 Guy Burak, *The Second Formation of Islamic Law: The Hanafi School in the Early Modern Ottoman Empire*, Cambridge University Press (2015). In his work examining the evolution of the Hanafi legal tradition in the Ottoman Empire, the author details the doctrinal debates among Hanafi jurists and their efforts to reconcile Islamic legal principles with modern governance demands. *See* also Roderic H. Davison, *Reform in the Ottoman Empire, 1856-1876*. Princeton University Press (1963). http://www.jstor.org/stable/j.ctt183q0cn.

In contrast, the Mecelle Commission, chaired by Ahmet Cevdet Pasha in the Ottoman Empire, provided a far earlier and more radical response to the same crisis. Awareness of law's temporally mutable character was enshrined in the Ottoman legal order at a much earlier date. The concept of *'experience'* formulated by Holmes in 1881 and the concept of *'ezmanın tagayyürü'* (the change of rules with the change of times) codified by Cevdet Pasha in 1869 address, in essence, the same sociological reality: law is not a frozen text; it must adapt to the rhythm of society. Yet Cevdet Pasha's approach emerged approximately twelve years before Holmes's and through an act of *‘legislative’* will. Article 39 of the Mecelle (‘*ezmanın tagayyürü ile ahkâmın tagayyürü inkâr olunamaz*’) (the change of rules with the change of times cannot be denied)[33] transformed this dynamism from a matter of philosophical discourse into a binding state order. Thus, while legal realism in the West remains at the level of a theoretical rupture, in the East it has been codified into written law by the state itself and rendered an inseparable component of judicial processes.

The crisis that emerged in the 19th century reflects profound philosophical differences between two fundamental approaches in legal philosophy. Conceptualism, or legal formalism, posits that law constitutes an autonomous system in which legal reasoning demands logical consistency and conceptual clarity. Functionalism, or legal realism, by contrast, contends that law is a social phenomenon and that legal decisions cannot be isolated from social, economic, and political factors. From an epistemological perspective, conceptualism maintains that legal knowledge can be obtained through logical deduction, whereas functionalism emphasizes empirical observation and social scientific methodology.

---

[32] Cohen, *supra* note 6. Cohen’s ‘*transcendental nonsense*’ theory comprises three fundamental elements: (1) the claim that traditional legal concepts such as corporation, property rights, and contract are ‘supernatural entities’ lacking verifiable existence; (2) the argument that judges justification of their decisions through these abstract concepts is misleading, and that decisions are in reality influenced by social, economic, and political factors; (3) the proposal to adopt a functional approach in resolving legal problems, namely examining the social and economic effects of legal decisions. Cohen’s functionalist approach advocated recourse to ‘facts, figures, graphs, statistical curves, and measuring tapes’ in solving legal problems. However, this approach was subjected to criticism from traditional law defenders such as Walter B. Kennedy. Kennedy argued that the functionalist legal approach was more vacuous and impractical than the traditional legal concepts it criticized; he claimed that the functionalists’ ‘reliable technique’ did not provide real solutions to legal problems. Kennedy further contended that the social sciences harbored conflicts and inconsistencies within themselves, and therefore it was unrealistic to expect definitive solutions from these fields for law. Finally, he emphasized that the common law had served for six centuries and possessed ‘the genius and capacity to embrace and assimilate both change and tradition’. *See* also Walter B. Kennedy, *More Functional Nonsense - A Reply to Felix S. Cohen*, 23 A.B.A. J. 625 (1937).

[33] *See* Ali Haydar Efendi, *supra* note 1.

Methodologically, conceptualism employs deductive reasoning and precedent analysis, while functionalism relies upon inductive reasoning and social scientific research.

In contemporary legal scholarship, an endeavor toward synthesis between these two approaches is discernible. Most legal academics acknowledge both the significance of legal concepts and the social context of law. It is widely accepted that legal reasoning encompasses both logical and pragmatic dimensions, necessitating both stability and flexibility. The semantic protocol proposed by Mecellem likewise embodies this quest for synthesis: it aims to preserve the ontological structure of legal concepts while systematically integrating their contextual and functional dimensions.

This rejection of formalism, as elaborated above, has paved the way for more dynamic and functional approaches grounded in experience. Among the most systematic exemplars of these approaches is Roscoe Pound's theory of social control, inspired by Émile Durkheim's conception of social facts. Durkheim demonstrated that societal phenomena such as suicide could be examined through a temporal lens, thereby rendering visible shifts in collective values and transformations in social structure. Pound adapted this methodology to the legal domain, arguing that judicial practice should draw sustenance not from abstract ideals of justice, but from the evolving texture of social reality. According to Pound, normative change occurs not through abrupt and ruptural interventions, but rather by foregrounding specific social facts in litigation, observing their effects over time, and reflecting their gradual transformations in case law. In this process, norms attain permanence and enforceability only when they achieve sufficient alignment with social reality.[34] This approach, by elevating social facts to the center of the courtroom, grounds norm production in more practical and material foundations, thereby distancing legal discourse from metaphysical debates over whether justice is abstractly attained. In other words, Pound, by demanding temporality, feedback mechanisms, and change within continuity in legal reasoning, has demonstrated the inevitability of a dynamic operation inherent in the nature of law.

Nevertheless, this dynamic and experience-centered approach, while strengthening law's engagement with social reality, leaves unanswered the questions of how such transformation can be systematically tracked and compared. The first systematic and deliberate response to this methodological lacuna emerges with the *jurimetrics* approach. This orientation, which not only diagnoses legal experience but also seeks to capture, classify, and render it systematically retrievable, acquires an institutional framework through the work of Loevinger. Loevinger contends that the true determinants of legal reasoning reside not in conceptual abstractions, but in decision patterns, probabilities, and actual applications; accordingly, he maintains that this experiential material can be meaningfully processed only through quantitative tools, logical symbolism, and systematic data-processing techniques. Thus, *jurimetrics* inherit Holmes's emphasis on experience, Cohen's critique of formalism, and Pound's recognition of law's inherently dynamic operation, staking a claim to reconstitute law for the first time as a measurable, comparable, and computable domain of knowledge. In this respect, *jurimetrics* manifest not merely as a novel set of technical instruments, but as a profound methodological reorientation in the production of legal knowledge.

---

[34] Roscoe Pound, *The Scope and Purpose of Sociological Jurisprudence (Concluded)*, 25 Harv. L. Rev. 489 (1912). *Also see* Roscoe Pound, *An Introduction to the Philosophy of Law*, Yale University Press (1922). One of the prominent figures in this debate is Karl Llewellyn. Llewellyn made a distinction between 'real rules' and 'paper rules' by emphasizing the indeterminacy of law and the importance of examining how law operates in practice: According to him, the actual practices of courts are more important than theoretical rules. *Also see* Karl N. Llewellyn, *A Realistic Jurisprudence—The Next Step*, 30 Colum. L. Rev. 431, 447-453 (1930); Karl N. Llewellyn, *Some Realism About Realism—Responding to Dean Pound*, 44 Harv. L. Rev. 1222 (1931).

Loevinger positions *jurimetrics* as an epistemological alternative to classical legal theory, thereby rendering the distinction between them stark: whereas classical legal theory engages in philosophical deliberations concerning the essence, purpose, and value of law, *jurimetrics* concentrates on generating observable, measurable, and empirically testable outputs.[35] In this approach, law's legitimacy is tethered not to abstract conceptual coherence, but to its capacity for testability and practical verifiability. In Loevinger's pithy formulation, the ultimate and unanswerable questions of life pertain to the domain of philosophy; classical legal theory (*jurisprudence*) constitutes the philosophy of law.[36] *Jurimetrics*, by contrast, suspends these philosophical deliberations, reducing law to the plane of measurable social facts and subjecting legal decision processes to experimental scrutiny. Thus, law ceases to be a normative system purged of metaphysical speculation; it transforms into a domain of practice that can be reconstructed through data, observation, and analysis.

### *B. From Measurement to Understanding: The Rise of the Semantic Plane*

This powerful methodological advance offered by *jurimetrics* renders a specific dimension of legal knowledge systematically visible while methodologically relegating another dimension of legal reasoning—particularly its contextual and semantic layers—to the background. By transforming law into a measurable, classifiable, and statistically analyzable object, *jurimetrics* has largely resolved the problem of '*access to knowledge*'; yet it leaves unresolved the question of what context, for what rationale, and within which network of conceptual relations the discovered knowledge acquires meaning. This circumstance elucidates why *jurimetrics* represented a compelling *next step forward* in its historical context, while simultaneously demonstrating why it cannot, on its own, encompass law's dynamic, multilayered, and interpretive nature.

It is precisely this methodological lacuna that has rendered visible a new imperative emerging immanently from *jurimetric* thought: rendering legal knowledge not merely discoverable, but meaningfully and contextually retrievable. Among the earliest and most systematic responses to this imperative appears in J. Russell Cades's seminal work, *Jurimetrics and General Semantics*. Cades contends that the functional classification of legal documents and normative sources is insufficient on its own; rather, conceptual and terminological structures must be explicitly defined to ensure their semantic coherence within the targeted context.[37] Thus, through Cades's contribution, *jurimetrics* ceases to be merely a quantitative program of information retrieval and begins to evolve toward a quest for semantic depth; one oriented toward questions of how legal meaning is organized, how context is constructed, and how knowledge is justified.

---

35 Lee Loevinger, *Jurimetrics: The Methodology of Legal Inquiry*, 28 Law & Contemp. Probs. 5, 8 (1963).

36 Id., at 7.

Cades analyzes the relationship between *jurimetrics*—the scientific examination of legal problems—and the '*general semantics*' approach, which investigates the influence of language and symbolic processes on human behavior, while discussing the role of computers in executing traditional legal operations. Ultimately, he concludes that computers will render the *jurimetrics* oriented jurist more semantically conscious; create enduring value as a research instrument; and, by eliminating routine workload, encourage the user toward deeper reflection and the exercise of wisdom.[38] The awareness articulated by Cades sixty years ago—in an era predating even the advent of the internet—signaled a methodological horizon far ahead of its time. Today, however, when considered alongside the heterogeneity of legal knowledge, its contextual density, continuously expanding decision-making ecosystems, and the integration of generative artificial intelligence and autonomous agents into legal processes, this approach has evolved from a complementary possibility into an indispensable frame of reference for the reorganization of legal thought.

When Loevinger first introduced the concept of *jurimetrics*, he posited that technology would not merely provide jurists with rapid access to documents, but would also offer the capacity to interconnect these documents within semantic structures that render their practical meanings visible. Indeed, technological developments have not only vindicated his predictions but have also transcended the limitations he could have foreseen. Rapid advances in disciplines such as computer science have demonstrated the applicability of empirical methodologies and technological tools to legal decision-making processes, thereby reshaping both the practice of law and its theoretical foundations.

Within this framework, the jurist's task is not merely to store data, but to classify, correlate, and, when necessary, functionally reconstruct documents within meaning-generating contexts. *Jurimetrics* thus transcends the status of a simple search or indexing technique, acquiring the character of a methodological approach that reconceptualizes how legal knowledge is organized, articulated, and rendered operational. The focus shifts from grounding law in ideal conceptual debates to accurately reconstructing law's concrete manifestations—the factual data and experiential elements upon which judicial decisions rest. In this regard, the jurist's fundamental task is not to pursue an inaccessible 'ideal justice' through abstract logical inferences, but rather to focus on the realization of 'justice in the present moment' by grasping and representing, with utmost precision, the factual reality of a particular case within its context.[39]

---

[37] J. Russell Cades, *Jurimetrics and General Semantics*, 22 ETC: A Review of General Semantics, no. 3, 279 (Sept. 1965).

[38] Id., at 291.

[39] Justice stands at the forefront of the most ultimate values that legal philosophy identifies as the purpose of law. However, the concept of justice is also one of the most frequently cited examples of the category of 'essentially contested concepts' introduced by W. B. Gallie in 1956. *See* also Walter Bryce Gallie, *Essentially Contested Concepts*, Proceedings of the Aristotelian Society, 56, 167-198 (1956).

This approach necessitates not merely the collection of legal documents, but their purpose-driven and context-sensitive retrieval. A functional system must be capable of presenting documents not in isolation, but within a holistic *semantic* field by establishing the relationships among them and activating distinct semantic layers. Thus, the data mass ceases to be a collection of disconnected fragments; it transforms into a coherent and contextual knowledge structure capable of guiding legal reasoning.

This functional and context-sensitive orientation is directly related to Loevinger's broader conception of science in relation to law. Indeed, according to Loevinger, science is not an activity that pursues ultimate and immutable truths; on the contrary, it is a dynamic process of inquiry that revises itself as new questions emerge, formulating provisional and testable answers[40]. This perspective represents a deliberate departure from the classical jurisprudence that conceives of law as an architecture of norms enclosed within its own internal coherence.

This departure becomes even more pronounced in the distinction Loevinger draws between logic and (*semantics*)[41]. According to him, logic is indispensable in legal and scientific reasoning; however, its function is essentially syntactic. Logic regulates the formal relationships among symbols; it monitors the internal consistency of propositions. Yet this formal consistency alone does not guarantee contact with reality. For the subject of legal proof is not merely the correctness of symbolic structures, but the accurate representation of the factual world. Loevinger therefore conceives the limits of logic in conjunction with the

---

https://doi.org/10.1093/aristotelian/56.1.167. What is meant here is the truth that certain concepts, despite being very widely used, cannot be reduced to a single definition and mode of understanding that is definitively and universally accepted. Such concepts are interpreted from different perspectives by different theoretical approaches and different philosophies, and the divergences between these interpretations can in no way be resolved through empirical evidence. For instance, justice has been addressed in the social context through different principles and within various frameworks such as equity, protection of rights, liberty, equality, equality of opportunity, equal distribution of resources, or maximization of social benefit and minimization of harm. One of the definitions related to justice is the epistemic dimension of justice, namely the understanding of epistemic justice. One of the pioneering contributions in this field is Miranda Fricker's work titled *Epistemic Injustice: Power and the Ethics of Knowing*. *See* Miranda Fricker, *Epistemic Injustice: Power and the Ethics of Knowing* (Oxford Univ. Press 2007). In this study, Fricker focused on the epistemic dimension of justice, drawing attention to the ethical and philosophical dimensions of the processes of production, sharing, and evaluation of knowledge. The 'justice' referred to within the framework of the Mecellem semantic protocol emerges precisely at this point in the context of justice's relationship with knowledge and truth; it signifies equity and equality in access to knowledge, and the establishment of a fair structure in the processes of production, verification, and distribution of knowledge. At the same time, the evaluation of knowledge must be conducted taking into account the context and conditions in which one finds oneself. According to this perspective, justice—that is, epistemic justice—is a process- and context-sensitive phenomenon rather than a static state. Since the truth and value of knowledge acquire meaning within a specific time and condition framework, this approach has been expressed as 'justice in the moment'. Thus, justice has been conceived as a dynamic concept that is continuously reshaped within changing social and epistemic conditions. *Also see* Susan Haack, *Evidence and Inquiry: Towards Reconstruction in Epistemology*, Blackwell Publishers (1993).

[40] Loevinger, *supra* note 35. *Also see* Lee Loevinger, *Jurimetrics: Science and Prediction in the Field of Law*, 46 Minn. L. Rev. 255 (1961);

[41] Lee Loevinger, *On Logic and Sociology*, 32 Jurimetrics J. 527, 529 (1992).

necessity of semantics. Semantics investigates the reality to which symbols refer and the substantive relationships within that reality. While syntax is concerned with formal relations, semantics establishes material and factual relations (*standards*). The two domains are not contradictory; rather, they operate on different planes. Logic constructs the structure; semantics establishes that structure's connection to the world.

Loevinger's debate with Kaye[42] also clarifies this point: Mathematical or logical form does not eliminate subjective elements; it merely transfers them to another plane of expression. Numerical or symbolic expression does not automatically objectify judgment. The ultimate foundation of legal proof remains the manner in which the human mind evaluates reality. For this reason, formal correctness and substantive correctness are not the same thing in legal reasoning.

The conclusion that emerges from this analysis is as follows: Both logic and semantics are necessary for the accurate representation of reality. Logic ensures the structural consistency of legal reasoning; semantics, however, connects law to the social and factual reality it seeks to govern. Unless *jurimetric'*s quantitative orientation is combined with this semantic foundation, it cannot encompass the dynamic nature of legal reasoning.

Yet even though *jurimetrics* and *semantic* structuring strengthen the technical infrastructure of legal knowledge, these architectures alone do not suffice to explain the decision-making and directive function that constitutes law's essential distinguishing characteristic. The question of how law not only is organized and finds meaning, but also how it learns, how it adapts, and how it transforms itself, demands a new theoretical depth beyond the *jurimetric* and *semantic* framework. It is at this juncture that the approach of Donald E. Elliott—who conceptualizes the operation of the common law directly as a learning process—comes into play. Elliott conceives of the common law not as a sum of static rules, but as a dynamic learning process that is fed by prior judicial decisions, generalizes through *analogies*, and continuously recalibrates itself in accordance with social expectations. In his view, legal reasoning operates not through chains of formal logical inference, but rather through the internalization of past experiences and the reweighing of those experiences in the face of new cases. Law is therefore not a closed system aimed at arriving at fixed normative truths, but rather a context-sensitive and evolutionary order of knowledge that advances by correcting its errors.[43]

Elliott's approach is one of the rare examples of legal theory that makes explicitly visible the *analogies* with artificial intelligence and learning systems, insofar as it treats legal reasoning as a structure that learns through analogies and updates itself through feedback. From this perspective, the common law operates as a system that processes past decisions as a kind of 'dataset'; finds direction by comparing new cases with these patterns; and revises its course through social feedback. Law thus becomes not merely a domain for the application of norms, but a historical process in which knowledge is produced, tested, and restructured. This learning and contextual depiction of law that Elliott sets forth also clearly demonstrates that legal reasoning cannot be reduced to problems of measurement, classification, and access; rather, it requires that decision-making itself be treated as a relational and interpretive process.

---

[42] D. H. Kaye, *Proof in Law and Science*, 32 Jurimetrics J. 3, 313 (1992).

[43] Donald E. Elliott, *Holmes and Evolution: Legal Process as Artificial Intelligence*, 1984 Wis. L. Rev. (1984). In the same vein, Loevinger aptly summarized the fundamental methodological approach of legal realism, which emphasizes adaptability and iterative improvement in legal analysis, by stating that 'the questions of science do not seek ultimate answers, but only momentary answers subject to further correction and modification as additional questions are formulated'. *Also see* Loevinger, *supra* note 35.

This evolutionary depiction of law that Elliott sets forth implies that legal reasoning is not merely a structural or semantic problem but is also ontologically a temporal phenomenon. For each of the concepts of learning, feedback, analogy, and adaptation presupposes that law operates within a determinate temporal horizon and that decision is not a fixed outcome but a movement that takes shape within continuity. Law is no longer solely a matter of the systematization of norms or the structuring of meaning; rather, it also encompasses the questions of how continuity is established within change, by what internal logic transformation occurs, and according to what principles movement is directed.

Within this framework, the issue cannot remain limited to describing law as a learning system; it transforms into the need to theoretically ground law's relationship with time and the movement-character of decision. Thus the discussion expands from semantic organization to temporal structure; from the construction of meaning to the becoming of being. It is this expansion that impels us to address the *ontodynamic*[44] structure of law—that is, its simultaneously existential and kinetic character.

---

[44] The term '*ontodynamic*' is used for the first time in legal literature in this article to denote, within the scope of this study, the structural tension between the ontological stability and epistemological dynamism of the legal order, and the continuous process of reconstitution produced by this tension. The term expresses that law can be grasped neither as an entirely static system of norms nor as a wholly fluid social reflex; rather, it must be understood as an order that contains its own internal motion, arising from the interaction between ontological fixation and contextual transformation. In the face of the acceleration of modern temporality, the widening distance between experience and expectation, and the intensification of social change, law must now be conceived not as a 'completed rational totality' but as a field of 'becoming' that can be reconstituted in each new case and processes within itself the tension between continuity and transformation. This conceptual framework constitutes one of the fundamental architectural principles of the *Mecellem semantic protocol*: the dynamic balance between ontological constants (*first intentions*) established through the *T-Box/A-Box* distinction and contextually variable factual assertions (*second intentions*) is the operational counterpart of the ontodynamic structure. Without ontological constants, context becomes chaotic; without contextual analysis, ontology becomes rigid. *Mecellem's dynamic knowledge networks* sustain this balance through the expansion of *T-Box/A-Box* and the restructuring of semantic relations each time a new document is added to the system. Each document upload signifies not merely the addition of data, but the creation of new (*nodes*) within the system's ontological framework, the updating of existing relations (*edges*), and the reweaving of the semantic network. Thus, in parallel with Luhmann's conceptualization of *autopoietic* systems, the system continuously reconstitutes itself through its own operations; meaning is not a fixed content loaded from outside, but a dynamic product reproduced in each query and each document entry through the interaction of the ontological framework with context. In this sense, the *ontodynamic* architecture is not merely a technical modeling choice; it is a constitutive principle that enables the epistemically reliable and methodologically operable reconstruction of legal meaning in the age of artificial intelligence. The theoretical framework and application domain of the term are addressed in detail in the subsequent sections of the article.

### *C. From Meaning to Temporality: The Ontodynamic Structure of Law*

The methodological transformation of modern legal thought is not solely a matter of how knowledge is produced or how meaning is constructed; it is also a matter of how time is conceived. In the classical understanding of law, time was imagined as a fixed and homogeneous ground upon which norms were placed. With modernity, however, time ceased to be a measurable line of progress and became a field of experience that accelerates, stratifies, and multiplies. While Comte's idea of progress configured time as a manageable historical direction, Hegel grounded historicity as the unfolding of a rational subject However, subsequent intellectual ruptures—particularly Koselleck's analysis emphasizing the tension between the *horizon of experience*' and the '*horizon of expectation*'—have shown that time no longer gains meaning through the accumulation of the past, but through projections toward the future. Time has become not a singular flow, but a field of intersection for plural temporalities possessing different velocities and intensities. This transformation reveals that law, too, must be positioned not through fixed normative continuities, but within changing fields of experience and expanding horizons of expectation.

#### *1. The Temporalization of Time*

Codification can be understood as a product of modernization that emerged at the end of the 18th century and, continuing for nearly a century, fundamentally transformed the legal landscape. The transposition of this modernization into the social sciences was initially systematized by Auguste Comte. Comte's approach, which can be summarized as a '*philosophy of progress*' was largely nourished by scientific theories developed in physics, particularly around the problematic of 'motion'. The transformation in theories of motion is not merely a revision within the natural sciences; it has also produced comprehensive consequences regarding how the relationship with time is established, how time evolves, and thus how reality is conceived.

For this reason, one of the constitutive claims of modernity is the rethinking of 'time' not merely as an external ground, but as a constitutive structure that itself makes change and transformation possible. Indeed, the age of artificial intelligence in which we live reveals the necessity of a specific (*temporalization*) that requires increasingly dynamic forms of connection among 'things': What makes change possible is no longer solely the succession of events, but fluid forms of connectivity in which relations can be rapidly reconfigured, and each reconfiguration can generate a new context. In this regard, advances in 'time theory' and the theory of 'motion' have facilitated a deeper understanding of complex phenomena not only in physics and chemistry, but also in the social sciences, laying the groundwork for the development of new approaches to persistent problems.

This new sensitivity to temporalization is powerfully depicted in Robert Musil's *The Man Without Qualities* through a striking image: *'The sequence of events is a train that lays its own rails ahead of itself. The river of time is a river that carries its banks along with it. The traveler moves on solid ground between solid walls; but the ground and the walls are* *also set in motion, imperceptibly yet very vividly, by the movements of his fellow travelers*'.[45] Musil's metaphor inverts the 'stable ground' assumption of everyday experience: The human being does not merely advance within time; time itself—together with space, including ground and walls—becomes part of the movement.

The increasing complexity of the relationship between time and phenomena has compelled thinkers such as Comte to confront the ontological and epistemological challenges posed by this entanglement. One of the fundamental outcomes of this intellectual endeavor is the 'pluralization of time': The idea that multiple temporalities can exist simultaneously, interact dynamically with one another, and cannot be reduced to a single schema of 'uniform time'[46], has become visible through the transformative impact of modern scientific discoveries. Indeed, today, while the effect of an economic decision is reflected in global markets within seconds, the social consequences of the same decision may emerge within a different historical rhythm that unfolds over years.

This conceptual rupture has given rise to a radically different understanding of time: Time is no longer viewed as a singular and uniform continuity, but as a field of discontinuous, fragmented, and interacting processes. This restructuring has enabled new belief structures centered around change, transformation, and progress, and has also strengthened the epistemic ground of Comte's philosophy of progress.

However, this transformation is not merely a change in the mode of thinking; it also transforms the ontological assumptions about time itself. *The pluralization of temporality* transcends being an abstract achievement of modern scientific thought and becomes a fundamental assumption that determines the functioning of the social, and consequently legal, order. As Hermann Lübbe points out[47], the temporal structure of historicity is not solely dependent on the subject's production of meaning; it is also inherent in the nature of all open and dynamic systems. This situation can be concretized through the process of precedent production by the supreme court: When a decision is rendered, not only the present dispute but also the accumulation of past precedents, the influence of doctrine, and expectations regarding consequences that may arise in the future are simultaneously in operation. The court does not decide within a single 'now'; it processes the layers of past, present, and future simultaneously. Thus, the legal system becomes a structure that organizes and reproduces different temporalities within its own movement, rather than being a structure that operates within time. For this reason, temporality begins to be understood not merely as the experience of the interpreting subject, but as an *ontodynamic* ground that the legal system itself produces and sustains within its own functioning.

---

[45] Robert Musil, *Der Mann ohne Eigenschaften* [The Man Without Qualities] (M. Kaygan trans., Yapı Kredi Publications 2022) (1930).

[46] For instance, Sandbothe has pointed to the dissolution of the idea of ontological unity presupposed by classical metaphysics in modern science. According to him, the holistic explanatory model grounded in nature or God in the metaphysical age has given way to a plural scientific structure in which different theoretical frameworks coexist. Although Auguste Comte diagnosed this transformation at an early stage, according to Sandbothe, his emphasis on the equivalence of mechanics and thermodynamics remains limited in fully explaining the internal theoretical plurality of modern science. Mike Sandbothe, *Pragmatic Media Philosophy: Foundations of a New Discipline in the Internet Age* 12, Online Publication: www.sandbothe.net 2005 (2001).

[47] Hermann Lübbe, *Im Zug der Zeit*, Verkürzter Aufenthalt in der Gegenwart 30, Springer-Verlag (1992).

### 2. Progress, Acceleration, and Modern Temporal Tension

The pluralization of temporality is not merely an ontological transformation; it also forms the foundation of the idea of 'progress' that constitutes modernity itself. As Reinhart Koselleck also notes, the concept of progress has opened up a future that transcends predictable, natural time and the space of experience, and this future, with its own dynamism, has made new and long-term projections possible.[48] Thus, the future ceases to be a continuation of the past; the future is no longer obliged to resemble the past.

However, Koselleck also emphasizes the tension inherent in the idea of progress: On the one hand, progress carries the claim of rational prediction and calculability, while on the other hand, it also contains a secularized version of the expectation of 'salvation' that is historically specific to religion—that is, the belief that the future will reach a better order through human intervention and planning. As the distance between experience and expectation widens, the modern subject increasingly turns toward a future that is 'not yet realized'.[49] On the one hand, this situation reveals the fundamental formula that determines the temporal structure of the modern age[50], 'the less experience there is, the more expectation increases.' On the other hand, as Koselleck puts it, self-accelerating temporality weakens the possibility of experiencing the present as 'the present'; the present time transforms into a horizon that constantly escapes toward the future.[51] For example, in a society where technological innovations are accelerating, new solutions, new software, and new systems are promised before past experiences have even been digested; decisions are now made not according to existing experience, but according to the anticipated future.

This acceleration is not merely a technical phenomenon, but also a transformation that reshapes the ontological architecture of modern society. Time is no longer an external framework. Time becomes an active force that shapes the social order, and reality not only changes; the speed of change also continuously increases. As Ricardo Prandini notes[52], modernity itself is a 'new age' (*neuzeit*); it is subject to the law of acceleration, and here everything changes faster than expected. This situation forms the basis of the phenomenon now called 'social acceleration.' The distance between experience and expectation gradually widens; the amount of innovation per unit of time increases; the duration of the present time narrows. The gap between social reality and the projections constructed about it becomes the defining feature of modernity.

---

[48] Reinhart Koselleck, Futures Past: On the Semantics of Historical Time (tr. Keith Tribe), Columbia University Press 22 (2004)

[49] Id. at 21.

[50] Id. at 274.

[51] Id. at 22.

[52] Riccardo Prandini, *The Future of Societal Constitutionalism in the Age of Acceleration*, 20 Ind. J. Global Legal Stud. 733 (2013).

As Hermann Lübbe points out[53], this acceleration also accelerates cultural obsolescence: In a dynamic civilization, elements that are still 'contemporary' can simultaneously be on the verge of becoming outdated. The 'non-contemporaneity of the contemporary' increases. Thus, the sense of continuity weakens; innovation ceases to be an exception and becomes the norm. In the information age, this acceleration has intensified even further. Digital technologies have accelerated not only communication but also decision-making, classification, and meaning-making processes.

Under these conditions, temporality is characterized not only by the idea of 'progress' but also by a structural tension. On one side, accelerating innovation; on the other, the search for stability; on one side, projection; on the other, experience. The modern legal order is also positioned within this tension: The distance between norm production and social change must be constantly readjusted. Therefore, the modern temporal structure requires thinking of law not merely as a system of rules but as an orientation mechanism attempting to position itself in the face of an accelerating reality.

### *3. The Factual-Normative Distinction and the Positivist Rupture*

The pluralization of temporality and the convergence of the idea of progress with acceleration have transformed not only the understanding of history but also the conception of reality. The most pronounced consequence of this intellectual transformation within legal theory is the separation between 'factual reality' and 'normative ideality,' which were previously intertwined. With modern thought, the distance between 'what is' and 'what ought to be' becomes institutionalized; law begins to be constructed as a domain conceptually separated from existing social reality, possessing its own normative systematicity.

This distinction becomes a methodological principle in positivist legal thought. Law is now defined not as metaphysical debates about what ought to be, but as the systematic analysis of valid norms. Thus, existing law (*lex lata*) gains an autonomous status vis-à-vis ideal law (*lex ferenda*). The temporal separation of historical experience makes it possible to conceive of a normative domain independent of the empirical. Koselleck's argument gains even greater significance here: The less one is bound to experience, the greater the projection toward the future; normative structure is legitimized through its own internal coherence rather than through lived reality.

The transition from an era defined by subjection to natural or divine authority to an era in which autonomous human reason assumes a constitutive role also reinforces this rupture. Hegel's conception of the 'end of history' and Comte's prediction that scientific methods would replace metaphysics, though emerging from different traditions, point to the same structural transformation: Normative order is now derived not from a transcendent source, but from the systematizing capacity of human reason.

---

[53] Hermann Lübbe, *The Contraction of the Present, in High-Speed Society: Social Acceleration, Power, and Modernity* (Hartmut Rosa & William E. Scheuerman eds., 2009).

However, this systematization produces a new tension in the modern world, where time accelerates and the distance between experience and expectation widens. While law maintains its formal coherence, social reality is in constant flux. This tension between form and substance becomes particularly visible in the rationalizing legal thought of the 19th century. Max Weber's analysis of law through formal rationality strengthens law's capacity for calculability and predictability; yet it also carries the risk of weakening the bond between norm and lived reality. It is precisely at this point that Oliver Wendell Holmes's emphasis on experience regains importance. Holmes's critique demonstrated that law cannot be explained by purely formal logical structures, and that the real determinants of decisions are often lived experience and social needs. Thus, the rigidity of the positivist distinction begins to be questioned.

This questioning reaches its culmination in the 20th century with the work of Lee Loevinger, who introduced a technological and methodological dimension to law. While Loevinger sought to make the factual dimension of law measurable and testable, he attempted to redefine the distance between the normative and empirical domains. Yet here too, a fundamental dilemma inevitably persists: The moment a norm is formulated, it carries the risk of obsolescence in the face of changing social reality. Thus, while the factual–normative separation made the systematization of law possible, it also rendered law fragile in the face of temporal transformation. Modern law must now explain not only its validity but also how it will adapt to the pace of change. This necessity leads us to the next stage: the quest to explain law's own internal movement and capacity for transformation.

## *4. The Fragility of Law in the Face of Movement: The Need for an Ontodynamic Order*

Modern legal thought, particularly after the positivist rupture, has oriented itself toward constructing law through formal coherence and systematic integrity. This orientation has enabled the codification of norms within a hierarchical order, the definition of concepts with technical precision, and the establishment of legal reasoning as a calculable structure. However, this formal rationalization has also made visible a fundamental problem that law faces: Within accelerating temporality, as reality changes, can form remain fixed?

This tension deepens the structural separation between essence and form. While a legal norm responds to the needs of the historical context in which it was produced, over time that context changes, new expectations emerge, and the norm gradually carries the risk of 'obsolescence'. In modernity's accelerated temporality, this process of obsolescence becomes more visible: Form produces stability through its fixity; yet that same fixity can turn into rigidity in the face of transforming reality.

Hegel's claim of the 'end of history' represents a philosophical culmination of this tension. For Hegel, history reaches rational reconciliation within the dialectical movement of common reason (*geist*); the state, as the institutional form of freedom, resolves contradictions and history realizes its purpose. That is, in Hegel, movement progresses toward an end; time is teleological, it does not flow randomly, it is purposeful; contradictions are resolved and closed. In this conception, a final harmony between form and essence is possible; movement completes itself by reaching its purpose. However, in the modern age of acceleration, such an idea of closure is not sustainable. The pluralization of temporality, the opening between experience and expectation, and the acceleration of social change demonstrate that law can never remain as a 'completed' rational totality. Law must now be conceived not as a dialectical endpoint, but as an order that is continuously reconstructed, existing within movement.

This perspective reveals that formal systematics alone is not sufficient; law must remain in continuous contact with essence, that is, with lived reality. Therefore, the issue is not only how law is formally organized, but through what mechanism form can track the change in essence. It is precisely here that the *ontodynamic* dimension becomes visible. Law can be grasped neither as a completely static system of norms nor as an entirely fluid social reflex. On the contrary, it is a structure that carries within itself the tension between continuity and transformation. Under the conditions of modern temporal tension, legal 'reality' requires a theoretical framework capable of explaining its own movement. Reality has now become not only a domain to be discovered but also one to be constructed. The problem is not so much the validity of the norm as the movement of the norm: How is law renewed? How does form internalize the change in essence? How does decision preserve the past while directing the future?

These questions compel us to think of law not merely as a systematic or semantic structure, but as an *ontodynamic* order with its own internal dynamism. Law does not close, complete itself, or freeze; it is reconstructed anew in each case. This situation emerges as the continuation of modernity's understanding of time and movement into the age of artificial intelligence.

In the age of artificial intelligence, the ability to interconnect different semantic domains has made it possible to classify concepts in a more flexible and functional manner. The necessity of a structure that will enable the reconstruction of legal meaning in each query and relate various semantic domains within a dynamic network is evident. This structure is compatible with the perspective of *autopoietic*[54] systems as defined by Niklas Luhmann: while each social domain produces its own internal logic, it simultaneously establishes structural connections with other systems. Law, too, within this networked order, interacts with the changing context while preserving its own normative code. In this context, the *dynamic knowledge* network should be understood not merely as a technical data organization, but as an *ontodynamic* structure in which different semantic layers are re-related according to practical context. Each query activates the network in a different way; meaning is not fixed, but a process constructed according to context.

Legal systems, as Ugo Mattei also emphasizes, are not static states of (being) but continuous processes of 'becoming'.[55] In this context, the issue is no longer solely the normative validity or systematic coherence of law, but how law itself enables its own movement. The pluralization of modern temporality, the opening between experience and expectation, and accelerated social transformation necessitate understanding law not as a fixed form, but rather in that dynamic structure which Mattei defines as the interaction of different layers within systems. Law, which preserves past experience while orienting itself toward the uncertainty of the future, dissolves within itself the tensions of producing stability while remaining open to transformation. Therefore, law is not merely a systematic sum of norms; it is an *ontodynamic* field of 'becoming' that incorporates time not as an external framework but as a constitutive element, processing within itself the tension between continuity and change.

---

[54] *Autopoietic* refers to the condition in which the parts and activities constituting a system are oriented toward the sustainability of the system itself. For example, a cell is an autopoietic system; it produces its own components, interacts with its environment, yet maintains its structural integrity. *See* Niklas Luhmann, *Law as a Social System*, tr Klaus A Ziegert, edited by Fatima Kastner, Richard Nobles, David Schiff and Rosamund Ziegert, Oxford: Oxford University Press (2004).

[55] Ugo Mattei, *Three Patterns of Law: Taxonomy and Change in the World's Legal Systems*, 45 Am. J. Comp. L. 5 (1997).

However, this *ontodynamic* depiction alone is not sufficient. Grasping the temporal movement of law is one thing; establishing a methodological framework that governs, tracks, operates, and allows objection to this movement is another. Therefore, the question is no longer merely what law is; it is how this dynamic structure can be concretized through which ontological assumptions, epistemological tools, and methodological arrangements. The next section will address this need in detail by examining the ontological, epistemological, and methodological architecture of *Mecellem*.

## *III. MECELLEM: CONTEXTUAL EPISTEMOLOGY AND THE NEW SEMANTIC PROTOCOL OF LAW*

The methodological trajectory of legal reasoning, evolving from experience-based functional approaches to jurimetrics, from jurimetric modeling to semantic structuring, and from there to a temporal and *ontodynamic* framework, now directs us to a more fundamental question: On what understanding of being is legal knowledge founded, how does this understanding of being produce a consistent knowledge regime, and by what method is this regime operated? The issue is no longer merely access to data, classification, or the technical organization of meaning; the issue is under what ontological commitments legal meaning is produced and how this production is epistemically justified.

*Mecellem*, precisely at this point, proposes a contextual legal epistemology and a new semantic protocol structured within the *ontology–epistemology–methodology* triangle. The first vertex of this triangle is ontological commitment. Every knowledge system contains explicit or implicit assumptions about which entities it accepts as ‘existent’. In the legal context, this is the question of how categories such as norm, right, obligation, person, institution, fact, relation, authority, and context are defined and how they are related to one another. *Mecellem* treats the legal domain not merely as a set of documents consisting of texts, but as a structured ontological field of relational entities. While *taxonomy*[56] constitutes

---

[56] As will be detailed in the following sections, the term ‘taxonomy’ here does not merely refer to the hierarchical codification of norms or concepts. Rather, it denotes a broader ontological ordering that determines under which entity categories legal reality will be grasped, through which relationships these categories will gain meaning in relation to one another, and at what level of abstraction they will be represented. In this context, taxonomy is not a static classification scheme; it is a relational and multi-layered organizational logic that enables the repositioning of concepts within context. *See* also Nicola Guarino,

the classification level of this field, ontology determines the philosophy of being of this classification. Thus, the system explicitly defines what types of objects it works with, what relations it recognizes, and at what semantic levels it operates.

The second vertex of the triangle is epistemic coherence. The ontologically committed domain of being requires a knowledge production regime that is internally consistent. *Mecellem*'s contextual epistemology acknowledges that legal meaning is founded on relational consistency and contextual balance rather than claims of absolute truth. The meaning of a norm emerges not only in its wording, but in the relationship it establishes with other norms, facts, case law patterns, and institutional practices.[57] Therefore, knowledge is not a fixed content, but a structure that is positioned within the semantic network and rebalanced with each query. Epistemic coherence means that the answer provided by the system is not only technically sound, but also consistent with its ontological assumptions and contextually justifiable.[58]

The third vertex of the triangle is methodological pluralism. *Mecellem* is designed as an architecture that does not operate with a singular and fixed method, but works on multiple methodological planes that sometimes complement, sometimes exclude, sometimes encompass, and even compete with one another in certain contexts. This plurality is not a theoretical eclecticism, but a dynamic methodological configuration in which different epistemic tools are activated according to the ontological and contextual characteristics of each query. *Mecellem*'s *neurosymbolic* structure brings together symbolic legal modeling and statistical learning approaches within the same epistemic framework. The symbolic layer preserves ontological commitment; it secures the categorical boundaries of concepts, the hierarchical order of normative relations, and the structural consistency of legal inference. In contrast, the neural layer provides epistemological flexibility through operations such as *embedding*, pattern recognition, semantic proximity measurement, and contextual reweighting. *Mecellem'*s *agentic* structure is the high-level decision architecture that integrates these two layers. The agent is not merely a passive computational mechanism; it is an organizational layer that governs goal-oriented, context-sensitive, and multi-step reasoning processes. The selection of *retrieval* strategy, the scope of *semantic chunking*, which nodes on the *knowledge graph* will be activated, and which contextual layers will be prioritized are determined under *agentic* control. This orchestration enables the model not merely to produce statistical outputs, but to generate structured decisions in line with normative purposes. *Model Context Protocols* (*MCP*) and similar context management protocols, within this architecture, regulate which context, which dataset, and which tool

---

Christopher A. Welty, *Supporting Ontological Analysis of Taxonomic Relationships*, Data & Knowledge Engineering, Vol. 39, 51–74 (2001), https://doi.org/10.1016/S0169-023X(01)00030-1.

[57] In accordance with the fundamental proposition in Searle's speech act theory, the meaning of an utterance derives not only from its lexical content, but also from the contextual conditions in which it is uttered. Legal norms, too, acquire their actual normative force only within the social and institutional context in which they are applied. John R. *Searle, Speech Acts: An Essay in the Philosophy of Language* 16–21 Cambridge Univ. Press (1969).

[58] Giovanni Sartori, *Legal Concepts as Inferential Nodes and Ontological Categories*, 21 Artificial Intelligence & L. 217, 217–251 (2005).

chain will be activated, thereby securing epistemic coherence and traceability. Thus, *Mecellem* is not merely a semantically enriched system; it is positioned as a legal architecture that is ontologically sensitive to conceptual constants, epistemologically flexible, and methodologically orchestrated.

Within this framework, we will first examine in detail the taxonomic and ontological foundation of *Mecellem* in the context of Fârâbian knowledge ordering; then we will examine in detail the dynamic reconstruction of *Mecellem* semantics and the operational counterpart of a new semantic protocol that regulates the contextual operation of legal reasoning in the age of artificial intelligence, structured within the *ontology–epistemology–methodology* triangle.

### ***A. Mecellem's Fârâbian Knowledge Order***

#### *1. The Dynamic and Functional Taxonomy of Mecellem*

In the history of law, *taxonomy* has often been identified, in a narrow sense, with the systematic codification of norms.[59] In this line extending from the Roman tradition to modern positivist legal systems, taxonomy refers to the transformation of legal norms into a consistent, hierarchical, and enforceable whole by a sovereign will. This framework, associated with natural law through the concept of 'rule' (*nomos*) in ancient Greek and Roman thought, becomes binding only when enforced by a political authority (*in foro externo*). In this context, taxonomy emerges as the fundamental tool that transforms abstract normative principles into a functional and manageable structure suited to the needs of the sovereign.

Taxonomy in this narrow sense enables law to produce stability through *codification*. Normative texts such as constitutions, statutes, or regulations are not merely technical arrangements; they are *political-performative* structures that reflect how we see the world at a particular historical moment, which relations we recognize legally, and which values we centralize. As Cohen states, order and chaos exclude each other while simultaneously being interdependent; order emerges from chaos.[60] Taxonomy, in this sense, structures not only legal norms but also the sovereign's conception of the world.

However, this classical meaning of taxonomy is insufficient to explain only the normative appearance of legal order. For legal order is not merely a hierarchical arrangement of norms; it is also an ontological activity of selection and naming that enables certain entities, relations, and events to be recognized legally. Therefore, taxonomy is not merely a technical classification system that organizes texts; it must also be understood as a deeper conceptual architecture that decides which layers of reality law makes visible and which relations it includes within the domain of legal meaning.

---

[59] For 'codification' in the narrow sense, *also see* Gunther A. Weiss, *The Enchantment of Codification in the Common-Law World*, 25 Yale J. Int'l L. 435 (2000).

[60] Cohen, *supra* note 6.

The definition of law as a tool of social engineering also gains meaning at this point. While taxonomy organizes legal norms in line with administrative objectives, it translates the political power's ideals concerning society into legal language. This process is as much political as it is technical: The concepts that constitute law are selected, categorized, and placed into a hierarchical order under specific historical and political conditions. Taxonomy, in this respect, does not merely regulate reality; it redefines and delimits it.

At this point, it is also seen that taxonomy is not only a tool that establishes normative order but also an epistemic structure that determines the mode of production and circulation of knowledge. Legal systems do not represent the world directly, 'as it is'; rather, they reconstruct the world through specific conceptual schemas. Taxonomy, precisely in this context, constitutes the epistemic architecture that determines how legal knowledge is organized, which concepts are placed at the center, and which relations are considered legally meaningful. Therefore, taxonomy is not merely classification; it means the establishment of the conceptual order in which legal meaning is produced.

However, in modern legal thought, the function of taxonomy has moved beyond this narrow normative framework. Particularly with technological developments, the sustainability of legal order now depends not only on how norms are codified, but also on how legal knowledge is accessed, how it is interrelated, and how legal knowledge is interpreted. At this point, the concept of taxonomy, in its broad sense, transforms into an epistemic and operational domain of organization that encompasses the organization of legal documents, precedents, concepts, and data.

As Loevinger emphasizes, 'one of the fundamental elements of data access in law is to find applicable, similar, or relevant precedential authority among published decisions of precedential nature for the resolution of a current problem.'[61] This observation reveals that the accuracy of legal reasoning is directly linked to the accessibility and relatability of legal knowledge, rather than to the normative content itself. In this context, the effective classification of legal documents, unlike codification in the narrow sense, expresses a broader taxonomic order that enables the analytical processing of legal knowledge. Such a taxonomic order requires the reorganization of legal sources in accordance with current research needs.

Loevinger concretizes this issue of taxonomy in the broad sense through the example of a blueprint.[62] In his approach, the issue is not merely to label a term, but to make its structural function visible by decomposing the concept into its logical components. Indeed, as Loevinger notes, when 'blueprint' is encoded, it is not as a singular object name; it is analyzed as (a) a form of information, (b) a regulatory tool, and (c) a schema that has an effect on the construction process. This analysis shows that taxonomy in the broad sense is not merely classification, but a process of revealing the ontological and functional layers of concepts. Blueprint here is not merely a technical drawing; it is a normative information structure that represents a specific domain of reality, regulates it, and produces concrete results. Loevinger's aim is to show that legal concepts, similarly, should be encoded not as superficial labels, but through the logical and functional elements that constitute them. Thus, taxonomy moves beyond being a classification activity in the narrow sense and transforms into an ontological practice of organization that structures meaning and makes reality processable. This practice aims not to recodify law, but to dynamically reorganize legal knowledge.

---

[61] Loevinger, *supra* note 35.

[62] Id. at 14.

At this point, current information technologies and artificial intelligence-based data processing methods are fundamentally expanding the function of taxonomy. Thanks to big data analysis, semantic data modeling, and artificial intelligence-supported knowledge extraction techniques, legal knowledge is no longer merely archived data; it is transforming into a knowledge ecosystem that can be interrelated across different contexts and dynamically reinterpreted. These technological capabilities are transforming taxonomy from being a static classification activity into an operational semantic infrastructure in which legal knowledge can be processed in real time.

Within this framework, taxonomy ceases to be merely a tool that serves the establishment of normative order; it becomes an epistemic framework used to understand and interpret reality according to the institution's own specific needs. Today, to an extent unprecedented in any period of history, we possess technological tools that make it possible to implement and sustainably apply *functional taxonomy*.[63] The act of 'naming the world' emphasized by Cohen is no longer static labeling; it is a dynamic design built upon who the user is, what question they are seeking to answer at which phase of work, and at what taxonomic depth this need can be met. *Mecellem*, by combining today's technical capabilities with this functional perspective, rescues legal knowledge from being perceived as a static closure and transforms it into a network of relations that flows synchronously with the institution's operational objectives.

Precisely on this ground, *Mecellem* conceives of law not as a codified closed totality of norms; but as a living knowledge ecosystem that centers the user and the context. In addition to the '*regulation tsunami*' being experienced, the '*data tsunami*'[64], growing exponentially with increasing digital communication has rendered traditional classification methods dysfunctional. *Mecellem*'s approach positions the question 'who is the user and what do they need in which context?' as a compass within this massive data pile. Today's advanced technological capabilities are conducive to the flexibility and speed to meet this need; therefore, taxonomy is no longer a structure that freezes categories, but a decision support navigation that is constantly in motion, integrating experiential evidence and institutional priorities into the system instantaneously.

---

[63] For the concept of '*functional taxonomy*' *see* Birger Hjørland, *Domain Analysis*. Knowledge Organization, 44. 436-464. 10.5771/0943-7444-2017-6-436 (2017); Claudio Gnoli, *Is An All-Purpose Classification Possible? Insights From Farradane's Approach To Knowledge Organization*, Synthese 205, 177 https://doi.org/10.1007/s11229-025-05011-9 (2025); Elaine Svenonius, *The Intellectual Foundations of Information Organization*, Cambridge, Mass.: MIT Pr. (Digital Libraries and Electronic Publishing) (2000); Elaine Svenonius, *The Epistemological Foundations of Knowledge Representations*, Library Trends, Vol. 52, No. 3 (Winter 2004); Geoffrey C. Bowker and Susan Leigh Star, *Sorting Things Out: Classification and Its ConsequencesUnavailable*, The MIT Press (1999) https://doi.org/10.7551/mitpress/6352.001.0001; Michel Foucault, *The Order of Things: An Archaeology of the Human Sciences*. New York: Pantheon Books, (1970).

[64] For '*regulatory tsunami*' *see supra* note 9; for '*data tsunami*' *see supra* note 8.

In this context, the real issue for today's legal order is no longer how many norms are produced; the essential problem is the ability to establish epistemic knowledge systems that can reveal legal meaning and process this data within the complexity created by rapidly proliferating normative and operational data. In this context, taxonomy is not merely a classification tool that establishes order; it also emerges as a guiding knowledge infrastructure in which legal meaning can be produced and operational decisions can be constructed within the complex data universe.

As Emily Shearwin points out[65], in an order where traditional tools are used, there are at least three reasonable ways to classify law. The first is *formal* classification, which organizes legal rules in a way that preserves the logical relationships between them; here the aim is to ensure consistency among normative categories and to establish a systematic structure. The second is *function-based* classification, which organizes rules according to the types of disputes they aim to resolve; this approach makes the social roles of law visible. The third is *logic-based* classification, which groups rules and decisions according to the more open-ended justifications that support them; here the rational basis behind the norm becomes the defining axis of classification.

*Mecellem*'s taxonomic architecture, however, is not merely the product of a theoretical classification model; it is also based on a 'corporate knowledge architecture' shaped by the systematic observation of companies' legal and governance needs within years of practical experience. During the work carried out with numerous companies operating in different sectors within the scope of the *Mecellem* project, it was observed that the legal and administrative activities of corporate life are concentrated in certain contextual domains. These observations revealed that the legal and governance needs of companies are organized around thirty-six distinct domains. These domains have been designed as separate '*knowledge domains*' representing different legal and operational universes of corporate life. For example, *Muhakemat* represents the legal universe in which the judicial processes to which companies are parties are conducted; *Mukavele* represents the contractual domain in which contractual relationships are established and managed; *Muamelat* refers to the transactional universe in which the legal, commercial, economic, and financial transactions to which companies are parties, as well as the related corporate and operational transactional relationships, are conducted. *Mahfuz* represents the corporate archival universe in which all types of documents, records, data, and information assets of companies are stored; *Mülk* represents the processes related to the management of all types of tangible assets they own; and *Mesai* has been structured as a separate contextual domain encompassing legal and administrative issues related to human resources and labor relations.

Each of these contexts has been treated as independent knowledge universes representing a different universe of corporate and legal life; *Mecellem*'s taxonomic structure has also been developed on the basis of this contextual distinction. Each of the thirty-six contexts within *Mecellem* has its own separate taxonomic arrangements and operational models. However, due to the scope and analytical framework of the article, only the taxonomic approach developed for *Mecellem*'s most fundamental knowledge domains will be addressed here.

---

[65] Emily Sherwin, *Legal Taxonomy*, 15 Legal Theory 25, 39 (2009).

At the central organizing core of *Mecellem*'s multi-layered contextual architecture is *Pusula*. Within the *Pusula* architecture, taxonomy is designed as a functional orientation mechanism that enables the user to determine within which contextual domain a particular legal or corporate transaction is positioned, at which operational stage the process is located, and which document types correspond to this stage. Within this framework, *Pusula* functions as a central reference platform in which the document types used in its contextual universe and the bibliographic information related to these documents are systematically defined.

Phased taxonomy assumes a critical function in this architecture. From the perspective of a competent professional in the field, a particular document type does not merely represent a document; it also functions as a semantic marker indicating at which stage the relevant legal or corporate process is located. In other words, document types emerge as contextual indicators that make visible certain threshold points in the operational timeline of the transaction or dispute being conducted.

Indeed, for example, the notification of a reasoned judgment in a lawsuit does not merely mean the recording of the judicial decision; it also indicates that the judicial process has entered a new stage with the commencement of the appeal period. Similarly, in contractual relationships, certain document types are critical threshold points indicating the transition from the negotiation phase of the transaction to the binding contract stage. For example, the signing of a letter of intent or a memorandum of understanding between the parties signals the transition from the free negotiation phase to the structured contract negotiation stage; the signing of the contract indicates that the negotiation process has been completed and a binding legal relationship has been established. In contrast, closing documents that demonstrate the fulfillment of conditions in certain transactions indicate the transition to a new process in which the contract has actually entered into force and the performance stage has begun. In a corporate merger transaction, the adoption of a board resolution indicates that the internal corporate approval stage of the transaction has been completed. Such documents are not merely recorded outputs of legal processes, but also threshold points that determine the chronological and operational progression of the relevant transaction.

By systematically defining these threshold points, the *Pusula* system makes visible what meaning the same concept or document type carries in the contexts of *Muhakemat*, *Mukavele*, or *Muamelat* and which operational stage it corresponds to. In this respect, *Pusula* is not merely a catalog listing document types, but functions as a central semantic and operational reference domain that determines in which contextual phase and at which work phase the system should make its assessment.

The taxonomy specific to *Mecellem*'s *Mahfuz* domain is designed not merely as a technical file organization system, but as a subjective classification framework that enables institutions to integrate their own commercial logics into the legal knowledge architecture. The way institutions conceptualize the world is largely determined by the operational and ontological requirements generated by their primary fields of activity. Therefore, corporate document taxonomies must also be structured in a manner consistent with the structural logic of these fields of activity. Indeed, for example, from the perspective of an airport operator, the fundamental distinction in documentation is shaped around the distinction between airside operations and landside operations, which define the physical and legal boundaries of the operation. In contrast, in a petrochemical facility, the main axis of legal documentation is based on different operational layers such as manufacturing, supply chain & logistics, and commercial operations. Similarly, in a beverage company, the classification of legal documents gains meaning within the framework of sector-specific commercial relationships such as brand portfolio management and distributor networks. In this context, *Mahfuz* offers, instead of a uniform legal template imposed from outside, a flexible taxonomic structure shaped in accordance with institutions' own operational terminologies. For the standardization of taxonomic categories in a manner detached from corporate reality would lead to the operational context in which legal knowledge is produced becoming invisible. *Mecellem*, by enabling institutions to integrate their own commercial terminologies into this taxonomic structure, aims to encode documents not merely as abstract legal categories, but as contextual semantic units of corporate practice.

*Mecellem*'s *Muhakemat* taxonomy, instead of treating the judicial process as static file tracking, is designed as a multi-layered classification system that models it as a dynamic flow and totality of phases. This structure, going beyond classical file identifiers such as '*branches of judiciary, parent-child file type, party capacities, and judicial unit,*' maps the judicial process through a chronological, procedural, and institution-specific taxonomy. Within this framework, not only at which stage—first instance, appellate review, or supreme court stages—a document was produced, but also how the relevant institution positions the case within its own operational reality becomes determinative for taxonomic signification. For example, from the perspective of an energy company, expropriation disputes can be classified under different institutional contexts according to the function of the investment: While lawsuits arising from land acquisitions related to the establishment of electricity transmission lines are evaluated under the category of 'transmission infrastructure expropriation lawsuits'; disputes arising from land acquisitions related to the establishment of energy production facilities can be classified as a separate lawsuit class in the form of 'expropriation lawsuits related to production facility investments.' The classification of documents under an institutional context differentiates not only the legal regime of the relevant documents but also their intra-corporate operational significance. However, *Mecellem*'s *Muhakemat* taxonomy offers a layered classification not only according to the institutional context of lawsuits, but also according to the procedural phases that determine the progression of the judicial process. Within this framework, a petition in a case file gains meaning not only in terms of its type, but also in terms of the time regime and procedural consequences generated by the judicial stage in which it is situated. For example, a petition related to an appellate application has a different taxonomic status from a petition submitted in first instance proceedings; for this document indicates not merely the production of a new document, but also that the judicial process has transitioned to a different procedural phase. Therefore, each new data entry made in the system is repositioned taking into account both the judicial hierarchy and the chronological progression of the judicial process. Through this approach, *Mecellem* models the case file not as a static pile of documents but as a living

‘litigation ecosystem’; thus transforming the ontodynamic structure of the judicial process into a representable technical reality within institutions’ own operational language.

*Mecellem*’s *Mukavele* taxonomy is designed not as a static system that classifies private law relationships solely according to the types of contract texts, but as a dynamic knowledge architecture that models the multi-layered semantic structure of the contract as it evolves over time. Within this framework, contracts are addressed within a phased structure that reflects the contract's life cycle and contextual document groups that reflect the institution's own commercial logic. Thus, the contractual relationship becomes visible not as a legal form fixed at the moment the text is signed, but as a living semantic regime continuously updated through the documents and practices produced by the parties over time.

Holmes’s observation that the life of the law has not been logic but ‘experience’ finds concrete correspondence in this taxonomic approach.[66] For a contract does not consist solely of ‘internal rules’ containing the wording at the moment of signature; notices, electronic correspondence, implementation instructions, and actual performance practices produced between the parties also constitute integral parts of this experience as the ‘external rules’ of the contract. While *Mecellem* unites these internal and external normative layers of the contract under the same root as complementary semantic rings, it provides the institution with the opportunity to integrate its own commercial terminology into the taxonomic structure instead of imposing a uniform template from outside.

In this architecture, phased taxonomy is particularly important. For each document produced within the contractual relationship carries a different epistemic weight not only in terms of the information it contains, but also in terms of at which stage and by which authority it was produced. Each document that emerges throughout the contract's life cycle constitutes a new contextual layer reflecting the evolution of the parties' wills over time. Although subsequent documents do not in every case abrogate previous arrangements, they must be evaluated with particular attention in terms of will updates that emerge within the institution's own commercial language. Mecellem, by placing this epistemic weighting at the center of the system as a taxonomic imperative, carries that ‘legal experience’ pointed to by Holmes to a traceable and coherent reasoning ground within institutions' own terminology.

*Mecellem*’s *Muamelat* taxonomy structures law not as a norm-centric or document-centric structure, but on a transaction-centric ontological plane and within the framework of a phased taxonomy discipline. This approach presents a knowledge architecture that grasps legal reality not through the static classification of individual documents, but through the dynamic flow of economic and commercial transactions conducted by institutions. *Mecellem*’s most distinctive feature in this universe is that it structures transactions from the outset in accordance with the institution's operational reality and international workflow practices. For example, when a merger and acquisition (M&A) transaction or project finance process is initiated, the system structures the transaction's specific phases—preparation, due diligence, negotiation and signing, closing, and post-closing integration stages—as a taxonomic backbone. Thus, what is presented to the user is not merely a classification space, but also the ideal workflow of the transaction in question as accepted in global practices, offered as a structural reference framework. While transactions are classified in a multi-layered manner taking into account the institution's commercial objectives, party configurations, and financial architecture, documents produced within the scope of the transaction also semantically indicate at which critical milestone of the relevant process one is situated. In this way, the taxonomy ceases to be a rigid classification mechanism that imposes an external order on transaction parties; it transforms into a broad ontological framework that grasps law from a transaction engineering perspective, integrates global workflow practices into institutional memory, and can be dynamically restructured along with the progression of transactions.

---

[66] *See* Holmes, *supra* note 12.

*Mecellem*'s *Malumat* context refers to a dynamic structure that transforms the system's taxonomic architecture from being a static data repository into a continuously self-updating living legal intelligence and compliance mechanism. In today's rapidly expanding regulatory environment, legal knowledge is not limited solely to legislative texts; judicial precedents, administrative authority decisions, *Official Gazette* publications, sectoral announcements, and various regulatory documents produced by public institutions are dispersed across numerous public sources. In addition, open-access scientific and professional publications such as academic journals, theses, lecture notes, and opinion-type works also constitute an important reference domain in terms of legal interpretation and application.

The *Malumat* context is designed precisely as a knowledge flow mechanism that systematically tracks this broad public knowledge universe and continuously feeds the taxonomic structure. Within this framework, legislative amendments, supreme court decisions, administrative regulations, and open-source academic or institutional publications incorporated into the system function not as static data entries that fix the taxonomic hierarchy; but as dynamic updating elements that enable the reinterpretation of existing categories in line with the new legal reality. Within this architecture, each regulatory text, decision, or publication is tracked not only at the content level, but also in terms of its normative effect and application context, and is positioned within the system at a micro-taxonomic depth. Thus, compliance activities cease to be a manually conducted tracking burden carried out in a dispersed manner; they become a natural reflex of the taxonomic structure. The *Malumat* context, by instantly transmitting legal and regulatory developments in the external world to taxonomic roots, ensures that the system remains continuously current and operational. In this respect, *Mecellem* conceives legal knowledge not merely as an archived data set, but as a dynamic knowledge architecture that brings together open-source legal and academic knowledge flow with institutional decision-making processes.

*Mecellem*'s knowledge architecture is not limited solely to classifying existing data; through the *Müellif* universe, it also brings the production process of legal knowledge under a taxonomic discipline. *Müellif* is conceived within the system as a 'document production universe' and refers to a dynamic production space that forms the foundation for contract types, document types, and file structures used in commercial life. This production process operates in an integrated manner with autonomous agent systems defined within *Mecellem*.

The taxonomic criteria defined for each file, document, or contract type constitute a guidance framework for these agent systems. When a document is being prepared or updated, the system detects legislative changes in the external world, new judicial decisions, or new customs emerging in commercial practices, associates them with relevant taxonomic headings, and integrates them into the document as contextual references. Thus, when a contract draft is being prepared in the *Müellif* universe, the taxonomy does not merely function as a template that divides the text into technical sections; it also incorporates current legal risks, relevant precedents, and legislative provisions pertaining to the contract type in question into the document's semantic structure at the moment of production. Through this approach, *Müellif* transforms legal drafting from being an individual skill domain into an institutional production ecosystem fed by taxonomic precision and system-supported currency. Thus, the document *becomes not a completed and fixed text, but a living knowledge asset* that is continuously updated and evolves through the taxonomic roots that feed it.

The integrated structure revealed by all these contextual universes demonstrates that *Mecellem*'s understanding of taxonomy is not a classification model aimed at passively storing legal knowledge within a static archival order. On the contrary, this approach is based on a functional taxonomy paradigm that aims to make legal knowledge accessible, producible, and recontextualizable in a manner that adapts to the pace of daily institutional operations. In *Mecellem*, taxonomy is conceived not merely as a technical shelf system that organizes documents; but as a multi-axial knowledge architecture that enables processes of access, production, comparison, and contextualization.

This structure, which brings together alphabetical accessibility, chronological depth, and typological axes indicating process phases, transforms legal knowledge from a library-like static storage order into a knowledge network that is navigable across multiple axes, regroupable, and restructurable according to context. Thus, classification ceases to be merely a preliminary stage of legal reasoning; it becomes an integral part of it. *Mecellem*'s taxonomic ground therefore offers not merely a technical organizing mechanism, but also a broad epistemic framework that enables legal knowledge to move between different contexts. In this respect, *Mecellem* redefines law not by fixing it within a single categorical order, but as a multi-layered, dynamic, and context-sensitive knowledge architecture.

### *2. From the Order of Being to the Ontological Design of Digital Entity*

*Mecellem*, which conceives the legal domain as a multi-layered classification network on the plane of dynamic and functional taxonomy, builds this structure not merely as a technical organizing model, but upon a deeper *ontological* design. This design rethinks the systematic understanding of being in Aristotelian doctrine through al-Fârâbî's approach to the order of knowledge and the ranks of being, and carries it into the context of the age of artificial intelligence. *Being* is not a collection of randomly heaped objects; it is a hierarchical, relational, and conceptually grounded order. *Mecellem* likewise positions legal reality—norms, transactions, institutions, and relations—within such an order: Each concept is placed in a specific ontological status, each relation in a specific contextual position, and each provision on a specific plane of being. Thus, taxonomy ceases to be a superficial classification technique; it transforms into an ontological architecture that determines how legal being is designed, positioned, and represented.

The questions of what being is and how it 'exists' have been among the most fundamental issues of human thought. In the history of philosophy, the answers given to these questions have not only established metaphysical systems, but have also formed the ground for inquiries into the nature of knowledge and the possibility of truth. Today, the fundamental issues that artificial intelligence systems encounter when working on large datasets (corpus) are in fact the counterparts of these ancient debates on the plane of artificial reasoning ecosystems: Questions such as what a data object is, what kind of ontological structure it possesses, and how knowledge can be derived from this structure are problematics not only in technical terms, but also in terms of the design of being and the construction of knowledge.

Within this framework, *Mecellem* has developed an approach that addresses the relationship that artificial reasoning ecosystems establish with knowledge not only at the level of technical analysis, but also together with its philosophical origins. In particular, the epistemological analyses concerning the distinction between existence and essence[67], and the relationship between knowledge and truth[68], shaped in the line of thought extending from Aristotle to Ibn Sînâ and al-Fârâbî in the classical metaphysical tradition, offer a theoretical source of inspiration in the classification and interpretation of digital data structures. In this context, al-Fârâbî's approach, which systematically interprets this metaphysical heritage, provides an analytical perspective to *Mecellem*'s ontological design. The idea of 'necessary being'[69], which occupies a central place in classical metaphysical literature, is addressed here not in its direct theological sense, but in terms of the structural logic it carries: While in this tradition of thought 'necessary being' is the ontological source and necessary foundation of all other beings; in *Mecellem*, the ontological framework is positioned as the epistemic prerequisite for consistent and verifiable knowledge production. In other words, it is not the metaphysical content of the concept of 'necessary being', but its structural principle pointing to the '*impossibility of groundlessness*' that has been adapted to digital environment ontology. This approach aims to establish meaning production processes not only through the structural organization of data, but through the conceptual presuppositions that make this organization possible; thus, through the analogical relationship it establishes with the philosophical tradition, Mecellem attains a methodologically distinct position within the knowledge systems of the age of artificial intelligence.

---

[67] Herbert A. Davidson (ed.), Alfarabi, Avicenna, and Averroes, on Intellect: Their Cosmologies, Theories of the Active Intellect, and Theories of Human Intellect, (New York, NY, 1992; online edn, Oxford Academic, 31 Oct. 2023), https://doi.org/10.1093/oso/9780195074239.001.0001. According to Davidson, al-Farabi set forth the existence-essence distinction in a systematic manner, but did not fully operationalize this distinction within a metaphysical system. Avicenna took this fundamental distinction of al-Fârâbî and developed the thesis that existence is 'added' to essence, and this thesis became one of the cornerstones of the ontology of existence in Islamic philosophy. Indeed, while in al-Fârâbî the existence-essence distinction was used more as an analytical tool, in Avicenna this distinction became a metaphysical system. In this context, although this distinction, which forms the foundation of *Mecellem*, is closer to Avicenna's systematic version, in this article the existence-essence distinction is attributed to al-Fârâbî by virtue of his being the founder of this distinction.

[68] Al-Fârâbî divides knowledge into two main categories: The definition of a concept (*taṣawwur*) and the rendering of judgment about that concept (*taṣdīq*) are different operations. The system must first define concepts ontologically correctly, and then produce epistemically valid judgments about the relations between these concepts. *See supra* note 17.

[69] Fârâbî, *Kitâbu'l-Hurûf* (Presidency of the Manuscripts Institution of Türkiye 2015).

This position is based on *an epistemological construct that prioritizes ontology*. *Mecellem*'s approach aims not at a relative mental design detached from being in the construction of 'legal reality', but at an *onto-epistemic order* grounded in being. [70] For law determines the relationship that a subject establishes with another subject and that a subject establishes with an object. Therefore, legal order is fundamentally an ontology of relations; the norm is the formalized expression of these ontological relations. Within this framework, the priority is the determination of data ontology; because epistemological inquiry can be conducted consistently only after the categories of being have been clarified.

Al-Fârâbî's system demonstrates that being is the foundation of knowledge and that knowledge can only be built upon the correct order of being, that is, according to him, the order of being is the order of knowledge. Being and knowledge are inseparable; the hierarchy of knowledge reflects the hierarchy of being. The epistemological order is the projection of the ontological order. In this respect, law's regulation of subject–subject and subject–object relations is not merely a normative activity, but an ontological positioning. *Legal reasoning* first determines which beings and which types of relations exist; then it interprets these relations and places them within a normative framework.

In *Mecellem*'s *digital corpus* work, this approach demonstrates that data ontology must form the foundation for knowledge inference processes. The tripartite understanding of being that al-Fârâbî sets forth in the *Book of Letters* (*Kitâb al-Ḥurûf*)[71] — being as a proposition in the mind, the object actually existing in the external world, and essence — offers a powerful theoretical framework in the analysis of the relationship that *Mecellem* establishes with the data object. A data object is not merely a textual unit; it is an ontological representation of a network of specific subjects, objects, and the relations between them. Therefore, legal knowledge production is based on the correct ontological classification of relations; *epistemological consistency* depends on the preservation of this *ontological commitment*.

### *3. Ontological Priority in the Construction of Knowledge: Epistemic Disruption in Ontological Void*

One of the fundamental ontological and epistemological issues encountered when working with digital *corpora* in the age of artificial intelligence is by which constitutive principles and in which systematic order the architecture of knowledge should be constructed. *Mecellem*'s point of departure has been shaped precisely around this problematic: to first question what exists and how it exists for the production of meaningful and processable knowledge in the data universe; then to establish the epistemic framework of this existence. In this respect, by prioritizing a philosophically grounded order of construction in the artificial reasoning ecosystem, *Mecellem* brings methodological discipline to meaning production processes.

---

[70] In classical legal thought, the hierarchy of norms (for example, in Kelsen's pure theory of law, the concept of the basic norm occupies a central place and is accepted as the fundamental element that guarantees the structural coherence of the legal system. *See* Kelsen, *supra* note 16. The approach proposed in this article, however, is a normative interpretation that envisages a fundamental transformation of the traditional hierarchical structure in legal theory. In particular, the integration of artificial intelligence with legal theory requires an ontological reform in which ontological structures will assume a constitutive role for the formal representation of legal knowledge. For a similar discussion regarding this proposed 'ontological transformation,' *see* Joost Breuker & Rinke Hoekstra, *Ontologies as Missing Link Between Legal Theory and AI & Law*, Legal Knowledge and Information Systems: *JURIX 2004: The Seventeenth Annual Conference* 125 (A. Lodder & L. Mommers eds., 2004).

[71] Fârâbî, *supra* note 69.

Behind this methodological orientation lies the question of how the relationship between being and knowledge is established in classical philosophy. While adapting Aristotle's ontology to the Islamic metaphysical framework, al-Fârâbî demonstrated that being is not a genus like essence and developed a conceptual distinction between being (*wujûd*) and essence (*mâhiyyah*): whether something 'exists' (*being*) and what that thing 'is' (essence) are different questions. Although the conceptual distinction between being and essence is inseparable in reality, making this distinction at the analytical level is necessary. This distinction represents one of the original contributions of al-Fârâbî's interpretation of Aristotle.

Al-Fârâbî reaches the conclusion that the existent has three meanings: being as a category in the sense of the *categorical classification of a data object*, *truth value* as a proposition that exists in the mind and corresponds to the external world, and the thing's possession of an essence outside the mind, that is, its *real-world counterpart*. To proceed to epistemological inquiries without laying this ontological foundation means constructing a structure without foundation. The hierarchy of being determines the hierarchy of data, and this hierarchy forms the foundation of all subsequent operations. The determination of being at the ontological level must come before how this being is to be known at the epistemological level. The priority intended here is not a temporal sequence, but a logical necessity.

This philosophical ground is not only conceptual, but also offers a methodological framework for how disciplines should be structured for systematic knowledge production. The classification of sciences that al-Fârâbî sets forth in his work the *Enumeration of the Sciences* (*Iḥṣâʾ al-ʿUlûm*)[72] is one of the most concrete examples of this framework. In the Enumeration of the Sciences, al-Fârâbî defines the science of logic as an instrument[73] that tests whether the intellect makes errors in perceiving what is real, while actually making a deeper proposition: Correct reasoning is possible only if it is based on a correct conception of being.[74] Logic is methodology; but the ground upon which methodology is built is ontology. Without knowing 'what' something is, correct knowledge about it cannot be produced. Therefore, knowledge production must begin with a sound understanding of 'being'; otherwise, epistemology remains without foundation.

---

[72] Al-Fârâbî, *Enumeration of the Sciences* (Türkiye İş Bankası Kültür Yayınları 4th ed. 2020). Al-Farabi's classification of the sciences is based on the ultimate purposes of the sciences ('theoretical sciences aim to know the truth, while practical sciences aim at living well'), whereas the *Mecellem* taxonomy and ontology are based on functional and operational criteria. While Al-Farabi's classification is *teleological* (purpose-oriented), *Mecellem*'s classification (and ordering) is function-oriented. This difference reflects a fundamental distinction between classical philosophy and modern *knowledge engineering*.

[73] Murat Uyanık, Ayla Aksoy, *Farabi'nin Medeniyet Tasavvuru ve İhsa al-Ulum*, İSAM Veri, https://isamveri.org/pdfdrg/D02420/2015_1/2015_1_UYANIKM_AKYOLA.pdf *(last visited Feb. 14, 2026)*.

In Islamic philosophy, the concepts of 'being' (*wujûd*), 'existent' (*mawjûd*), and 'conscience' (*wijdân*)[75], constitute a multi-layered semantic field that touches not only ontological, but also epistemic and moral dimensions. The word 'mawjûd' is derived from the Arabic root w-j-d (وجد) and carries the meanings of 'that which has been found', 'that which is present', 'that which is perceived'; 'wijdân', which comes from the same root, contains the meanings of finding, becoming aware, and inner consciousness. This etymological proximity points to a structural isomorphism between ontology and inner perception: The existent is not merely an object located in the external world; it is also a phenomenon that is 'found' in consciousness, that is witnessed. In this framework, to deny being is not merely an ontological negation, but can also be read as a form of epistemic blindness and moral insensitivity. For if 'wijdân' is the capacity to find what 'exists' and to bear witness to it, to deny being means the suspension of this capacity. Thus a conceptual parallelism can be established between ontological denial and consciencelessness: The denial of being is not merely a metaphysical position, but can also be evaluated as the rejection of responsibility toward what 'exists'. This isomorphic reading makes visible the intrinsic connection between ontology and ethics; it asserts that there is a structural continuity between accepting being and developing an ethical sensitivity toward it.

This intrinsic connection is indispensable for *agentic* structures today. An artificial intelligence agent, when making decisions, cannot produce consistent decisions without clearly determining which categories of being it works with, which types of relations it recognizes, and which ontological order it accepts. If the ontological status of the data object is uncertain, the *agent* operates only through superficial patterns; this produces not a 'decision' in the true sense, but a statistical prediction. Therefore, in *agentic* systems, epistemological accuracy is directly dependent on ontological clarity. Al-Fârâbî's understanding of the unity of being and knowledge thus becomes a current technical principle for *Mecellem*'s digital corpus and legal knowledge architecture: Order requires rules, rules require being; every system established without defining being weakens its own epistemic foundation. Epistemic justice, in fact, today seeks to ensure the establishment of the balance between being and conscience.

[74] This proposition is directly related to Al-Farabi's distinction between necessary (*apodictic*) and probable (*dialectical*) reasoning. Necessary reasoning provides the principles of metaphysics and physics because these principles derive from the necessary structure of being. Therefore, correct reasoning depends on the correct apprehension of being. Al-Farabi's epistemology posits that 'truth can be attained through the orderly arrangement of reason'; this arrangement corresponds to the ontological order of being. *See* Ahmet Yusuf Yılmaz, *Fârâbî'nin Epistemolojisi, Ontolojisi ve Felsefeye Katkıları,* 15(7) Turkish Studies 3167, 3171 (2020). This approach is being reinterpreted in AI systems through the integrity of concept-context. *Also see* Nadja Germann, *Al-Farabi's Metaphysics*, Stanford Encyclopedia of Philosophy (Summer 2024), https://plato.stanford.edu/archives/sum2024/entries/al-farabi-metaphysics/reviving-sciences.html *(last visited Feb. 14, 2026)*.

[75] Ali Durusoy, Mahmut Kaya, *Vücûd*, Türkiye Diyanet Vakfı İslâm Ansiklopedisi, (https://islamansiklopedisi.org.tr/vucud--felsefe) *(last visited Feb. 16, 2026)*.

### *4. The Ontological Architecture of Mecellem*

Today, especially in the age of artificial intelligence, the flow of legal information has turned into a vast ocean of data consisting of millions of documents, minutes, contracts, emails, and communication records. However, this quantitative increase does not ensure the more robust construction of legal meaning; on the contrary, as data multiplies, the risk of the fragmentation of meaning increases. Because meaning does not arise from mere data abundance; it arises from the ontologically consistent, rational, and systematic relating of data. The fundamental paradox of the modern artificial intelligence age emerges here: while information increases, meaning fragments.

The solution to this crisis must be sought in ontology before epistemology.[76] Al-Fârâbî's theory of knowledge offers a constitutive framework precisely at this point. Al-Fârâbî asserts that being and knowledge must be grasped on two different planes and distinguishes the positions of objects in the mind and in the external world. This distinction forms a strong epistemological ground for the duality of *T-Box* (*terminological structure*) and *A-Box* (*assertional claims*) in *description logic*[77] systems in modern computational ontology.[78] Al-Fârâbî's effort to separate universal logical rules from the particular grammar of language is an early philosophical precursor to the ideal of universal knowledge representation in Tarskian semantics.[79] While logical rules are based on a universal structure, linguistic expressions are contingent and context-dependent.[80]

---

[76] For ontology engineering, *see* Grüninger, Mark & Fox, Mark S., *Ontological Foundations for Enterprise Modelling*, in Design and Implementation of Intelligent Manufacturing Systems (Hadj-Alina *et al*. eds., 1995).

[77] John F. Sowa, *Knowledge Representation: Logical, Philosophical, and Computational Foundations*, Brooks/Cole (2000). *Also see* Marvin Minsky, *A Framework for Representing Knowledge*, MIT Artificial Intelligence Laboratory Memo No. 306 (June 1974), http://hdl.handle.net/1721.1/6089.

[78] For '*Description logics*' *also see* Giuseppe De Giacomo & Maurizio Lenzerini, *TBox and ABox Reasoning in Expressive Description Logics*, in Principles of Knowledge Representation and Reasoning: Proceedings of the Fifth International Conference (KR '96) 316 (Luigia Carlucci Aiello *et al*. eds., 1996).

[79] Tarskian semantics, developed by logician Alfred Tarski, is a foundational theory that defines with mathematical precision what it means for a statement in a formal language (e.g., logic or mathematical language) to be 'true'. To formalize this idea, Tarski employs the concepts of interpretation (model), which specifies which objects correspond to which symbols in a structure, and satisfaction of formulas under that interpretation. For further details, *see* Alfred Tarski, *The Semantic Conception of Truth and the Foundations of Semantics*, *Philosophy and Phenomenological Research*, Vol. 4, No. 3, 341-376 (Mar., 1944).

[80] It should be noted that Al-Farabi's distinction between existence and essence is presented as a metaphysical necessity, whereas the distinction between *T-Box and A-Box* emerges as a functional modeling choice. For Al-Farabi, this distinction reflects a reality inherent in the nature of being, while in digital ontologies, this distinction is a design decision made for the efficiency and consistency of knowledge representation.

Al-Fârâbî's distinction between *ḥurûf* (elements that constitute logical structure) and *names* (elements that designate beings) resembles the distinction between conceptual structures and concrete entities in contemporary legal ontology. In this context, the *first intelligibles* (*al-maʿqûlât al-ûlâ*) denote the conceptual constants with universal validity that the intellect obtains by abstracting from objects in the external world. In *Mecellem*, this plane corresponds to the *T-Box*: the ontological core of legal concepts is defined here.

On the legal plane, concepts such as 'lesion', 'ownership', 'acknowledgment of debt', or 'public interest' constitute the ontological constants of law and the stable axis of meaning. In his '*Core of Certainty*' thesis[81], one of the cornerstones of legal positivism, H.L.A. Hart asserts that legal concepts contain an indisputable core for the vast majority. There is a strong analogical parallelism between this *normative stability* defined by Hart and Al-Fârâbî's *first intelligibles* with universal validity that the intellect obtains by abstracting from the external world. Although they represent different intellectual approaches, both thinkers accept that the legal order is built upon certain '*semantic invariants*'. The existence of s*emantic invariants* is a necessity for the construction of meaning and the correctness of judgment. Without these constants, neither normative stability can be achieved nor can artificial intelligence systems make safe and consistent inferences.

The ontological core is the precondition for the processability of legal meaning in the age of artificial intelligence and only the initial step in the construction of legal meaning. Real and applicable meaning emerges in the relationship that these constants establish with events, facts, and contextual elements.[82] The plane that al-Fârâbî calls the *second intelligibles* (*al-maʿqûlât al-thâniya*) expresses precisely this relational knowledge layer, that is, the layer where knowledge about knowledge resides.[83] The second intelligibles concern the relationship between objects rather than the object of knowledge.[84] In al-Fârâbî's words, 'every knowing is knowing the relation of one thing to another.' Therefore, the depth of legal meaning lies not only in what concepts are, but in how and under what conditions they are related to one another. *Mecellem* takes this one step further. In *Mecellem*, the knowledge plane, in addition to concepts and how they are related, also encompasses logical categories, linguistic forms of expression, and epistemological meta-concepts. This plane corresponds to the *A-Box*: events, facts, actors, spatio-temporal coordinates, terms, jargons, and metadata are positioned here. Thus, the *Mecellem ontology* functions as a '*legal knowledge map*' that defines the essence of legal concepts, all their modes of existence, and the relations among them.

---

[81] H.L.A. Hart, *The Concept of Law*, Oxford Univ. Press 3d ed. (2012). *Also see* Robert S. Summers, *Professor H.L.A. Hart's Concept of Law*, 1963 Duke L.J. 629 (1963) (book review).

[82] Melanie Mitchell, *Artificial Intelligence: A Guide for Thinking Humans*, Farrar, Straus and Giroux (2019).

[83] Ömer Mahir Alper, *Akledileni Akletmek: İslâm Mantık Geleneğinde "İkinci Akledilirler"in Yorumu ve Osmanlı'da Alımlanması Üzerine Bir İnceleme*, 1 Nazariyat: J. Hist. Islamic Phil. & Sci. 35 (2015).

[84] Aristotle's ten categories, which classify the fundamental forms of being in the external world (substance, quantity, quality, relation, place, time, position, possession, action, and affection), have been reinterpreted by al-Farabi by placing them at the center of the science of logic in the context of 'second intelligibles' (logical forms). Al-Farabi defines these categories not merely as a list of physical entities, but as logical tools that show how the mind conceives these entities and expresses them linguistically; thus, he establishes the bridge between 'first intelligibles,' which correspond to individual entities in the external world, and 'second intelligibles,' which are universal concepts built upon them, through this categorical schema. *See* Al-Farabi, *Al-Farabi's Commentary and Short Treatise on Aristotle's De Interpretatione* 54-58 (F.W. Zimmermann trans., Oxford Univ. Press 1981); *also see* Fârâbî, *el-Makulat: Kategoriler* 35-40 (Mübahat Türker-Küyel çev., Türk Tarih Kurumu Yayınları 1990).

The operational threshold of this ontological architecture is the *decomposition* process in *Mecellem*. Decomposition makes the construction of meaning possible by breaking down the legal text into its ontological layers. The system analyzes legal reality, first, on four fundamental planes: (i) *concept*, (ii) *event*, (iii) *fact*, (iv) *term and jargon*.

'*Event*' is the concrete interaction that occurs in a specific time and place. '*Fact*' is the provable proposition derived from the event and subject to legal evaluation. '*Concept*' is the abstract and universal category of this proposition. '*Term and jargon*' is the linguistic concretization of the concept in a specific context. This decomposition ensures that legal meaning is constructed not superficially, but in an ontologically grounded manner. Meaning is not ready-made within the data; it is reconstructed within context through correct ontological decomposition.

The second dimension of decomposition consists of *metadata and NER* (*named entity recognition*) processes. *Metadata* structures compliant with *Dublin Core* standards[85] are extracted from documents; in addition, bibliographic information with legal specificity, actor relations, citations, and spatio-temporal coordinates are also meticulously determined. Who produced a document, in which process it is situated, and with which normative network it is associated are fixed on the ontological plane. This stage is not a simple information management technique, but an epistemological necessity. Because access to context is only possible when ontological coordinates are correctly determined.[86]

This two-stage structure—*ontological fixation* and *contextual decomposition*—represents a level of epistemic achievement that conventional large language models often cannot reach. Unlike systems that only perform pattern inference from the linguistic surface, *Mecellem* makes the ontological layers of legal knowledge visible and explicitly reveals how the decision progresses from which conceptual constant to which relational context. For the correctness of a legal decision is directly dependent on the capacity to access current and accurate context.

---

[85] *Dublin Core* is an international metadata standard that facilitates the discovery and management of digital resources. This standard provides a set of 15 core and flexible elements such as title, creator, and date to describe a resource, aiming for interoperability across systems. For more detailed information, *see* Weibel, S. L., & Koch, T. (2000). *The Dublin Core Metadata Initiative: Mission, current activities, and future directions*, D-Lib Magazine, *6*(12). https://www.dlib.org/dlib/december00/weibel/12weibel.html (*last visited Mar. 1, 2026*). *Also see* Godby, Schrum, & Smith, *The Dublin Core Metadata Initiative: An Overview, 2202*

[86] Semantic legal metadata provides information that aids in the understanding and interpretation of legal provisions. Such metadata is of great importance for the systematic analysis of legal requirements and their adaptation to changing contexts. *See* Amin Sleimi, Nicolas Sannier, Mehrdad Sabetzadeh, Lionel Briand, Marcello Ceci & John Dann, *An Automated Framework for the Extraction of Semantic Legal Metadata from Legal Documents,* arXiv preprint (January 30, 2020) *https://arxiv.org/abs/2001.11245* *See* also Franz Baader, Diego Calvanese, Deborah L. McGuinness, Daniele Nardi & Peter F. Patel-Schneider (eds.), *The Description Logic Handbook: Theory, Implementation, and Applications,* Cambridge Univ. Press, 2d ed. (2007).

Therefore, the decomposition of legal reality into the layers of *concept, event, fact, term, jargon, metadata*, and *NER* is not merely a technical operation; it is the guarantee of accuracy, transparency, and especially the principles of explainable artificial intelligence (*XAI*). [87] This layered structure makes the decision process retrospectively traceable; the transition from ontological constants to the network of contextual relations can be clearly followed. Thus, meaning is sought neither solely within the data nor in the abstract norm itself; meaning is constructed in the balance between *ontological stability* and *contextual dynamism*.

In this framework, *ontology* expresses a broader plane than *taxonomy*. Ontology does not merely classify; it also establishes qualitative, hierarchical, and logical connections between *concepts*.[88] *Concepts* function as building blocks; however, the ability of these building blocks to form a meaningful and coherent whole depends on correct ontological positioning. Without ontological placement, conceptual integrity cannot be achieved; without contextual analysis, this integrity cannot remain alive.

The ontological architecture of *Mecellem* therefore rests on both a normative foundation and operational precision: While the *first intelligibles* represent the ontological core and conceptual constants of law, the *second intelligibles* represent the contextual movement and relational dynamism of legal meaning. The plane of *concept* constitutes the stable order of the *first intelligibles*; *events, facts, terms and jargons, metadata, NER*, and the relational network they establish constitute the dynamic field of the *second intelligibles*. Decomposition is the *epistemic bridge* between these two planes.

Without ontological constants, context becomes chaotic; without contextual analysis, ontology becomes rigid. Through the application of the *T-Box/A-Box* structure to legal ontology, the system can simultaneously verify both the consistency of fundamental legal concepts and the reality in concrete events. *Mecellem*, by establishing precisely this delicate balance between ontological stability and epistemological dynamism, presents an *ontodynamic* architecture that enables the reliable reconstruction of legal meaning in the age of artificial intelligence.

---

[87] On how layered structures strengthen the foundations of explainable artificial intelligence, *see* John Grant, *Explainable Reasoning with Legal Big Data: A Layered Framework*, 30 Artificial Intelligence & Law 435 (2022); Finale Doshi-Velez & Been Kim, Towards A Rigorous Science of Interpretable Machine Learning, arXiv:1702.08608 (Feb. 28, 2017), https://doi.org/10.48550/arXiv.1702.08608.

[88] Beyond simple taxonomic classification, ontologies offer rich structural relationships that encompass multiple classification, inheritance, attribute-value relationships, and connections between concepts. Through these features, they can represent the multiple facets inherent in legal terms and provide a robust framework for the parsing and clustering of documents. In contrast, simple *lexicons* or relational databases cannot provide this ontological depth. *See* Joost Breuker & Radboud Winkels, *Use and Reuse of Legal Ontologies in Knowledge Engineering and Information Management,* 30–31 (Univ. of Amsterdam 2003); Rinke Hoekstra, Joost Breuker & Alexander Boer*, The LKIF Core Ontology of Basic Legal Concepts,* Legal Knowledge And Information Systems: Jurix 2007: The Twentieth Annual Conference 43, 43–52 (A. R. Lodder & L. Mommers eds., 2007). *Also see* Nicola Guarino, Formal ontology, conceptual analysis and knowledge representation, International Journal of Human-Computer Studies, Volume 43, Issues 5–6, (1995) https://doi.org/10.1006/ijhc.1995.1066.

### B. *Ontodynamic Reconstruction of Semantic Context*

#### 1. *Semantic Fluidity – Mecellem's Dynamic Knowledge Networks*

*Mecellem*'s taxonomic order and ontological architecture offer a framework that conceptually fixes the legal domain of being; however, this fixation does not yet produce knowledge itself. Ontology determines 'what exists' and within what structural order of relations these beings are situated. Epistemology, on the other hand, raises the question of 'how and under what conditions true knowledge about this being can be attained.' Therefore, the semantic layer is not merely an extension of ontology; it is the operational plane of the transition from ontology to epistemology. Meaning emerges here not as a fixed content, but as the product of the dynamic encounter between ontological constants and contextual interaction.

Gadamer's hermeneutic approach philosophically grounds this transition. Truth is neither objective data enclosed in the text nor the arbitrary construction of the subject; it is a disclosure that 'comes to language' within historical consciousness. In this sense, Gadamer's conception of truth draws both from the Heideggerian idea of unconcealment (*aletheia*) and from the Hegelian emphasis on historical consciousness.[89]

*Mecellem*'s semantic layer establishes precisely the digital counterpart of this hermeneutic understanding: the legal text is placed into a conceptual domain of being fixed by ontological categories; however, knowledge is produced through the contextual processing of this fixation. Semantics, in this respect, plays a decisive role in reflecting the most accurate reality of concepts. [90] As Johnson observes, semantics makes us 'wise against words.' In other words, semantics takes us beyond literal interpretation to teleological interpretation that takes into account the purposes in the enactment of law. Here, the ontology–epistemology distinction is determinative. Ontological constants (*first intelligibles; T-Box plane*) constitute the invariant core of legal concepts. However, epistemological production (*second intelligibles; A-Box* and relational layer) arises from the configuration of these concepts within concrete events, actors, time, space, and normative context. Epistemological truth depends not only on the correct definition of the concept but on the correct contextual placement of that concept.[91] For it is not the 'concept' but the 'context' that carries meaning. For example, while the concept of 'garden' in everyday language expresses merely an element of property, in zoning law it transforms into a technical category determining construction conditions, in inheritance law into an asset element included in the estate, and in neighbor law into a relational legal status positioned at the center of boundary, encroachment, and interference disputes; that is, its meaning arises not from the concept itself, but from the normative context in which it is placed. Therefore, the semantic layer locates the source of knowledge not in purely textual content but at the intersection of ontological structure and the network of contextual relations.

---

[89] Hans-Georg Gadamer, *Truth and Method* (Joel Weinsheimer & Donald G. Marshall trans., Continuum Publ'g Gr. 2d rev. ed. 2004) (1975).

[90] Wendell Johnson, *People in Quandaries: And Why They Are There,* ETC A Review of General Semantics 1 No 2, 79 (1946).

[91] The emergence of meaning in contextual embedding confirms Wittgenstein's fundamental thesis in the philosophy of language: The meaning of a word is its use in the language. Legal concepts, too, acquire their real meanings only within concrete contextual relations; abstract definitions, when detached from the context of use, carry no epistemic value. *See* Ludwig Wittgenstein, *Philosophical Investigations* § 43 (G.E.M. Anscombe trans., Basil Blackwell 1953).

*Mecellem*'s *dynamic knowledge networks* constitute precisely the infrastructure of this epistemic production process. Legal meaning emerges not from a linear data flow but from a multi-layered network of relations established among concepts, norms, actors, events, and metadata, both ontologically and temporally. This network structure enables *semantic fluidity*: the meaning of concepts is reweighted according to context and temporality; relations are dynamically updated; new data is positioned within the existing ontological and temporal framework, preserving epistemic consistency. Thus, the system neither fixes meaning nor distributes it in an uncontrolled manner. Meaning is reproduced within a fluid yet disciplined structure.

This point is of critical importance from the perspective of the *agentic AI* framework. A goal-oriented agent does not merely match data; it produces decisions. Decision production is an epistemological claim: an acceptance that a particular judgment is correct in a specific context and time. If ontological constants are not clear, the agent miscategorizes concepts; if semantic context cannot be dynamically reconstructed, the agent may render judgments based on superficial similarities and ignoring temporality. *Mecellem*'s *dynamic knowledge networks* both ontologically fix the agent's decision process and contextually and temporally update it.[92] This enhances epistemic reliability and explainability.[93]

---

[92] The operational effectiveness of micro-agents relies on the coordinated functioning of functional sub-components such as routing agents and research agents. In this context, *playground* environments constitute a critical experimental space for testing micro-agentic structures, developing use cases, and conducting iterative improvement processes. However, the functionality of this system can only be realized if platform-based AI architectures can be integrated into the agentic structure in question. In other words, platform-based AIs can assume a meaningful role within agentic orchestration to the extent that they possess the capacity to respond to this system. For this integration to be achieved, it is imperative that the system be connected to a network infrastructure. Database structures may remain static on their own; however, the dynamic nature of agentic interaction necessitates that the system be configured through a flexible and continuously updated network architecture. Therefore, designing micro-agentic systems within a dynamic network structure emerges as a structural necessity both for ensuring the continuity of data flow and the consistency of inter-agent interaction. *Mecellem*'s dynamic knowledge networks respond to this necessity.

[93] In this context, Floridi and Cowls, based on their comparative analysis, have emphasized the necessity of a new principle in addition to existing AI ethical principles. The authors have defined this new principle as '*explicability*' and stated that it encompasses both intelligibility in the epistemological sense (as an answer to the question 'how does it work?') and accountability in the ethical sense (as an answer to the question 'who is responsible for it working this way?'). Within this framework, it has been stated that the principle of *explicability* plays a critical role in ensuring the transparency and trustworthiness of AI systems. *See* Luciano Floridi & Josh Crowls, *A Unified Framework of Five Principles for AI in Society*, 1(1) Harv. Data Sci. Rev. (2019), https://doi.org/10.1162/99608f92.8cd550d1.

Another important distinctive contribution of *Mecellem* is that it reorganizes the legal order from being a static codification system into a dynamic knowledge network. The conceptual counterpart of this choice is the deliberate adoption of the concept of '*network*' instead of '*platform*' in software architecture. While the term '*platform*' evokes a certain fixed structure and unidirectional operational space, '*network*' denotes a structure that operates through continuous interaction, reweighting, and evolutionary connections. The production of legal meaning occurs within this network through nodes (concepts, norms, actors, events) and relations (contextual, logical, temporal connections). This network structure is not merely a technical tool designed to reduce legal uncertainties; it makes visible to the decision-maker (both human subjects and autonomous cognitive agents) the ontological constants and semantic relations behind the norms. Thus, the decision process proceeds not through mere textual matching but through conceptual positioning and contextual verification. The social function of law is also redefined here: law does not merely establish order, it also produces meaning and reproduces this meaning within changing contexts.

While the classical legal order operates largely within the fixed boundaries of codification, *Mecellem* integrates taxonomy with ontology, ontology with the semantic layer, and semantics with AI-supported dynamic knowledge networks. This integration preserves both the ontological stability and epistemological flexibility of legal knowledge within the same architecture. *Mecellem* positions law not as a frozen set of norms but as a network of meaning that moves together with social and economic transformations, capable of continuously reconstructing itself contextually and temporally.

Al-Fârâbî's approach that 'knowledge is continuously reconstructed under the guidance of the active intellect' finds a concrete and operational counterpart in *Mecellem*'s temporally-dimensioned update architecture in the age of *agentic* artificial intelligence. *Mecellem* not only fixes data at a historical moment; it also continuously tracks them with the parameter of 'temporal validity.' Thus, meaning becomes not a fixed and closed content but a dynamic process redefined together with normative change, jurisprudential evolution, and socio-economic transformations. Since legal texts continuously evolve through multiple temporal versions, this necessity is of vital importance for reducing the risk of temporal inaccuracy in artificial intelligence systems.[94] For example, a contractual provision may gain new layers of interpretation with the change of legislation in force[95] or the differentiation of economic conditions;[96] similarly, a court decision also redefines its position and weight in the legal system in light of subsequent jurisprudence. *Mecellem* not only detects this temporal and contextual mobility; it reprocesses it in every query through *agentic* mechanisms and reconstructs meaning within the current legal order. This temporal modeling protects the system against the risk of temporal inaccuracy faced by large language models (*LLMs*), ensuring that context remains current and accurate. Each unit of knowledge is ontologically added to the system together with the period and conditions in which it is valid; the system distinguishes with precision at the semantic level the difference between past and current law, thus 'true knowledge' is now constructed with temporal awareness.

---

[94] The context of multi-temporal versioning and current validity ensures that a legal text or precedent is temporally accurate when applied to a particular case; this reveals the necessity of *time-stamping* and *version control* in legal knowledge management. *See* Giovanni Sartori, *Legal Reasoning: A Cognitive Approach to the Law*, 5 Treatise of Legal Phil. & Gen. Juris. (2005).

[95] The dynamism of legal context and the necessity of temporal alignment are of vital importance especially in statutory interpretation processes; the meaning of a statute may evolve along with the changing social and technological conditions from the moment it enters into force. *See* William N. Eskridge Jr., Dynamic Statutory Interpretation (Harvard University Press 1994).

[96] The change in economic conditions is directly related to principles such as '*clausula rebus sic stantibus*' in the interpretation of contractual provisions; this reinforces the legal significance of temporality. *See* Reinhard

In conclusion, *Mecellem*'s semantic layer serves as the bridge between ontological constants and epistemological production. Truth is positioned here neither as merely ontological data nor as the product of pure interpretation; truth is disclosed within the temporal balance of ontological structure and contextual interaction. *Dynamic knowledge networks* and *semantic fluidity* are the technical counterpart of this disclosure. Thus, *Mecellem* reconstructs legal meaning within an epistemic architecture that is both faithful to the ground of being and historically and contextually mobile. This framework provides a solid theoretical ground for discussing the source of knowledge, the criterion of truth, and problems of epistemic legitimacy in subsequent sections.

### *2. Ontological Commitment and Epistemic Coherence*

*Mecellem*'s claim to dynamically reconstruct semantic context concerns not only contextual flexibility but also the question of on what ground and within what limits this flexibility will occur. For the continuous reconstructability of context, if it does not rest on an ontological constant and epistemological coherence, *semantic fluidity* can easily turn into arbitrariness. Therefore, the dynamic reconstruction of semantic context is not merely a network logic that expands relations; it is also a control mechanism that ties every contextual update back to the order of being. Ontological commitment and epistemological coherence are two fundamental principles that normatively and cognitively secure *Mecellem*'s semantic dynamism.

The fundamental aim is not merely to define knowledge but to construct it in structural harmony with the order of being it represents.[97] This structural harmony expresses the principle of *ontological commitment*, which is *Mecellem*'s constitutive principle.

---

Zimmermann, *The Law of Obligations: Roman Foundations of the Civilian Tradition* (Oxford University Press 1996).

[97] The concept of *ontological commitment* determines the existential consequences of our logical statements. For instance, when we say 'there exist prime numbers greater than one million,' we commit ourselves to an ontology that includes numbers; we commit ourselves to an ontology that contains the things over which our bound variables must range. *See* Willard Van Orman Quine, *On What There Is*, 2 Rev. Metaphysics 21 (1948), reprinted in From A Logical Point of View, Harvard Univ. Press (1953). *See* also Thomas Hofweber, *Logic and Ontology*, The Stanford Encyclopedia of Philosophy (Spring 2013 Edition), Edward N. Zalta (ed.), URL=<https://plato.stanford.edu/archives/spr2013/entries/logic-ontology/> (*last visited Mar. 1, 2026)*; Phillip Bricker, *Ontological Commitment*, The Stanford Encyclopedia of Philosophy (Winter 2016 Edition), Edward N. Zalta (ed.), URL = <https://plato.stanford.edu/archives/win2016/entries/ontological-commitment/>.

Knowledge cannot be constructed in a manner detached from the domain of being it represents; otherwise, every inference produced rests on abstractions that lack ontological correspondence and misleads the legal decision-making process.

Ontological commitment requires not only the necessity of establishing contact with being but also that the knowledge based on this contact be organized in a coherent manner within itself. Any unit of knowledge can be reliable only to the extent that it remains faithful to the structural patterns of the world of being it represents, and only if it can establish a solid, coherent, and authentic connection with the ontological network of relations in which it exists can it rise to the status of 'true knowledge' in the genuine sense. Conceptual expansions without ontological correspondence lead to inaccurate results in legal reasoning. Therefore, *Mecellem* is designed as an architecture that ties knowledge representation back to the order of being at every stage, continuously monitoring its ontological references. If a datum, for example, contradicts the conceptual family to which it belongs or if its relational context has lost its integrity,[98] the system immediately labels it as '*out of context*' and prepares a comprehensive inconsistency report.[99] This rigorous approach constitutes the essence of the critical processes expressed as '*consistency checking*' and '*inconsistency handling*' in knowledge graph and ontology-based systems.[100] This control process, beyond being merely a technical validation, is the direct and concrete reflection of al-Farabi's teaching of epistemological coherence in the age of artificial intelligence. For the order of every knowledge must exhibit absolute harmony with the order of being it represents; otherwise, the integrity of meaning is shattered and the reliability of knowledge is fundamentally undermined. Therefore, in *Mecellem*, '*truth*' is a multi-layered concept encompassing not only the substantive reality of the data but also the authenticity of the relations established. When a context is severed, the holistic epistemic structure of the system also inevitably suffers a deep wound. In line with this vital necessity, every update within the knowledge graph undergoes a rigorous '*ontological validation*' process[101], which is a kind of legal accounting operation made inevitable by the age of artificial intelligence.

---

[98] The preservation of contextual integrity is an adaptation of the principle of 'contextual integrity' in Nissenbaum's privacy theory to legal epistemology. According to Nissenbaum, the legitimacy of information flow depends on its conformity with the contextual norms to which the information belongs; similarly, the epistemic accuracy of legal knowledge can only be ensured when it is consistent with its ontological and relational context. *See* Helen Nissenbaum, *Privacy in Context: Technology, Policy, and the Integrity of Social Life* 127–157 (Stan. L. Books 2010).

[99] Horrocks' work explains how logical inference systems used in ontology languages such as OWL (Web Ontology Language) can detect such inconsistencies. The system produces reports of 'unsatisfiable class' or 'contradictory assertion' informing the ontology engineer which classes are logically inconsistent. This is a critical part of ontology quality control. The work notes that such inconsistencies typically arise from 'subtle interactions between axioms.' Yet the ontology designer may not have made a logical error when defining each rule or constraint separately. However, when these rules come together, they may contradict each other in unforeseen ways. *See* Ian Horrocks, *Ontologies and the Semantic Web*, Communications of the ACM (2008).

[100] Baader *et al*., *supra* note 86.

[101] In legal information systems, when ontological validation of normative concepts is not performed, the same concept may acquire contradictory meanings in different documents. *See* Hoekstra, Breuker & Boer, *supra* note 88. *See* also Natalya F. Noy & Deborah L. McGuinness, *Ontology Development 101: A Guide to*

*Epistemological coherence* in *Mecellem* concerns not only the internal consistency of knowledge but also its acquisition from the right source, by the right method, and on the right contextual plane. For example, the correctness of a legal norm is directly related to which *corpus* you retrieve it from. If an explicit and codified rule is at issue, the public normative corpus is the primary reference source; attempting to derive the same rule from the enterprise corpus produces an epistemic error. Conversely, if the norm is not explicit and its meaning has been shaped through case law, the method automatically shifts to the judicial precedent layer. Coherence here is not about adhering to a single source; it is about selecting the knowledge production method appropriate to the nature of the source. Similarly, epistemological coherence in contractual relations requires not only examining the signed text but also taking into account how the parties' wills have actually manifested. If a contractual provision has been subsequently amended via email, relying solely on the main contract text produces an ontologically incomplete representation of reality. In this case, the correspondence in the enterprise dataset becomes part of the living context of the contract. An epistemologically coherent system evaluates the normative text together with the actual network of relations; otherwise, it misses the current form of legal reality.

Therefore, epistemological coherence in *Mecellem* is not about imposing a uniform method; it is about remaining within a principled framework while selecting the methodology appropriate to the context in each query. [102] The method may vary, but the criterion for selecting the method does not change. It is determined which knowledge belongs to which ontological layer, and then the retrieval and evaluation protocol appropriate to that layer is executed. This approach establishes a balance between contextual flexibility and structural discipline. Therefore, *Mecellem* does not reduce the correctness of legal knowledge merely to its formal source or normative authority; taking into account the relative value that different knowledge sources carry within context, it constructs meaning through coherence, justificatory capacity, and contextual appropriateness. Epistemological coherence here is not ‘knowing everything from the same place’; it means ‘knowing everything from the right place and by the right method’, that is, ‘knowing the truth’.

This epistemological discipline is not merely a theoretical criterion of correctness; it is also decisive for the reliability of decision-making processes. Particularly in *agentic AI* architectures, the system possesses the capacity to choose among different knowledge sources, to determine which *corpus* to consult, and to decide which methodology to apply. If

---

*Creating Your First Ontology*, Stan. Knowledge Sys. Lab’y Tech. Rep. KSL-01-05 (2001). Modern knowledge graphs automate such ontological checks with ‘reasoner’ components, providing continuous validation. *See* Baader *et al*., *supra* note 86. In the field of law, this process prevents the misclassification of norms or their decontextualized use, contributing to the construction of justice with semantic consistency. *See* Sartori, *supra* note 58.

[102] In Habermas’ theory of communicative action, the legitimacy and validity of norms depend on their being contextually justifiable within communicative rationality. Norms are considered legitimate only when they are contextually validated in discursive processes involving relevant stakeholders. *Mecellem*’s methodological pluralism carries this principle of contextual rationality into digital legal systems, ensuring that each query is addressed with an epistemic method appropriate to its own context. *See* Habermas, *Between Facts and Norms: Contributions to a Discourse Theory of Law and Democracy* 107–118, 287–328, William Rehg trans., MIT Press (1996).

this selection process is not grounded in the principles of ontological commitment and epistemological coherence, the agent may produce an inference that appears correct from the wrong source but is contrary to the context. For example, turning to the case law-oriented interpretation layer when an explicit statutory provision exists, or relying solely on the textual surface while disregarding the living context of the contract, undermines the accuracy of the decision. Within *Mecellem*'s *ontodynamic* semantic network, epistemological coherence functions as a governance principle that regulates which knowledge layer the agent will consult and when. Thus, the decision is not only technically producible; it becomes ontologically grounded and contextually justifiable.

### *C. Epistemic and Methodological Regime: The Semantic Protocol of Legal Meaning*

Once the ontological constants and the ontodynamic reconstruction of semantic context have been established, the inevitable question is: How, and within what methodological order, can we know a piece of legal knowledge? *Mecellem*'s epistemic regime answers precisely this question. This regime does not reduce legal meaning merely to the formal authority of the normative source or to textual expression; it proposes a layered epistemological order that takes into account from which *corpus*, with what contextual weighting, and within what retrieval logic the knowledge is obtained. Thus, knowledge ceases to be an abstract citation of a norm or a fixed content; it becomes an output that is contextually validated, whose ontological references are verified, and which is methodologically justified.[103]

---

[103] The concise observation expressed as '*our failure to arrive stems from our lack of method*' may appear in legal thought as a statement pointing to a practical deficiency, yet it actually exposes a much deeper ontological deficiency. The issue here is not merely failing to reach a goal; the issue is that the ontological and epistemological ground for how to construct the knowledge leading to the goal has not been established. For in the legal tradition, the concept of *usûl* (method/procedure) does not merely denote a technical procedure. The root of the word, 'asıl' (*asl*), means the foundation, essence, and ontological ground upon which something rests. Therefore, *usûl* is not simply the sum of steps to be followed; it is a constitutive epistemic architecture that shows from which principles knowledge arises, upon which conceptual foundations it is built, and by which method it gains meaning. In this sense, there is a deep connection between *method* (*methodos*) and asıl/ontology: method is not merely a research technique, but the order of principles concerning how being is to be grasped. Thus, the absence of *usûl* is not merely a methodological deficiency; it is the failure to establish the ontological arrangement of being and knowledge. For this reason, '*lack of method*' (usulsüzlük) in legal thought often arises not from the misapplication of rules, but from the insufficient establishment of the ontological and epistemological foundation upon which legal knowledge rests. In this regard, usûl is not only the path of legal thought, but also the epistemic bridge extending from being to knowledge. Indeed, in modern legal philosophy and epistemology literature, this relationship between method and the ontological ground of knowledge has been discussed through different concepts. Kelsen's *pure theory of law* demonstrates that for legal knowledge to establish a coherent normative structure within itself, it must rest upon a hypothetical 'basic norm' (grundnorm), showing that legal methodology is actually founded upon a certain ontological assumption. *See* Kelsen, *supra* note 16. The concept of '*rule of recognition*' developed by Hart in *The Concept of Law* similarly points to a meta-level principle that determines which norms are to be considered valid in a legal system; this principle functions as a constitutive methodological framework that determines how legal knowledge is to be constructed. *See* Hart, *supra* note 81. Dworkin, in *Law's Empire*, treats law not merely as the application of rules, but as a holistic structure of

Classical legal epistemology operates largely on two axes: it either relies on the source of the norm (such as statute, case law, doctrine) or leans on the method of interpretation (literal, systematic, teleological). However, in the age of artificial intelligence, legal reasoning has transformed into a contextual problem-solving process that moves among multiple *corpora*, taking into account versioned texts, institutional data flows, and temporal change together. *Mecellem*'s epistemic regime therefore focuses on the *verifiability of context* and *the coherence of method* rather than the authority of the source. The fact that a norm is explicitly regulated in statute versus implied through case law requires different epistemic paths; the legal effect of a contractual provision versus an email amendment between the parties cannot be resolved on the same epistemic plane. Epistemic coherence here is not fidelity to a single method; it is the systematicity in selecting the method appropriate to the context.

At this point, *Mecellem* combines epistemology with methodological pluralism. The Mecellem architecture does not operate with a single and fixed method; it is designed as an architecture that works on multiple methodological planes that sometimes complement, sometimes exclude, sometimes subsume, and even in certain contexts compete with one another. *Mecellem*'s *neurosymbolic configuration* unites symbolic legal modeling and statistical learning approaches within a single and holistic epistemic framework. The *symbolic layer*, while preserving ontological commitment, secures the categorical boundaries of concepts, the hierarchical structure of normative relations, and the internal coherence of legal inference. In contrast, the *neural layer* produces a flexible and context-sensitive capacity for meaning-making through operations such as *embedding*, pattern recognition, semantic similarity analysis, and contextual reweighting. What integrates these two layers is *Mecellem*'s *agentic decision architecture*: The agent is not a passive computational unit; it is a higher-level orchestration layer that is goal-oriented, analyzes context, and organizes multi-step reasoning processes. Within this architectural structure of *Mecellem*, legal knowledge graphs make visible the ontological positions of concepts and entities and the relations among them; *semantic chunking* and *embedding* techniques

---

meaning that emerges through the interpretation of legal principles and values, and argues that the principles underlying legal interpretation actually constitute the ontological fabric of law. *See* Ronald Dworkin, *Law's Empire*, Cambridge (Massachusetts)/London (England): Belknap Press of Harvard University Press (1986). Similarly, in the field of philosophy of science, Karl Popper emphasizes that scientific knowledge arises not only from observational data, but from certain methodological assumptions that center on the principle of '*falsifiability*,' thereby revealing that method is the constitutive principle of knowledge. *See* Popper, *supra* note 25. Thomas Kuhn's concept of '*paradigm*' shows that scientific knowledge is shaped not only by methodological techniques, but by fundamental assumptions that determine within which ontological and conceptual framework research is to be conducted. *See* Thomas S. Kuhn, *The Structure of Scientific Revolutions*, Foundations of the Unity of Science (Volumes I—II of the Encyclopedia) (Second ed. 1970). In the hermeneutic tradition, Gadamer's work *Truth and Method* argues that method is not merely a technical tool, but is itself the ontological horizon within which understanding and interpretation occur, thereby clearly revealing the connection between method and the understanding of being. *See* Gadamer, *supra* note 89. In this context, the approach that usûl in legal thought should be understood not merely as a procedural technique, but as a constitutive principle that establishes the ontological and epistemological architecture of legal knowledge, finds strong counterparts in modern legal theory and epistemology literature as well.

restructure legal elements dispersed across the textual surface into contextual clusters; and *retrieval* mechanisms dynamically assign weight to the relevant *corpus* according to the nature of the query. Under agentic control, which retrieval strategy will be selected, which nodes on the *knowledge graph* will be activated, and which normative source and which contextual parameter will be prioritized are systematically determined. Thus, epistemology ceases to be an abstract theory of knowing; it acquires the character of *an operational semantic protocol* that governs data layers, normative references, and contextual variables by weighting them in each query. Through this orchestration, the model does not merely produce statistical output; it is oriented toward the production of legal decisions that are structured with normative purpose, ontologically grounded, and contextually justified.

The distinctive aspect of *Mecellem*'s epistemic and methodological regime is that it makes this process formulable and traceable. Legal meaning is not a fixed content; it is an output that manifests within the dynamic equilibrium of ontological constants, contextual data, and retrieval strategies. Therefore, the system operates with a logic of *semantic equilibrium* that recalculates *corpus* weights, temporal validity, and relational density in each query. The *agentic* structure, by managing this equilibrium in a goal-oriented manner, determines which knowledge source is epistemically more authoritative and makes the decision process traceable. Thus, legal knowledge is transformed not merely into the correct answer, but into the correct answer reached by the correct method—that is, into 'genuine knowledge.'

### *1. Ontological Mapping of Meaning: Legal Knowledge Graphs*

As stated above, *Mecellem* is not merely a software architecture; it is an epistemic discipline that continuously tests under what ontological conditions meaning acquires legitimacy. Drawing from the Fārābian ordering of knowledge, the system addresses meaning not merely at the level of 'what does the text say?' but through the question: 'to what order of being, to what normative structure, and to what relational totality does this expression correspond?' Thus, interpretation at the hermeneutic level is grounded at the ontological level. Each data element is evaluated not as a content in itself, but within the conceptual ordering, normative position, and contextual network to which it belongs.

When this ordering inherent in the nature of legal texts, hierarchical reference chains, inter-normative citation structures, temporal version layers, and contextual relations are taken into account, the superiority of the *legal knowledge graph* over classical relational databases stems not merely from a technical difference, but from an ontological one.[104] Relational databases store data through tables, rows, and columns; relations are limited to predefined schemas and are often represented by linear connections between two entities. However, legal meaning arises from multi-layered, cross-referenced, and contextually reweighted relations. The knowledge graph represents this multidimensional structure through a *node* and *edge* architecture; thus, the system ceases to be merely a structure that stores data; it transforms into a representational order that constructs meaning within relational patterns.

[104] André Valente, *Types and Roles of Legal Ontologies, in Law and the Semantic Web: Legal Ontologies, Methodologies, Legal Information Retrieval, and Applications* (V. Richard Benjamins *et al*. eds., Springer 2005). *Also see* Pompeu Casanovas, Monica Palmirani, Silvio Peroni, Tom van Engers & Fabio Vitali, *Semantic Web for the Legal Domain: The Next Step*, Semantic Web Journal, Vol. 7, No. 3, 213–227 (2016), https://doi.org/10.3233/SW-160224

In the deep layer of ontological mapping in *Mecellem*, the distinction al-Fârâbî makes between existence (*wujûd*) and essence (*mâhiyyah*) provides an analytical framework. A legal entity—for example, a contract or a court decision—is an '*existent*' on the ontological plane; what it is, that is, its *essence*, is determined through the properties defined in the knowledge graph. However, the *identity* of the entity is not constituted solely by its essence; that entity becomes concrete through the position it acquires within a specific context, within a specific network of relations, while answering a specific question. While the properties defined in the knowledge graph correspond to *essence*, the relational positioning of that entity in a specific query, its weighted connections, and its active context constitute its identity. Thus, each legal entity acquires meaning not only through 'what it is,' but through 'how it is positioned in what context.'

*Mecellem* transforms al-Fârâbî's distinction between *first and second intelligibles* into a contemporary *legal knowledge graph* architecture, thereby endowing the analysis of legal texts with both theoretical depth and operational precision. In *Mecellem*, not only concept families, *metadata* layers, and *named entity recognition* outputs, but also *events, facts, terms, jargon*, and the contextual relations they form are essential elements of the graph structure. This layer, which al-Fârâbî defines as *second intelligibles*, focuses not on the object of knowledge, but on the relations among objects, conditions, and modes of positioning. In *Mecellem*, each of these is potentially represented as a *node* or an *edge*. A norm, a contractual clause, a party statement, an email, a court interpretation, or a technical term—all take their place within the contextual network. This structure acknowledges that legal meaning arises not only from conceptual constants, but from the dynamic connections these constants establish with concrete events and relations.

By being built upon a knowledge graph architecture, *Mecellem* constitutes one of the most concrete and distinctive examples in terms of *neurosymbolic AI* and *agentic* AI. Because at the foundation of knowledge graphs lies not merely the storage of data, but the modeling of relations among concepts and meaning through a semantic layer—that is, the ontological and taxonomic representation of data. This approach does not merely relate data; it renders meaning, together with hierarchy and context, processable and inferential by both human subjects and autonomous cognitive agents. Especially in fields such as law that require complex, multi-layered, and multi-hop reasoning, the coexistence of conceptual ontology and taxonomy reveals that *Mecellem* offers the most conducive framework in terms of both *neurosymbolic AI* and *agentic* AI. *Agentic framework*s without ontology or with weak taxonomic foundations experience limitations in meaning production and contextual inference from dense and complex datasets. Because in *open-ended* and dynamic legal networks, contextual and semantic relation maps are as necessary as fixed rules. The semantic modeling of *Mecellem*'s knowledge graphs and its robust ontological scaffolding enable agents not only to operate on data, but also to take into account inter-conceptual relations, normative context, and epistemics.[105]

Moreover, this mapping in *Mecellem* is not static. Each query reconfigures the relations on the graph; which nodes will be activated, which connections will gain priority, and which contextual layers will be weighted are determined according to the nature of the question.[106] In other words, with each new query, a different *semantic topography* emerges from the same dataset. Therefore, *Mecellem*'s ontological map is not a fixed schema; it operates as a context-sensitive, dynamic, and self-reorganizing structure. The map does not change; however, the paths on the map, focal points, and relational densities are reorganized with each query. This transforms the system into a highly dynamic mapping architecture capable of reconstructing meaning each time.

Behind this dynamic mapping architecture lies the multi-layered epistemic structure that *Mecellem* constructs. The *epistemic map* created in *Mecellem* consists of three fundamental planes functionally integrated with one another: the conceptual layer, the contextual layer, and the semantic layer[107] The *conceptual layer* corresponds to al-Fârâbî's

---

[105] *Knowledge graphs* and *neurosymbolic integration* provide critical grounding for *agentic AI* systems. While *agentic AI* systems based purely on large language models achieve impressive results in many tasks, emerging research demonstrates that significant benefits are gained from the integration of symbolic knowledge representations, particularly knowledge graphs, with neural components. *See* Artur d'Avila Garcez and Luís C. Lamb, *Neurosymbolic AI: The 3rd Wave*, arXiv preprint (2020) https://arxiv.org/abs/2012.05876; Brenden M. Lake, Tomer D. Ullman, Joshua B. Tenenbaum & Samuel J. Gershman, *Building Machines That Learn and Think Like People*, Behavioral and Brain Sciences (2016) https://arxiv.org/abs/1604.00289; Jiayuan Mao, Chuang Gan, Pushmeet Kohli, Joshua B. Tenenbaum & Jiajun Wu, *The Neuro-Symbolic Concept Learner: Interpreting Scenes, Words, and Sentences From Natural Supervision*, International Conference on Learning Representations (ICLR) (2019) https://arxiv.org/abs/1904.12584. Knowledge graphs represent structured knowledge as networks of entities (nodes) and relationships (edges), thereby explicitly encoding how different concepts are related to one another. When integrated with *agentic AI* systems, knowledge graphs provide grounding, anchoring abstract language model reasoning to concrete, organizational knowledge that reflects domain-specific relationships and constraints. *See* also Timothy Berners-Lee, James Hendler & Ora Lassila, *The Semantic Web*, Scientific American, Vol. 284, No. 5, 34–43 (May 1, 2001), https://doi.org/10.1038/scientificamerican0501-34.

[106] Each query or event input creates a semantic clustering in specific regions of the knowledge graph. The system, by quantitatively measuring these clusterings, determines in which legal topics knowledge and interpretation are concentrated, and in which areas meaning becomes sparse and ambiguous. This dynamic state technically reflects the dynamic nature inherent in legal interpretation: If, during a particular period, for example, the usage intensity of the concept of 'force majeure' and the number of related decisions increase, this increase indicates that new judicial decisions, legislative changes, or social events are adding new and critical coordinates to the universe of meaning. The system maps these movements in real time; whenever the context changes, legal meaning is also repositioned and settled anew. Thus, *Mecellem* ceases to be a static knowledge repository or a static knowledge graph, and becomes a semantic ecosystem that continuously reinterprets itself and undergoes evolution.

[107] *Mecellem*'s epistemic mapping approach is currently being tested through experimental applications in different legal fields, and its competence level is being evaluated: Contract Management (Mukavele): In trials conducted on international contracts with complex structures such as FIDIC and EPC, the system performed semantic inference with high accuracy by defining each provision within its own contextual framework and relational network. Judicial Decision Analysis: 'Similar case density maps' have been created from large datasets consisting of thousands of judicial decisions; similarities in meaning and interpretation have been

plane of *first intelligibles* and constitutes the system's *T-Box* structure; here, fundamental legal concepts such as 'contract,' 'property,' 'act,' 'obligation' are defined with ontological precision, their hierarchical relations are determined, and they are grouped within concept families. The *contextual layer*, in turn, reflects al-Fârâbî's plane of *second intelligibles* and represents the *A-Box* structure; metadata belonging to concrete documents, party relations, factual elements, evidence, and temporal and spatial variables are positioned and interrelated on this plane. The *semantic layer* is the surface that dynamically unites these two planes; it represents the operational plane where meaning is computed in real time, where each new query or information input creates a unique '*semantic path*.' Thus, while each piece of information in the system is ontologically anchored in its respective layer, it can make logical transitions to other layers; this movement, which can be described at the philosophical level as a '*process of intellection*,' is operationalized at the technical level through a *directed graph traversal* mechanism. This tripartite structure renders the flow of meaning computable, traceable, and auditable; thus, dynamic topography and ontological stability are preserved together within the same architecture.

This dynamic structure aligns with Niklas Luhmann's conceptualization of law[108] as a system (autopoietic system) that produces meaning through its own operations and reconstructs itself through its own internal references. Law is not an external repository of meaning; it is a structure that continuously reproduces meaning within its own codes and relational structures. Within *Mecellem*'s *dynamic knowledge* networks, instead of carrying a fixed meaning 'loaded' from outside with each new piece of data, the system produces meaning by reorganizing the relations within the existing network.

However, this reorganization is not a passive process that occurs spontaneously. In *Mecellem*'s '*agent factory*' [109] section, decision-oriented agents operating on the knowledge

---

visualized intuitively through graphical representations. Institutional Document Management (*Mahfuz*): By extracting the *Dublin Core* metadata fields of internal documents, each document has been positioned on the relational network through the ontological positioning method, and its correct contextual framework has been identified automatically and reliably. These experimental studies strongly confirm the system's capacity to generate meaning from context and its high compatibility with the Farabian knowledge architecture.

[108] Luhmann, *supra* note 54.

[109] Within the *Mecellem* architecture, the processes of data production, decomposition, and processing by artificial intelligence agents are structured through two complementary institutional frameworks: the *data factory* and the *agent factory*. The *data factory* layer establishes the system's semantic data pipeline, enabling the decomposition of the corporate data universe at ontological and functional levels. This layer consists of two distinct functional modules. The first of these, the *Mecellem data factory*, atomizes the corporate data incorporated into the system by decomposing it into its semantic components on the basis of a legal perspective. In this process, documents, events, actors, concepts, and normative relationships are decomposed and positioned within *Mecellem*'s legal taxonomy and knowledge graph structure; thus, the data ceases to be merely textual content and is transformed into an ontological representation layer conducive to legal meaning production. The second module, the *NewMind data factory*, approaches the same data universe from a corporate governance perspective and re-atomizes the data in a manner that makes visible the contextual relationships needed particularly by C-level executives in their decision-making processes. Thus, the data is transformed into a functional knowledge structure not only for legal interpretation but also for strategic management and corporate decision-making mechanisms. In contrast, the *Agent Factory* layer represents the architecture that determines at which thresholds, to which datasets, and under which contextual conditions

graph make choices regarding which nodes will be activated in each query, which relations (edges) will be strengthened, which contextual layers will be prioritized, and which *corpus* will gain weight. The regular monitoring and optimization of these choices, together with the robustness of the ontological and epistemological order, ensures both the rapid and economical provision of the correct decision—one of the other distinguishing features of the Mecellem system. Meaning does not reside in the fixed structure of the map; it emerges in this continuous reconfiguration process managed by agents.[110]

Thanks to this structure, the system does not merely recognize the term 'force majeure' appearing in a judicial decision at the lexical level; it can analyze, within relational integrity, how it has been interpreted in which contract types, in which historical periods, in which supreme court precedents, and under which contextual conditions. Meaning arises not from superficial matches, but from the balance between ontological constancy and contextual dynamism. Legal knowledge thus ceases to be a static collection of texts; it transforms into a dynamic semantic universe that is topologically organized, reshaped with each query, and produces meaning within the relation of *existence–essence–identity*.

After the ontological architecture and semantic fluidity of legal knowledge graphs have been established, the question arises of how this structure functions not only as a representational but also as an effective retrieval mechanism. For a knowledge graph produces epistemic value only to the extent that it enables the return of the right context at the right time to human subjects or epistemic cognitive actors. At this point, *Mecellem*'s *retrieval* architecture constitutes the second critical layer that transforms ontological mapping into operational context reconstruction.

---

the artificial intelligence agents operating on the *knowledge graph* will have access. This layer functions as an *'agent orchestration layer*' that designs the threshold mechanisms and orchestration rules regulating the agents' behaviors of data retrieval, interpretation, and processing on the knowledge graph. Through this holistic architecture, data is produced on a semantically structured ontological ground; and artificial intelligence agents are able to operate on this structured knowledge universe in a controlled and context-sensitive manner.

[110] The decision-oriented agent architecture discussed in this study and the processes of contextual prioritization, relationship weighting, and query-based reconfiguration conducted on the knowledge graph are implemented by an application called *Muayene* developed within the *Mecellem* ecosystem. *Muayene* is designed as an agent-based decision support layer that analyzes the query context and dynamically configures the relevant node-relationship structures, thereby enabling the continuous development of meaningful inferences on the knowledge graph.

*Mecellem's* retrieval capability transcends the classical *Retrieval-Augmented Generation (RAG)*[111] approach and is based on a *GraphRAG-based*[112] topological retrieval

---

[111] For *RAG*, *see* Patrick Lewis, Ethan Perez, Aleksandra Piktus, Fabio Petroni, Vladimir Karpukhin, Naman Goyal, Heinrich Küttler, Mike Lewis, Wen-tau Yih, Tim Rocktäschel, Sebastian Riedel, and Douwe Kiela, *Retrieval-Augmented Generation For Knowledge-Intensive NLP Tasks*, Advances in Neural Information Processing Systems, 33, 9459–9474 (2020) https://arxiv.org/abs/2005.11401. Recently, with the expansion of context windows in large language models such as *Claude* (*100K tokens*) and *Gemini* (*1M+ tokens*), the adoption of simple *grep* and file search methods by *agentic code* tools, and the rise of new paradigms such as the Model Context Protocol (MCP), the debate '*Is RAG dead?*' has resurfaced in academic and industrial circles. For these debates, *see* Caroline Boudier, Christina Hsiao, Vivien Tran-Thien, *With Context Windows Expanding So Rapidly, Is RAG Obsolete?*, Dataiku, GENAI&AGENTS (Sep. 26, 2024), https://www.dataiku.com/stories/blog/is-rag-obsolete (last visited Mar. 9, 2026). For the view that in the *agentic* era, old terms such as RAG and *prompt engineering* have become obsolete and traditional *RAG pipelines* have gone out of fashion, *see* Reliable Data Engineering, *RAG Is Dead*, Medium, Jan. 11, 2026, https://medium.com/@reliabledataengineering/rag-is-dead-and-why-thats-the-best-news-you-ll-hear-all-year-0f3de8c44604 (last visited Mar. 10, 2026). However, these arguments overlook the context-sensitive functional transformation of *RAG*: *RAG technology* is evolving from being a simple document access mechanism into a multi-layered knowledge management system. Indeed, according to Ruiwen, '*finding the needle in the haystack*' in a context of 10 million *tokens* is much more difficult compared to a context of 10,000 tokens; *RAG* structurally solves this problem by placing this needle at the point where the model's attention is sharpest. *See* Ruiwen Zhang, *RAG Isn't Dead: Why Long Context Windows Will Not Replace Retrieval-Augmented Generation*, Medium, Mar. 6, 2026, https://medium.com/@rivenzhang_8674/rag-isnt-dead-why-long-context-windows-will-not-replace retrieval-augmented-generation-62ae557dd097, (last visited Mar. 11, 2026). Current technical analyses also confirm this view: while *RAG* systems generate responses in approximately 1 second, long-context models processing millions of *tokens* take 30-60 seconds per query. This makes *RAG* much more suitable for real-time applications. The difference in terms of cost is even more striking: while a typical *RAG* query costs approximately 0.0004 USD per 1,000 *tokens*, the cost of processing large contexts can reach hundreds or even thousands of dollars per month at enterprise scale. *See* Jim Allan Wallace, *RAG Vs Large Context Window: The Real Trade-Offs For AI Apps*, Redis Blog, Feb. 6, 2026, https://redis.io/blog/rag-vs-large-context-window-ai-apps/ (last visited Mar. 11, 2026); Atai Barkai, *RAG vs. Context-Window in GPT-4: Accuracy, Cost & Latency*, CoPilot Kit, Dec. 5, 2023, https://www.copilotkit.ai/blog/rag-vs-context-window-in-gpt-4 (last visited Mar. 11, 2026).

[112] The fundamental distinction of *GraphRAG* from traditional vector-based retrieval methods lies in its capacity to process structured relational data through knowledge graphs. The structural limitations inherent in flat vector representations, particularly their inability to encode contextual relationships and to resolve semantic ambiguities, are overcome by the *GraphRAG* architecture through three core functional competencies: multi-hop reasoning, entity disambiguation, and relationship understanding. These competencies significantly enhance the accuracy and reliability of knowledge retrieval, particularly in domains requiring deep contextual comprehension such as legal and biomedical information processing. It is reported that *GraphRAG* increases accuracy rates by 15-20% in complex domains such as law and healthcare. As of 2025, the *Agentic RAG* paradigm has come to the forefront; it is reported that through the integration of autonomous agents, hallucination rates have decreased by approximately 68% and retrieval accuracy in complex scenarios has doubled. *See* Darren Edge, Ha Trinh, Newman Cheng, Joshua Bradley, Alex Chao, Apurva N. Mody, Steven Truitt & Jonathan Larson, *From Local to Global: A Graph RAG Approach to Query-Focused Summarization*, arXiv:2404.16130 (Apr. 24, 2024), https://arxiv.org/abs/2404.16130; Boci Peng, Yun Zhu, Yongchao Liu, Xiaohe Bo, Haizhou Shi, Chuntao Hong, Yan Zhang & Siliang Tang, *Graph Retrieval-Augmented Generation: A Survey*, arXiv:2408.08921 (Aug. 15, 2024), https://arxiv.org/abs/2408.08921; Tharun Sekar, Kushal, Supprethaa Shankar, Sabah Mohammed, Jinan Fiaidhi, *Investigations On Using Evidence-Based Graphrag Pipeline Using LLM Tailored For USMLE Style Questions*, MedRxiv, The Preprint Server for Health Sciences, May 05, 2025, https://doi.org/10.1101/2025.05.03.25325604; Denise Gosnell & Vivien de Saint Pern, *Improving Retrieval Augmented Generation Accuracy with GraphRAG*, AWS Machine Learning

model.[113] The system operates by matching the relations among concept, event, fact, term, jargon, metadata, and *NER* nodes previously positioned on the knowledge graph with the contextual vector of the query. Thus, the *retrieval* process ceases to be a statistical search based solely on textual similarities;[114] it transforms into a context selection process in which normative and relational ties are activated according to semantic resonance distances. The retrieved content is determined by following an ontologically grounded path on the knowledge graph.

The distinguishing aspect of this approach is that retrieval is integrated with the internal logic of the knowledge graph. Since the knowledge graph represents multiple knowledge layers such as legislation, case law, doctrine, and institutional documents with their semantic weights, the system contextually determines which *corpus* will gain priority in each query. Thus, *retrieval* can distinguish among heterogeneous legal domains; in a normative question, it can prioritize legislation and case law, while in a contractual dispute, it can prioritize party intentions and the relevant transactional context. This ensures that legal meaning is constructed not by chance, but through an ontologically positioned retrieval process.

Another determinant of *retrieval* success is the contextual integrity provided during the data processing and digitization phase. In *Mecellem*, documents are processed through domain-specific small language models capable of recognizing document type and terminological patterns; *OCR* outputs are corrected with semantic context taken into account and ontologically embedded into the knowledge graph structure. As a result, the content retrieved during the *retrieval* phase is not only technically accessible, but also semantically coherent and normatively positioned. Experimental observations demonstrate that this context-sensitive approach significantly increases both the scope of access and the accuracy of answers compared to classical *OCR* and superficial *RAG* combinations.[115]

---

Blog (Dec. 4, 2024), https://aws.amazon.com/blogs/machine-learning/improving-retrieval-augmented-generation-accuracy-with-graphrag/; Aditi Singh, Abul Ehtesham, Saket Kumar & Tala Talaei Khoei, *Agentic Retrieval-Augmented Generation: A Survey on Agentic RAG*, arXiv:2501.09136 (Jan. 15, 2025, rev. Feb. 4, 2025), https://arxiv.org/abs/2501.09136.

[113] *For the combined use of retrieval-augmented generation* (*RAG*) and knowledge graph architecture, *see* Xiangrong Zhu *et al., Knowledge Graph-Guided Retrieval Augmented Generation (KG²RAG)*, arXiv (Feb. 2025), https://arxiv.org/abs/2502.06864; Haoyu Han, Yu Wang, Harry Shomer, Kai Guo, Jiayuan Ding, Yongjia Lei, Mahantesh Halappanavar, Ryan A. Rossi, Subhabrata Mukherjee, Xianfeng Tang, Qi He, Zhigang Hua, Bo Long, Tong Zhao, Neil Shah, Amin Javari, Yinglong Xia, Jiliang Tang, *Retrieval-Augmented Generation with Graphs (GraphRAG)* (2025) arXiv:2501.00309. Also, for the problems that *RAG* approaches may encounter due to the lack of semantic guidance, *see* Chenguang Fang *et al., Towards Escaping from Language Bias and OCR Error: Semantics-Centered Text Visual Question Answering*, arXiv (Mar. 24, 2022), https://arxiv.org/abs/2203.12929.

[114] For the view that the transfer of contextual nodes and relationships to large language models through knowledge graphs provides deeper inferences compared to pure vector search, *see* Shirui Pan, Linhao Luo, Yufei Wang, Chen Chen, Jiapu Wang & Xindong Wu, *Unifying Large Language Models and Knowledge Graphs: A Roadmap*, IEEE Transactions on Knowledge and Data Engineering Vol 36 Issue 7 (2024).

[115] Our studies have revealed that in cases where the semantic representation of data is insufficient, task-oriented agents are unable to read eighty percent of the datasets (corpus) created with classical data preparation methods when performing tasks related to legal matters. This situation demonstrates that meticulous

In conclusion, *Mecellem*'s *retrieval* architecture, by combining *ontological mapping* with *semantic fluidity*, transforms legal meaning into a structure that can be recalled and reconstructed as needed. Meaning is no longer merely a structure represented and 'interpreted' on the graph; it is an epistemic output that is topologically activated with each query and reorganized within normative context. In this respect, *Mecellem* is positioned not merely as a system that stores legal knowledge, but as a holistic semantic system that finds, selects, and reconstructs it according to context.[116]

### *2. Vectorial Mapping of Meaning: Contextual Chunking and Vectorial Representation*

The correct construction of meaning is directly related to how data is segmented and transformed into contextual chunks. Contextual chunking (*semantic chunking*) is not the superficial division of a text into equal-length segments; it is the decomposition of the text into meaning units appropriate to its ontological function and epistemic role. In classical chunking approaches, texts are segmented from beginning to end according to certain technical segmentation rules; document type, the normative function of the document, or its institutional role are often not taken into account. *Mecellem* presents a distinct difference here: Based on *Mecellem*'s functional taxonomy, each document type is evaluated within its own ontological class, and semantic chunking is performed according to a schema specific to that class. Legal meaning emerges through the segmentation of the text in a manner appropriate to its document type and epistemic function; for the same sentence can produce different normative consequences in different document types.

For this reason, in *Mecellem*, for example, a contract text is divided into segments around structural axes such as parties and definitions, performance and obligations, breach and corrective mechanisms, termination and consequences. In contrast, an administrative decision is structured in terms of procedure and authority, cause element, legal basis, reasoning, conclusion, and avenue of appeal. This chunking is not merely a textual

---

segmentation in data processing procedures (such as the textualization of the knowledge base constituting the *corpus* with advanced technologies like *VLLM*s) is of vital importance for task-oriented agents to successfully perform the legal tasks entrusted to them. Seals, handwriting, or signatures on documents may sometimes obscure critical parts of the text, causing an important sentence or word to remain under a seal, signature, or stamp. Similarly, tables, images, or maps contained within texts can also create serious difficulties for task-oriented agents in accurately extracting textual meaning. The meticulous processing of such elements, in accordance with the understanding stated above, is a prerequisite for task-oriented agents to fully perform the tasks assigned to them.

[116] The *retrieval* architecture presented in this section is a conceptual description that outlines the general principles of *Mecellem*'s *ontodynamic* and *semantic* framework. The structure described should not be understood as the fixed and singular form of a specific technical implementation; rather, it should be understood as a high-level architectural framework that aims to maintain the epistemic balance between *ontological stability* and *contextual dynamism*. At the implementation level, the types of models used, vector indexing strategies, graph weighting mechanisms, and fine-tuning processes can be adapted with different technical configurations according to the nature of the context and the requirements of the domain. The purpose of these explanations is not to present a specific technical recipe; rather, it is to make visible the structural role of ontological mapping in the production of legal meaning. The same framework also applies to all technical sections, particularly the section addressing the vectorial fragmentation of meaning.

organization, but the extraction of the ontological map of 'legal reasoning.' Segmentation performed at the micro level (article, paragraph, clause) is repositioned at the macro level (institutional context, process, normative structure). Thus, each chunk preserves both its subjective and objective meaning as well as its position within the system. This multi-layered chunking enables the semantic restructuring of context with each query.

*Semantic fluidity* emerges precisely at this point. First, the text is chunked according to ontological and functional criteria; then these chunks are reassembled according to context with each new query. In other words, the system performs both a deductive and an inductive function: It decomposes into chunks, then constructs a new whole according to the normative and contextual coordinates of the query. Chunks remain as fixed data units within the system; however, which chunk will be associated with which ontological node and which other chunks is redetermined with each query. Thus, each query actually produces a new semantic map over the same data pool. Meaning is not extracted from a static data repository; it is constructed through contextual reorganization.

At this point, the dataset on which the model is trained and domain-specific architectures come into play. Because how the system will read the chunks and with which contextual patterns it will associate them depends on in which semantic universe the model has been trained. What a model can 'understand' is directly related to which structured data universe has been shown to it. The concept of domain-specific *resonance* also gains meaning here: The more intensively the model has been exposed to the ontological and terminological patterns of a particular legal domain, the more accurately it can reconstruct the meaning patterns of that domain.

For example, the sentence 'the request for extension of time has been rejected,' when read in the context of construction law, is interpreted in a context where the contractor may be held liable for delay, a discussion of fault and risk allocation will arise, and guarantee mechanisms may be triggered. The same expression, from an administrative law perspective, means that an administrative act has been established, the decision has entered into force, and the avenue for objection or annulment action against it has been opened. Semantic chunking divides the text into chunks appropriate to these contexts; the domain-specific model, however, reads these chunks within the established meaning patterns of that domain. Thus, the semantic potential produced by chunking is transformed into actual legal understanding through *embeddings* and domain-specific models.

The *embedding* layer constitutes the vectorial infrastructure of this process. The same word can have different neighborhoods in different *domain embedding* spaces. For example, the concept of 'guarantee' shows proximity to the security instruments of the credit debtor in banking law; to the performance guarantees of the contractor in construction law; and to public receivable collection mechanisms in administrative law. The *embedding* space is not a neutral mathematical field; it carries the semantic topology of the domain in which it has been trained. Thus, the distance between concepts is based not only on statistical co-occurrence, but on domain-specific ontological patterns.

A symbolic analytical representation of this integrity can be expressed as follows; $S_{(chunk)}$ represents the contextual contribution of *semantic chunking*; $E_{(embed)}$ represents the semantic depth contribution arising from the domain-specific training of *embedding* models:

$$\delta \times Model_((training\ data)\ ) \times [S_chunk, E_embed]$$

Here, δ represents the semantic contribution coefficient of the dataset on which the model is trained:

$$\delta = w(Contextual\ Richness)\ \times\ q\ (Domain\ Purity)$$

In this representation, the contextual richness of the training data is expressed by w(*contextual richness*), and the extent to which the data belongs homogeneously and purely to the relevant legal domain is expressed by q(*domain purity*). When extracting legal meaning, the model is weighted by these coefficients derived from model training. The system can detect whether the query belongs to a single legal domain or has a heterogeneous and multi-disciplinary structure; accordingly, it decides which *embedding* space and which *domain-specific model* will operate more dominantly. Thus, meaning production is calibrated according to contextual richness and domain purity.

In conclusion, *semantic chunking* and *embedding* are two complementary mechanisms that carry legal meaning from surface to depth. *Chunking* reveals semantic potential by decomposing the text in a manner appropriate to its ontological function; *embedding* transforms this potential into conceptual depth by positioning it within the domain-specific vectorial topology. Meaning is concentrated in the dynamic interaction of these two mechanisms and is reconstructed with each query.

### 3. *The Semantic and Methodological Weighting Equation of Authentic Meaning*

In *Mecellem*, the construction of legal meaning is not a natural output of a singular technical method or the statistical prediction capacity of large language models.[117] On the contrary, meaning is the result of a dynamic equilibrium system in which ontologically structured knowledge layers, epistemic source planes, and methodological tools are weighted according to context.[118] Therefore, the framework proposed here is not merely a '*retrieval*

---

[117] The inadequacy of reducing large language models to mere statistical prediction capacity in legal reasoning is comprehensively addressed in the current literature. Eljas and Tuula Linna emphasize that in the evaluation of legal AI systems, epistemic foundations and knowledge quality must be measured separately from statistical accuracy, and demonstrate that systems that produce purely statistical outputs are insufficient for legal reasoning. *See Eljas Linna, Tuula Linna, Challenges for Generative AI in Legal Reasoning*, arXiv:2508.18880v2 [cs.AI] (Jan., 24 2026)

[118] The view that legal meaning is produced within a multi-layered and dynamic equilibrium system is supported by Villa's dynamic pragmatic theory of interpretation. Villa argues that legal interpretation involves both discovering and constructing meaning; and that meaning evolves through the interaction of case-specific contexts and interpretive roles. This approach demonstrates that meaning is not a fixed content, but rather a dynamic output of contextual equilibrium. *See* Vittorio Villa, *A Pragmatist Theory of Legal Interpretation*, 9 Revus 89 (2010).

*strategy*'; it is a holistic *semantic protocol* concerning from which source, by which method, and with which epistemic priority legal knowledge can be known.[119]

In legal systems generally, legal knowledge is organized at least at four fundamental normative levels: (i) legislation, (ii) case law (jurisprudence), (iii) doctrinal literature,[120] and (iv) the contextual knowledge layer (*enterprise corpus*) consisting of institution-specific practices, contracts, correspondence, factual cases, and inter-actor relations.[121] *Mecellem* treats the knowledge layers at these four levels not as mere datasets, but as meaning layers carrying different epistemic densities. Therefore, legal meaning is produced not according to a fixed hierarchy of these layers, but by activating them with weights that vary according to the ontological and conceptual nature of the query.[122] Epistemic weighting takes into account not only the source type, but also that source's position in the legal system, its temporal currency, and its normative effect.

---

[119] In legal information systems, context-aware query interpretation and the understanding of semantic protocols are being established within an operational framework in the current literature. Context-aware systems treat legal queries not merely as a technical retrieval operation; but as a meaning production process that is dynamically adjusted according to case facts, legal norms, and interpretive objectives. *See* Ramanathan Udayakumar, Haider Abbas, R. Velmurugan, A. Vanathi, N. Sundaram, Ishimbaev Nailevich, *Context-Aware Query Interpretation in Legal Information Systems*. Indian Journal of Information Sources and Services (2025). 15. 122-130. 10.51983/ijiss-2025.IJISS.15.3.14.

[120] The concept of 'doctrinal literature' expressed here does not encompass the inclusion of contemporary works protected by copyright into the system; rather, it comprises texts that have been made accessible in accordance with academic conventions, have entered the public domain, and knowledge sources found in open-source scientific repositories. *Mecellem*'s *dynamic knowledge network* is structured on the basis of full compliance with intellectual property rights; and the sources constituting the system's dataset consist solely of documentation whose legal accessibility has been confirmed and which has been shared through open-source methodology. In this context, *Mecellem*'s semantic parsing process targets not the ownership of protected original works, but the epistemic and ontological organization of publicly accessible legal knowledge.

[121] This multi-layered normative structure of legal knowledge aligns with Mommers' conceptualization of legal knowledge as a structured ontology that incorporates criteria of accuracy, consistency, and reliability. Mommers demonstrates that legal knowledge consists of layers with different epistemic statuses, and that these layers should be evaluated not independently, but within a relational totality. *See* Laurens Mommers, *Applied Legal Epistemology: Building a Knowledge-Based Ontology of the Legal Domain* (Leiden Univ. Press 2002). The view that knowledge layers carry different epistemic densities and that these densities need to be weighted is grounded on a mathematical basis in Wright's work. Wright proposes an update model for legal evidence and knowledge sources that includes weighted authority, citation scores, and temporal decay parameters; this model demonstrates that different knowledge sources do not carry equal epistemic value and need to be weighted according to context. Craig S. Wright, *Bayesian Epistemology with Weighted Authority: A Formal Architecture for Truth-Promoting Autonomous Scientific Reasoning*, (Jun. 19, 2025) https://doi.org/10.48550/arXiv.2506.16015.

[122] The notion that legal meaning should be produced through context-aware dynamic weighting rather than a fixed hierarchy is supported by formal modeling studies. For the formalization of legal semantics as functions of multiple variables; and the dynamic weighting of literal, systematic, teleological, and historical interpretation methods according to case-specific facts and normative priorities, *see* Silvana Castano, Alfio Ferrara, Emanuela Furiosi, Stefano Montanelli, Sergio Picascia, Davide Riva, Carolina Stefanetti, *Enforcing Legal Information Extraction Through Context-Aware Techniques: The ASKE Approach*, Computer Law & Security Review, Volume 52 (2024) https://doi.org/10.1016/j.clsr.2023.105903.

For example, there is a difference in epistemic weight between a *Supreme Court of Appeals Grand Chamber Decision* (*Yargıtay İçtihadı Birleştirme Kararı*) and a local court decision on the same matter; similarly, epistemic density also differs between a precedent decision cited in hundreds of subsequent decisions and a decision referenced only once. The system dynamically determines the contextual weight of a source by taking into account how frequently a source is cited, at what level of courts it is referenced, and how temporally current it is. In this way, sources with high citation impact and normatively binding status are prioritized according to the nature of the query; sources with low citation impact or that have lost temporal currency lose epistemic weight. An analytical representation of this framework at the knowledge base level can be expressed as follows:

$$\text{Legal Meaning} = (\alpha \times legislation) + (\beta \times case\ law) + (\gamma \times doctrine) + (\theta \times institutional\ corpus)$$

Here, the coefficients α, β, γ, and θ represent the *semantic weight* carried by the relevant knowledge layer in a specific query context. For example, if there is a clear and explicit provision in legislation, the coefficient α becomes dominant; if there is a normative gap or interpretive ambiguity, the coefficients β (case law) and γ (doctrine) increase.[123] When a question concerns institutional contracts, party correspondence, or institution-specific technical documents, the coefficient θ—that is, the relevant knowledge layer at the institutional level—can become determinative. In the context of case law-based systems (*common law*), the case law layer gains normative weight, while in the context of Continental European-based systems (civil law), legislation may be prioritized. Thus, legal meaning is produced not according to 'where it is written,' but according to 'which knowledge layer carries the highest epistemic density in that context.'[124]

However, *Mecellem*'s methodology weights not only knowledge sources, but also methods of accessing knowledge and producing meaning. Knowledge graphs (*KG*), ontological inference mechanisms, vector-based similarity searches (*vector database - Vector DB*), classical textual matching (*elastic/keyword search*), and each of the retrieval techniques such as *RAG/GraphRAG* are positioned not according to a fixed order, but according to the structural nature of the query.[125] In a structural and deterministic query (for example, the contractual status of a specific company), the knowledge graph and ontological inference may be almost sufficient on their own; *RAG* is either disabled or operates with minimal weight.[126] Conversely, in an analogical, interpretive, or precedent-based query, semantic proximity and vectorial representation mechanisms gain higher weight. This methodological layer can also be represented analytically within a weighting system as follows:

$$Methodological\ Layer = (\lambda_1\ x\ KG) + (\lambda_2\ x\ VectorDB) + (\lambda_3\ x\ RAG) + (\lambda_4\ x\ TextSearch)$$

---

[123] In cases of normative gaps and interpretive uncertainty, the gaining of weight by the layers of case law and doctrine is grounded philosophically in Dworkin's interpretive theory of law. Dworkin argues that legal principles carry weight based on moral and political justification; he contends that in normative gaps, the judge must find the best interpretation by weighting principles. *See* Dworkin, *supra* note 103; Fanny de Graaf, *Dworkin's Constructive Interpretation as a Method of Legal Research*, Law and Method 12 (2015).

[124] The view that legal meaning should be determined not by the literal text, but according to epistemic density, also finds correspondence in the *Stanford Encyclopedia of Philosophy*'s comprehensive analysis on evidence evaluation in the legal context. This analysis reveals that the weighting of evidence according to relevance, probative value, and source reliability constitutes the foundation of epistemic evaluation; it emphasizes that which knowledge source carries higher epistemic authority in a particular context must be systematically determined. Hock Lai Ho, *The Legal Concept of Evidence*, The Stanford Encyclopedia of Philosophy (Winter 2021 Edition), Edward N. Zalta (ed.), URL = https://plato.stanford.edu/archives/win2021/entries/evidence-legal/. *Also see* Floris J. Bex, *Arguments, Stories and Criminal Evidence: A Formal Hybrid Theory*, Law and Philosophy Library, Springer (2011), https://doi.org/10.1007/978-94-007-0140-3.

Here, the λ coefficients represent the relative importance of retrieval methods determined according to context. Therefore, what is weighted in *Mecellem* is not only knowledge; it is the mode of knowing itself.[127] For the correct context is not the mechanical retrieval of a pre-fixed set of knowledge; it is produced by the dynamic rebalancing of epistemic layers according to the ontological nature of the query. This balance determines not only which source will be used, but also to what extent which methodological tool will be activated. Epistemic weighting therefore enables the *retrieval* process to cease being merely a statistical matching, and to transform into a *semantic protocol* in which meaning is ontologically reconstructed within a normative context.

The third determinative element is the semantic coefficient derived from the dataset on which the model is trained. As stated in the previous section, the model's capacity to produce meaning is directly related to the contextual richness and domain purity of the data on which it is trained.[128] Therefore, the coefficient δ is defined as follows:

$$\delta = w(Contextual\ Richness) \times q\ (Domain\ Purity)$$

---

[125] In a multi-stage retrieval framework that combines lexical matching, semantic ranking, and large language models, it has been empirically demonstrated that the dynamic weighting of methodological tools according to the structural characteristics of the query significantly enhances retrieval accuracy and interpretability. This study supports at a practical level the argument that λ coefficients should be positioned not as fixed values, but in a context-aware manner. *See* Hoang-Trung Nguyen *et al., Multi-Stage Legal Case Retrieval Framework*, arXiv (2025), https://arxiv.org/html/2509.08025v1.

[126] The assumption of a decisive role by semantic similarity mechanisms in analogical and precedent-based queries is supported by the *ACM* study on case-based reasoning systems. This study demonstrates that case-based reasoning systems using similarity weights produce legally relevant arguments and ensure the alignment of artificial intelligence outputs with legal precedents; this explains why vectorial representation mechanisms should gain higher weight in analogical queries. *See* Morgan Gray, Li Zhang, and Kevin D. Ashley, *Generating Case-Based Legal Arguments with LLMs*, Symposium on Computer Science and Law (CSLAW '25), March 25–27, 2025, München, Germany. ACM, New York, NY, USA (2025) https://doi.org/10.1145/3709025.3712216. *See* also Kevin D. Ashley, *Artificial Intelligence and Legal Analytics: New Tools for Law Practice in the Digital Age*, Cambridge University Press (2017), https://doi.org/10.1017/9781316761380.

[127] The view that methodological pluralism is not merely a theoretical eclecticism, but an epistemic necessity that mandates the selection of different methods according to the contextual characteristics of each query, is comprehensively addressed in Sekula's doctoral thesis, which models legal pluralism through a game-theoretic approach. Sekula examines coordination among overlapping legal systems by assigning weights to competing rules and interpretive strategies based on social, economic, and institutional factors; this supports at a methodological level the argument for 'the weighting of the mode of knowing itself.' *See* Pawel Sekula, *Legal Pluralism and Formal Coordination Models* (Ph.D. thesis, Univ. of Turin 2021).

If the question belongs to a single legal domain, is homogeneous and narrow in scope, the domain-specific architecture receives higher weight. Conversely, in a multi-domain, heterogeneous, and interdisciplinary question, the broader-scope model layer is activated.[129] Thus, the system also weights the effects from model training in a context-sensitive manner.

All these layers—the epistemic source plane, the methodological toolset, and the model training coefficient—are dynamically determined by agents that perform query-specific context analysis.[130] The *agentic* structure analyzes the ontological type of the question (normative, institutional, or interpretive), its conceptual density, and its multi-step reasoning requirement, and recalibrates the weights of both the knowledge layers, the methods, and the model coefficient.[131] By automating this multi-layered weighting and contextual reconstruction process, the agents structurally guarantee the epistemic consistency and normative reliability of legal reasoning. However, this guarantee is not merely a technical improvement; it is the expression of a radical transformation in the theoretical foundation of legal knowledge. Therefore, in *Mecellem*, meaning is not a pre-fixed result; it is a semantic equilibrium problem that is resolved anew in each query. [132] For the correct legal answer is now possible not only through access to the correct knowledge, but by bringing that knowledge into context with the correct epistemic weight, in the correct ontological position, and with the correct methodological tools. In this framework, the dynamic adaptation of epistemic weighting is an indispensable prerequisite for legal AI systems to produce correct answers; because context can fulfill its critical function only within this dynamic equilibrium and can enable the authentic construction of legal meaning. This holistic framework can be represented in expanded form as follows:

$$\text{Legal Meaning} = f\ [(\alpha \text{Legislation} + \beta \text{Case Law} + \gamma \text{Doctrine} + \theta \text{Relevant Corpus})\ x\ (\lambda_1 KG + \lambda_2 \text{Vector DB} + \lambda_3 RAG + \lambda_4 \text{TextSearch})\ x\ (\delta\ x\ \text{Model}_{training\ data}, S_{chunk}, E_{embed})]$$

---

[128] The view that the contextual richness and domain purity of the dataset on which the model is trained directly determine its capacity for meaning production is supported by the multi-method integration approach of the LogiKEy framework. LogiKEy, while supporting the orchestration of case-specific weighting and interpretive strategies, reveals that the symbolic layer secures the categorical boundaries of concepts by preserving ontological commitment; whereas the neural layer carries the semantic topology of the data universe on which it is trained. LogiKEy, *Framework for Multi-Method Integration in Legal Reasoning*, 2 Law, Governance & Tech. 3 (2023), https://www.mdpi.com/2813-0405/2/1/3.

[129] For the view that context-dependent source prioritization in legal artificial intelligence systems directly affects decision quality, *see* Li Shangyuan, Shiman Zhao, Zhuoran Zhang, Zihao Fang, Wei Chen, Tengjiao Wang, *Basis Is Also Explanation: Interpretable Legal Judgment Reasoning Prompted By Multi-Source Knowledge*, Information Processing & Management, Volume 62, Issue 3 (2025) https://doi.org/10.1016/j.ipm.2024.103996. This study reveals that the weighting of legal norms according to their relevance and authority must differ between narrow-scope and broad-scope queries; it supports the epistemic rationale for the dynamic transition between the domain-specific architecture and the general model layer.

[130] Explainable artificial intelligence studies on the management of context analysis and dynamic weighting in legal reasoning by *agentic* architecture present a formal model that dynamically weights literal, systematic, teleological, and historical methods of interpretation according to case-specific facts and normative priorities; they operationalize an understanding of *agentic* orchestration that enables adaptive and transparent decision-making. In this framework, *see* Joe Collenette, Katie Atkinson, Trevor Bench-Capon, *Explainable AI Tools For Legal Reasoning About Cases: A Study On The European Court Of Human Rights*, Artificial Intelligence, Volume 317 (2023) https://doi.org/10.1016/j.artint.2023.103861. This study also examines explainable AI tools in European Court of Human Rights cases, arguing that legal reasoning must simultaneously satisfy the criteria of both technical scalability and epistemic defensibility. *Also see* Henry Prakken, An Abstract Framework for Argumentation with Structured Arguments, Argument & Computation, Vol. 1, No. 2, 93–124 (2010), https://doi.org/10.1080/19462160903564592.

[131] The modeling of attack and support relations with weights in computational models of legal argumentation is addressed in detail in the work of Bench-Capon and colleagues. This study formalizes how argument evaluation and dispute resolution can be realized through weighted relations in multi-step reasoning processes; it supports at the computational level the calibration function of *agentic* architecture based on ontological type analysis. *See* Trevor Bench-Capon, Katie Atkinson, Floris Bex, Henry Prakken, Bart Verheij,

In this expanded analytical representation,

- $\alpha, \beta, \gamma, \theta$ represent epistemic layer weights;
- $\lambda_1$–$\lambda_4$ represent methodological tool weights;
- δ represents the semantic coefficient derived from model training;
- *Schunk*; represents the contribution of document-type-specific semantic chunking;
- *Eembed*; represents the embedding depth aligned to the relevant domain;
- *f*; represents the agentic orchestration function (dynamic contextual calibration).

This representation is the theoretical counterpart of the conceptual architectural representation expressed as the *semantic weighting equation of legal meaning* in *Mecellem*: model performance and hardware capacity work together with the epistemic value of the knowledge base and the context-based weighting of retrieval methods.[133] The resulting

*Computational Models of Legal Argument*, Journal of Applied Logics — IfCoLog Journal of Logics and their Applications, Vol. 12 No. 3 (2025). *Also see* Henry Prakken & Giovanni Sartor, *Law and Logic: A Review from an Argumentation Perspective*, Artificial Intelligence, Vol. 227, 214–245 (2015), https://doi.org/10.1016/j.artint.2015.06.005.

[132] The view that legal meaning is not a predetermined outcome, but a dynamic equilibrium problem resolved anew in each query, is supported at the level of legal philosophy by Alexy's balancing theory. Alexy formalizes the weighing of conflicting principles through a mathematical weight formula; this approach, which integrates normative and empirical assessments, reveals that legal decision emerges from the weighted balance of multiple factors and that this balance must be reconstructed in each case. *See* Robert Alexy, *On Balancing and Subsumption. A Structural Comparison*, 16 Ratio Juris (2003). *See* also Lars Lindahl, *On Robert Alexy's Weight Formula for Weighing and Balancing*, Liber Amicorum de José de Sousa e Brito, Almedina (January 2009).

[133] How legal knowledge graphs support relevance ranking and automated reasoning through weighted relations is comprehensively addressed in a study published in 2026. This study demonstrates that the encoding of

answer (*Q_answer*) is not merely a computational output; it is a legal meaning that is ontologically appropriate, epistemologically justifiable, and methodologically traceable.[134]

Therefore, in *Mecellem*, meaning is not derived from a single source; it is constructed within the dynamic equilibrium of multi-layered knowledge planes, multiple methodological tools, and the semantic resonance derived from model training. This equilibrium, the *semantic weighting equation*, is the fundamental principle that makes legal reasoning both technically scalable and epistemically defensible. Undoubtedly, the most critical point here is that the legal knowledge base has been systematically designed from the initial stage to incorporate epistemic weighting. As explained in detail in various sections of the article, the atomization of different sources of law through specific methods and the structuring of these elements on a *knowledge graph* architecture through epistemic weighting metrics constitutes one of the most fundamental elements of the semantic weighting approach. In this framework, merely classifying or relating legal data is not considered sufficient; additionally, the epistemic value, contextual power, and normative impact of each piece of knowledge are also measured and incorporated into the network structure. Indeed, the distinctive founding rationale of the *Mecellem* system emerges precisely at this layer; this approach, which integrates the semantic organization of legal knowledge with its epistemic prioritization within the same architecture, constitutes the foundation of the system's theoretical and technological originality.

### *D. Epistemic Justice and Contextual Accuracy*

Classical legal philosophy mostly defines *justice* through the principle of 'giving each their due' (*suum cuique tribuere*) and identifies justice with the correct distribution of rights. However, 'giving someone their due' first requires correctly knowing what that right is, to whom it belongs, and within which context it gains meaning. Therefore, normative justice presupposes an epistemic order that precedes it. Al-Farabi's principle that 'if the order of knowledge is corrupted, the order of justice is also corrupted' makes precisely this connection visible: The violation of justice often stems not from a volitional deviation, but

---

weighted relations among legal entities, courts, and norms directly affects model performance and recall quality; it presents the theoretical foundations of the knowledge graph architecture that constitutes the technical infrastructure of the *semantic weighting equation*. *See* Zhang, M., Zhao, N., Qin, J. *et al, A Comprehensive Framework For Legal Dispute Analysis Integrating Prompt Engineering And Multi-Dimensional Knowledge Graphs*, Sci Rep 16, 679 (2026). https://doi.org/10.1038/s41598-025-30306-9.

[134] Oren L. Gazit, *Constitutive Knowledge Sources: An Institutional Approach To Epistemic Trust In Opaque AI Systems*. AI Ethics 6, 58 (2026). https://doi.org/10.1007/s43681-025-00930-2. The necessity for legal outputs produced by AI systems to transcend being mere computational outputs and acquire an epistemically justifiable and transparent character emerges as a central argument in Gazit's work addressing epistemic trust in AI systems. Gazit proposes institutional models that evaluate the reliability of AI outputs based on opaque sources; he emphasizes that transparency and normative trust are necessary conditions for the epistemic legitimacy of the meaning produced. In the same vein, *see* Shen Jiaxin, Xu Jinan, Hu Huiqi, Lin Luyi, Ma Guoyang, Zheng Fei, Meng Fandong, Zhou Jie, Han Wenjuan, *A Law Reasoning Benchmark for LLM with Tree-Organized Structures including Factum Probandum, Evidence and Experiences*, Findings of the Association for Computational Linguistics: ACL (2025)

from the incorrect ordering of knowledge.[135] The erroneous establishment of the relationship between right, fact, and norm inevitably leads to the deviation of the judgment as well. Thus, *justice* is not merely a distributive activity realized in the outcome, but primarily an epistemic structure dependent on the process of producing, organizing, relating, and auditing knowledge in the correct context.

The *Mecellem semantic protocol* systematizes this process and explicitly grounds justice on an epistemic foundation. In *Mecellem*, *epistemic justice* rests on three fundamental and unshakable principles. According to the principle of *ontological consistency*, every norm and concept must absolutely conform to the order of being and institutional reality it represents. This also expresses the obligation to preserve the contextual integrity, conceptual consistency, and normative references of legal knowledge. Each query directed to the system does not merely surface the relevant pieces of knowledge; it also meticulously verifies the robustness of the context, the consistency of the relationships, and the integrity of the entire chain of meaning simultaneously. Every legal recommendation or inference produced by artificial intelligence is evaluated on two fundamental axes for the sake of ensuring *epistemic justice*. On the axis of *factual accuracy*, the recommendation is expected to fully reflect the substantive and factual reality, and the raw data is expected to be error-free and reliable; for *epistemic justice* encompasses not only equality of access to knowledge, [136] but also the verification of accuracy and consistency in every link of the knowledge chain. On the axis of *contextual validity*, it is necessary to ensure that the meaning produced is correctly, reliably, and appropriately positioned within the unique legal or social context to which it belongs. In high-risk artificial intelligence applications, the mandating of *context verification*, such as not only accuracy but also contextual appropriateness and the transparency of knowledge graph relationships, underscores the critical importance of this axis;[137] for a piece of evidence positioned in the wrong context, an incompletely related fact,

---

[135] This principle is based on al-Fārābī's understanding of the necessary connection between political theory and epistemology. According to him, social order is the earthly reflection of the order (*niẓām*) of being and knowledge; when the hierarchical arrangement of knowledge is disrupted, that is, when degrees of truth, purposes, and values are not properly positioned, the political and moral order is inevitably shaken as well. Therefore, injustice arises not always from deliberate evil, but most often from a misapprehension of truth or from the failure to position knowledge in its proper place. The model of the virtuous city that al-Fārābī sets forth in *al-Madīna al-Fāḍila* argues precisely that this epistemic order must constitute the foundation of political order; the ruler can establish justice to the extent that he is one who knows the truth correctly and arranges knowledge properly. *Abū Naṣr al-Fārābī, Al-Farabi on the Perfect State: Abū Naṣr al-Fārābī's Mabādiʾ Ārāʾ Ahl al-Madīna al-Fāḍila: A Revised Text with Introduction, Translation, and Commentary* (Richard Walzer ed. & trans., Clarendon Press 1985).

[136] In Rawls's theory of justice, equal access to knowledge and epistemic equality in decision-making processes are fundamental conditions of a just social order. The parties behind the 'veil of ignorance' can agree on just principles only when they are in an epistemically equal position. *See* John Rawls, *A Theory of Justice* 118–123, 136–142, Belknap Press of Harv. Univ. Press (1971). *Mecellem*'s contextual epistemology also carries this principle of 'epistemic equality' into digital legal systems, aiming for structural justice in access to knowledge and meaning production.

[137] McInerney, in discussing how algorithmic systems construct epistemic hierarchies through contextual frameworks, focuses particularly on issues of hermeneutical injustice, the locking of context, the entrenchment of epistemic authority, and the reproduction of existing knowledge hierarchies by digital

or a disconnected normative reference is not merely a technical deficiency; it is a rupture in the epistemic chain of justice and can create *epistemic harm*.[138]

The second principle underlying epistemic justice in *Mecellem* is *dynamic meaning flow*. Accordingly, legal meaning is not static; it is continuously repositioned with new documents, norms, judicial decisions, evidence, and conditions added to the system. Thus, in *Mecellem*, the legal system ceases to be a fixed and immutable system of norms; it becomes a living, continuously self-updating, meaning-based justice ecosystem.

Another prerequisite of epistemic justice is *epistemic verification*. According to the principle of *epistemic verification*, every judgment and decision produced must pass the knowledge-context alignment and temporality test, and this verification must be reportable. Knowledge-context alignment requires that the responses generated by models be supported by the relevant *normative corpus* and *institutional corpus* drawn from *Mecellem* legal knowledge graphs. The temporality test, on the other hand, requires that the *normative* and *institutional corpus* grounding the response be not only substantively relevant, but also factually and legally valid in terms of enactment, amendment, and repeal dates. This test requires that legal knowledge be treated not as a static dataset, but as a dynamic normative structure that is continuously updated and gains meaning within historical context. Therefore, epistemic verification includes a traceable temporal validation mechanism that verifies whether each reference used accurately reflects the normative status valid at a specific point in time.

The ultimate vision of the *Mecellem semantic protocol* is not only to build a consistent and accessible knowledge architecture, but also to establish a framework that enables *epistemic justice* in the complex legal ecosystem of the artificial intelligence age.[139] This vision is realized by preventing not only problems such as *testimonial injustice* and *hermeneutical injustice*[140] that may arise during the presentation of knowledge in digital

---

systems. Tomás McInerney, *The Algorithmic Construction of Epistemic Injustice* (SSRN Working Paper No. 5172655, 2024), https://ssrn.com/abstract=5172655 (*last visited Feb. 22, 2026*). Kasirzadeh, on the other hand, in examining the ethics of knowledge management in AI-supported systems, emphasizes that the decontextualization of data can lead to epistemic injustices. Amir Hossein Kasirzadeh, *Algorithmic Fairness and Epistemic Justice*, Univ. of Bristol (2022), https://research-information.bris.ac.uk/en/publications/algorithmic-fairness-and-epistemic-justice. (*last visited Feb. 22, 2026*). According to Kasirzadeh, ensuring justice in algorithmic systems is concerned not only with statistical equality criteria, but also with epistemological questions such as how knowledge is produced, in what context it is used, and whose epistemic authority is recognized.

[138] Warmhold Jan Thomas Mollema, *A Taxonomy Of Epistemic Injustice In The Context Of AI And The Case For Generative Hermeneutical Erasure*, https://doi.org/10.48550/arXiv.2504.07531 (2024).

[139] For the proposition that the proper processing of knowledge is fundamental to justice in the digital age, *see* Floridi *et al*., *supra* note 93.

[140] Testimonial injustice, as defined by Fricker, leads to the disregard of the statements of marginalized groups in legal systems through the deflation of a speaker's credibility due to prejudices, while hermeneutical injustice results in the inability to properly articulate legal meaning as a consequence of individuals' lack of access to conceptual tools for making sense of their experiences. The concept of epistemic justice is defined in Fricker's work as a framework that examines how individuals are harmed in the processes of knowledge production and sharing. *See* Fricker, *supra* note 39.

systems, but also new types of epistemic injustice specific to artificial intelligence systems,[141] based on the accuracy of data and the complete preservation of the context in which knowledge is presented. In this framework, *justice* ceases to be merely an abstract normative principle; it transforms into a concrete value measured and verified by the uninterrupted and transparent preservation of the principle of contextual accuracy in meaning production processes.

Another link in epistemic *justice* is to secure the '*contestability*' of knowledge. Al-Farabi's '*theory of demonstration*'[142], forms the epistemic foundation of contestability, shedding light from a thousand years ago on the issue of verifiability and questionability of knowledge in digital systems. According to him, *certain knowledge* (*yaqīnī knowledge*) has three fundamental conditions: the belief must be true, the subject must have complete conviction regarding this truth, and most critically, it must be known that the contrary of the belief is in no way possible.[143] This third condition is of vital importance for *epistemic justice* and offers an important framework for how the principle of *contestability* should be structured in artificial intelligence systems. Contesting an artificial intelligence decision should provide not only the possibility of rejecting the outcome, but also the possibility of supporting the objection with evidence and justifications that can question the epistemic foundations of that outcome. In artificial intelligence systems, this means: when a user contests a result produced by the system, the system must be able to clearly display not only the result, but also the data sources it used to reach that result, the contextual relationships, the chain of inference, and the normative references.

---

[141] Recent literature expands Fricker's binary distinction to reveal new types of *epistemic injustice* specific to artificial intelligence systems. Notable among these types are: generative hermeneutical ignorance, the epistemic invisibility arising from the insufficient representation of marginalized groups' experiences in training data; generative hermeneutical access, the restriction of users' access to knowledge and capacity for meaning-making by AI outputs; generative manipulative testimonial injustice, the devaluation of individuals' testimonies through the production of disinformation; and generative amplified testimonial injustice, the reinforcement of existing prejudices and false narratives by AI. One of the most critical dimensions in this context is the concept of generative hermeneutical erasure. The 'view from nowhere' approach adopted by AI systems tends to universalize Western-centric epistemologies, thereby gradually eliminating local knowledge systems and conceptual tools. This phenomenon constitutes an integral part of the process referred to in the literature as *epistemic colonization*: the fact that training data is predominantly of Western origin (WEIRD – Western, Educated, Industrialized, Rich, Democratic) renders AI's mode of thinking and epistemic horizon one-dimensional, thereby also jeopardizing the polyphony of legal decisions. Epistemic colonization is not limited to the loss of cultural or linguistic diversity, but also manifests in the legal domain. In particular, the deliberate disregard or restriction of access to certain open-source legal texts or portions of the literature directly undermines epistemic pluralism. For instance, if an AI-based legal research system processes only the sources of certain publishers or paid databases, the systematic exclusion of diverse academic perspectives or critical literature occurs. This situation corresponds not only to inequality in data access, but also to the deliberate narrowing of the context pertaining to legal interpretation. Consequently, decision recommendations that are ostensibly based on 'correct' knowledge, but are in fact incomplete and manipulative, deepen epistemic injustice.

[142] Fârâbî, *supra* note 17.

[143] According to al-Fârâbî, a demonstrative inference must not only be a formally correct (valid) syllogism, but its premises must also be certain and necessary. The premises of demonstration must possess the following characteristics: (i) They must be true and certain, (ii) They must be the cause of the conclusion (causal connection), (iii) They must be necessary and universal, (iv) They must be self-evident or demonstrated. Al-Farabi, following Aristotle, identifies knowing with knowing the 'cause' of a thing. Merely knowing that a fact exists is 'assent' (tasdik); however, knowing why that fact is so and why it cannot be otherwise is 'demonstrative knowledge' (burhanî ilim). *See* al-Fârâbî *supra* note 17. In this respect, contemporary epistemological research positions al-Farabi as a 'moderate evidentialist.' *Evidentialism* holds that a belief should be accepted only to the extent that it is supported by rational and empirical evidence. Al-Farabi, by defending the absolute authority of reason and demonstration especially in theoretical philosophy, exhibits full alignment with this modern current. *See* Anthony Robert Booth, *Al-Farabi and Moderate Islamic Evidentialism, Analytic Islamic Philosophy Chapter 4, Routledge* (2023); Jalal Paykani, *Evidentialism in Farabi's Epistemology,* Theosophia Islamica, 5(2), (2025) https://doi.org/10.22081/jti.2025.72006.1080;

The effectiveness of the *contestability* mechanism depends on the epistemic transparency of the system. By ensuring this transparency, the *Mecellem semantic protocol* enables the criteria for evidence evaluation to be applied consistently in both al-Fârâbî's '*theory of demonstration*' and modern epistemology. The system not only produces accurate knowledge, but also clearly demonstrates how this knowledge was produced, what epistemic standards it is based on, and why it is contestable. This constitutes one of the most important contributions that *Mecellem* offers for *epistemic justice*: The contestability of knowledge becomes not only a procedural right at the individual level, but also an indicator of epistemic quality at the institutional level.[144] In this context, context verification in legal knowledge systems is not merely a technical requirement, but also a profound epistemic responsibility that forms the foundation of justice.

---

Deborah L. Black, Al-Farabi on the Conditions of Certitude, The Cambridge Handbook of Religious Epistemology, Cambridge Univ. Press (2023). Deborah Black emphasizes that the first three conditions al-Farabi established for certitude evoke the definition of 'justified true belief' (*JTB*), which occupies a central place in contemporary epistemology. In traditional epistemology, knowledge has generally been defined by three components since Plato: first, the subject's belief in a proposition (*belief*); second, that this proposition is actually true (*truth*); third, that this belief is justified, meaning that the subject has good reasons or sufficient evidence for holding this belief (*justification*). Black's observation is that the conditions of certitude that al-Farabi developed in the eleventh century offer a framework similar to this classical tripartite structure. This similarity demonstrates how relevant and sophisticated al-Farabi's epistemological thought is to contemporary debates. However, Black also reveals a critical difference in al-Farabi's approach. As Black notes, in al-Farabi, the traditional 'justification condition' is replaced by a 'knowledge condition.' What does this mean? In the traditional *JTB* definition, for a belief to count as knowledge, it is sufficient to have good reasons or adequate evidence for holding that belief. However, in al-Farabi's approach, the situation is more complex. For al-Farabi, believing in a proposition and having evidence for this belief is not sufficient. Instead, the subject must actually know that proposition. This means that knowledge itself becomes a necessary component within 'certitude.' In other words, al-Farabi demands not merely having good justifications but possessing 'genuine knowledge' in order to attain 'certitude.' Thus, al-Farabi positions the concept of 'certitude' as a higher epistemic status than 'knowledge.'

[144] Epistemic justice is of critical importance not only in interpersonal relations, but also at the institutional level. The principles of representation and participation emphasized in the theories of justice of Rawls and Habermas constitute the institutional dimension of epistemic justice; for a just social order requires individuals' equal access to knowledge production processes and their equal representation in these processes. *See* John Rawls, *A Theory Of Justice (1971); Jürgen Habermas, Between Facts And Norms: Contributions To A Discourse Theory Of Law And Democracy* (1992) (William Rehg trans., MIT Press 1996). In this context, epistemic justice encompasses not only the recognition of individual knowledge claims, but also the just structuring of societal knowledge production mechanisms.

Consequently, when an artificial intelligence output is contested, the system must be able to answer the following questions within the framework of the aforementioned principles: (1) Are the norms and concepts used ontologically consistent? (2) Is the meaning correctly positioned within the current context? (3) Has the chain of inference undergone epistemic verification and is it reportable? These questions grant the *Mecellem semantic protocol* the capacity to offer an effective solution to the 'black box' problem of artificial intelligence models and provide a significant contribution, especially for legal professionals, in legal reasoning processes. The lack of transparency in the decision-making processes of artificial intelligence systems, that is, the inability to explain by which data, algorithm, or chain of logic a decision was made, constitutes a serious obstacle in a field such as law that requires a high degree of justification. *Mecellem*, through *semantic layers, dynamic knowledge networks*, and *epistemic verification*, overcomes this problem by making the decision-making processes of artificial intelligence visible and verifiable within a contextual and relational framework. The system offers not only the result of a decision, but also a detailed logical and semantic map that answers the questions of why and how. This approach enables legal professionals not only to reach a conclusion in AI-assisted analyses, but also to deeply understand the epistemic and contextual foundations of that conclusion. Thus, the *Mecellem semantic protocol*, by strengthening the principles of transparency and accountability in the legal reasoning processes of artificial intelligence, enables legal professionals to adopt a more reliable and justified approach in their decision-making processes.

In this framework, al-Fârâbî 's understanding that the harmony between humanity and the universal order is established through knowledge[145] is carried into the law of the artificial intelligence age and is concretized in the chain established by the *Mecellem*

---

[145] In al-Fârâbî's teleological understanding of politics and knowledge, happiness (*saʿāda*) is positioned as the ultimate end of both the theoretical and practical perfection of the human being; the virtuous city model establishes a necessary parallelism between the proper ordering of knowledge and political justice. *See supra* note Fârâbî 135. The concept of teleology in its classical sense is derived from the Greek terms *telos* (*purpose*) and *logos* (*reason, explanation*), and denotes a mode of explanation that takes as its basis the reason for a thing's existence and the ultimate end toward which it is directed. *See* The Ethics Centre, *Ethics Explainer: Teleology* (Apr. 4, 2022), https://ethics.org.au/ethics-explainer-teleology (*last visited Feb. 22, 2026*). In legal theory, teleological interpretation centers not on the literal structure of norms but on the purpose they aim to achieve, and in this respect is closely related to the natural law tradition. Indeed, in contemporary legal philosophy, '*legal teleology*' is defined as an approach that ties the legitimacy of law not only to positive regulation or social acceptance, but also to the conformity of norms with their purposes. In this framework, the bridge al-Farabi establishes between the cosmic order and the individual-moral order is based on a conception of teleological wholeness in which epistemic order precedes political order. In the age of artificial intelligence, goal-oriented AI legal models similarly take as their basis not only formal correctness, but the alignment of system outputs with normative purpose; this approach can be regarded as the technological projection of teleological law. The teleological conception of justice, as in the Aristotelian and natural law traditions, treats justice not only as a procedural category but also as a consequential one. *See* Byron Kaldis, *Laws versus Teleology*, Encyclopedia of Philosophy and the Social Sciences (Byron Kaldis ed., SAGE Publications 2013), https://doi.org/10.4135/9781452276052.n203. Therefore, the chain of 'correct knowledge – correct context – correct meaning – correct judgment – just society' can be read as a contemporary interpretation of classical teleology that relates the ultimate end of knowledge to social justice.

*semantic protocol*[146]: Accurate knowledge – accurate context – accurate meaning – accurate judgment – just society.[147]

## *IV. ARTIFICIAL INTELLIGENCE AGENTS AS EPISTEMIC SUBJECTS IN MECELLEM*

Traditional human-centered legal practice has entered a process of structural transformation in the wake of developments in artificial intelligence and, more particularly, in the field of *agentic AI*. Where legal knowledge is systematically structured along the taxonomic–ontological–semantic triangle, it appears feasible to position artificial intelligence agents not merely as analytical instruments that extract meaning from texts, but as cognitive actors participating in the execution of legal processes. Within a properly architected legal knowledge framework, the agent's generative cognitive capacity may yield action-oriented outcomes such as drafting decision proposals, classifying evidentiary corpora, weighting normative sources, and governing procedural workflows.

When extract, transform, load (*ETL*) tools that have emerged over the past two decades, advanced hardware architectures and chips[148], speech-to-text and text-to-speech

---

[146] In this chain, justice becomes operative at three interconnected levels that are immanent to one another. At the *epistemic level*, the production of knowledge in a correct and coherent context is ensured; at the *hermeneutic level*, the interpretation and understanding of legal meaning within the correct context is realized; at the *normative level*, the establishment of the judgment to be rendered on the basis of the correct context and meaning that have been produced is secured. These three levels are tightly bound to one another; a breakdown in any one of them leads to the deviation of the others from coherence as well. Therefore, justice can no longer be reduced merely to the exercise of judicial power and authority; it must also be understood as a moral and constitutive function of a reliable, coherent, and contextually integrated knowledge management.

[147] Here, reference is made to al-Farabi's knowledge-justice nexus, that is, to his fundamental premise that correct knowledge makes a correct social order possible. However, this article takes as its basis not the principle that the knowledge-justice relationship al-Farabi establishes in *al-Madīna al-Fāḍila* rests on an ontological hierarchy—that is, the placement of each being and each epistemic subject in a position appropriate to its nature—but rather the teleolojik premise that the proper ordering of knowledge is necessary for social order. In other words, what is borrowed from al-Farabi is the ontological foundation of knowledge and the argument that epistemic corruption leads to social corruption; not the political model that positions epistemic subjects according to a hierarchical ordering of virtue. Observing this distinction makes it possible to leave outside the scope of this article the tension pointed out by Syukri and colleagues, namely the conflict between al-Farabi's hierarchical conception of society and modern egalitarian epistemic justice perspectives; for the claim of this article is not to reproduce al-Farabi's political model, but to draw upon the structural insight in his philosophy of knowledge. For the tension between al-Farabi's virtue-based, hierarchical model of justice and contemporary demands such as equality, human rights, and democratic participation, and the epistemological challenges of adapting metaphysical frameworks to rights-based paradigms, *see* Syukri, Adenan, Syahminan, *Reinterpreting Justice in Al-Farabi's Political Philosophy: Relevance to Contemporary Islamic Human Rights Thought, MILRev: Metro Islamic Law Review*. 4, 489-516 (2025). 10.32332/milrev.v4i1.10466.

[148] The determinative role of GPUs (graphics processing units) in the training and inference processes of artificial intelligence models has been extensively documented in the deep learning literature. The parallel computing capacity provided by GPUs through thousands of cores has enabled the scalable training of large language models and neural networks. Chip market studies reveal that GPU-based AI accelerators have become a

technologies, *RAG* and *Graph-RAG-based* knowledge retrieval models, function calling, blockchain infrastructures, *MCPs* (*model context protocols*), legal design methodologies, and domain-specific model architectures are considered collectively, the capacity for legal data processing has expanded both quantitatively and qualitatively. This technological convergence demonstrates that artificial intelligence agents have evolved beyond merely transforming raw legal data into secure, traceable, and contextually meaningful insights, reaching a point where they can effectively execute legal workflows.[149]

In a knowledge system so architected, artificial intelligence integrates the pattern recognition capacity of statistical language models (such as large language models and their derivatives) with the principles of logical consistency, normative hierarchy, and explainability afforded by legal ontology. This approach, referred to in the literature as *neurosymbolic artificial intelligence*, brings together the intuitive inferential power of deep learning and the rule-based reasoning of symbolic systems within a single epistemic framework. This synthesis aims not only to produce more accurate outcomes, but also to render traceable the normative premises, conceptual classifications, and inferential chains upon which the outcome rests. Thus the system moves away from the '*black box*' character and transforms into an accountable, transparent, and auditable legal assistant.

The developments currently taking place in the field of *agentic AI*, which are accelerating exponentially, and the *Mecellem semantic protocol* itself, provide strong evidence that this transformation can occur at an institutional scale. As current technological limitations, particularly constraints related to context window capacity and scalability

---

fundamental component of data center infrastructure as of 2025. For the determinative effect of hardware architecture on the cognitive capacity of AI systems, *see* Alex Krizhevsky, Ilya Sutskever & Geoffrey E. Hinton, *ImageNet Classification with Deep Convolutional Neural Networks*, 25 ADVANCES IN NEURAL INFO. PROCESSING SYS. 1097 (2012); Sparsh Mittal, *A Survey of Techniques for Optimizing Deep Learning on GPUs*, 99 J. SYS. ARCHITECTURE 101635 (2019); Ming Li et al., *Deep Learning and Machine Learning with GPGPU and CUDA: Unlocking the Power of Parallel Computing*, arXiv:2410.05686 (Oct. 2024) issues[150], are substantially overcome in the near term, even the redesign of the organizational and operational structure of justice institutions becomes possible and probable within the scope of this transformation accelerated by the *AI-first* movement.[151]

[149] The multi-step reasoning, tool-calling, and parallel task execution capacity of agentic artificial intelligence systems requires a heterogeneous computing infrastructure beyond uniform processor architecture. While CPUs manage orchestration logic, GPUs handle deep learning tasks, TPUs undertake matrix multiplication-intensive inference operations, and NPUs assume structured inference tasks. This division of labor maximizes both computational efficiency and energy savings. For the importance of heterogeneous computing architectures for AI systems and hardware-software co-design, *see* Cristina Silvano et al., *A Survey on Deep Learning Hardware Accelerators for Heterogeneous HPC Platforms*, 55 ACM COMPUTING SURVS., no. 3, June 2025, at 1; Renchao Yu, *Application of CPU in AI and Machine Learning* (ACM, Nov. 2024). On the other hand, the multi-step task execution, tool-calling coordination, and real-time inference capacity of agentic AI systems necessitates the use of specialized hardware accelerators (ASIC, FPGA, TPU). General-purpose CPU architectures create bottlenecks in agentic workflows where tool processing can account for up to 90 percent of total latency. While FPGAs dynamically adapt to different model sizes and task types through their reconfigurable architectures, TPUs offer distinct efficiency advantages in matrix multiplication-intensive inference tasks. For the role of hardware accelerators and design frameworks for agentic systems, *see* Norman Jouppi et al., *In-Datacenter Performance Analysis of a Tensor Processing Unit*, at 1 (2017); Stavros Kalapothas, Georgios Flamis & Paris Kitsos, *Efficient Edge-AI Application Deployment for FPGAs*, 13 INFO., no. 6, May 2022, at 279; Aybars Yunusoglu et al., *A Reconfigurable Framework for AI-FPGA Agent Integration and Acceleration*, arXiv:2601.19263 (Jan. 2026).

However, the legitimacy of such a transformation cannot rest solely on technological capacity. For the *agentic* structure to be integrated into legal practice, the principles of ontological commitment, epistemic consistency, and methodological weighting must be implemented fully. Otherwise, agents may become mechanisms that merely generate speed but weaken legal meaning and carry the risk of *hallucination*.[152] *The Mecellem semantic*

---

[150] The self-attention mechanism of the Transformer architecture exhibits a quadratic relationship ($O(n^2)$) with the length of the input sequence in terms of computational complexity. This structural limitation makes it difficult for large language models to incorporate lengthy legal texts, comprehensive case law collections, and multi-page case files entirely within the context window. Research reveals that models perform well at the beginning and end of the context window while experiencing significant accuracy loss in the middle; this phenomenon is termed 'lost in the middle'. *See* Ashish Vaswani et al., *Attention Is All You Need*, 30 ADVANCES IN NEURAL INFO. PROCESSING SYS., arXiv:1706.03762 (2017); Nelson F. Liu et al., *Lost in the Middle: How Language Models Use Long Contexts*, TRANSACTIONS ASS'N FOR COMPUTATIONAL LINGUISTICS, arXiv:2307.03172 (2023). Norman Paulsen demonstrates that there are significant discrepancies between the advertised maximum context window of large language models and the maximum effective context window they can actually utilize in practice. The author shows that accuracy decreases as context width increases across different problem types, that models can sometimes effectively use only as few as one hundred tokens of content, meaning there can be chasms between theoretical limits and practical effectiveness. This study constitutes a current and concrete reference discussing the gap between both the theoretical limit and practical usage of the 'context window' concept. In summary, the context window refers to the number of *tokens* (words/fragments) that can be held in view simultaneously in large language models, and there can be significant differences between the theoretical value of this limit and how much the model can effectively utilize in practice. *See* Norman Paulsen, *Context Is What You Need: The Maximum Effective Context Window for Real World Limits of LLMs* (preprint, arXiv, 2025), https://arxiv.org/abs/2509.21361 (last visited Oct. 13, 2025).

[151] The AI-first concept refers to an approach in which artificial intelligence occupies a central position in the design of digital platforms and enterprise information architectures. The term was first articulated in the context of corporate strategy by Sundar Pichai at the 2016 Google I/O developer conference when describing Google's orientation as a '*transition from mobile-first to AI-first,*' and subsequently began to be employed as a conceptual framework in technology policy and digital transformation literature. *See* Sundar Pichai, CEO, Google, Keynote Address at the Google I/O Developers Conference: From Mobile-First to AI-First (May 18, 2016), https://www.youtube.com/watch?v=862r3XS2YB0 (last visited Mar. 9, 2026). The AI-first approach envisions systems being designed from the outset around data-intensive learning models, prediction mechanisms, and autonomous decision processes. In this context, the paradigm in question is associated in the literature with discussions of data-driven organizations, platform economy and algorithmic coordination, and AI-supported corporate transformation strategies. These studies demonstrate that artificial intelligence has become not merely an analytical tool but a constitutive infrastructure layer that guides decision processes and shapes the architecture of digital ecosystems. *See* Yann LeCun, Yoshua Bengio & Geoffrey E. Hinton, *Deep Learning*, 521 NATURE 436 (2015), https://doi.org/10.1038/nature14539; Erik Brynjolfsson, Daniel Rock & Chad Syverson, *Artificial Intelligence and the Modern Productivity Paradox: A Clash of Expectations and Statistics*, *in* THE ECONOMICS OF ARTIFICIAL INTELLIGENCE: AN AGENDA 23 (Ajay Agrawal, Joshua Gans & Avi Goldfarb eds., 2018), https://www.nber.org/papers/w24001 (last visited Mar. 9, 2026).

[152] The deployment of agentic AI systems in legal contexts requires addressing certain challenges that distinguish legal work from many other professional domains. Transparency and explainability emerge as paramount concerns; legal professionals must understand not only what an agentic system recommends but also why

*protocol* enters precisely at this point: by ensuring that meaning is constructed within ontological constants, semantic context, and methodological orchestration, it enables the new digital subject to make decisions in a manner that is fair, transparent, justifiable, and accountable.

Therefore, the *agentic* transformation is not merely a technological innovation; it is a normative and epistemological reconstitution concerning how legal meaning is to be produced, how it is to be justified, and how it is to be supervised. *Mecellem*'s semantic orchestration model presents both the theoretical foundation and the operational protocol of this new institutional horizon.

### *A. The Positioning of Agents as Legal Subjects: The Subject of Knowledge and Action*

The artificial intelligence agent, rising upon the ontological, epistemological, and methodological infrastructure of *Mecellem*'s *semantic protocol*, is being positioned as a new '*de facto subject*' in legal practice. Here, the concept of '*subject*' encompasses not merely a legal personality in the sense of being a bearer of rights and obligations, but also the meanings of being the *subject of knowledge* (one who knows, understands, and reasons) and the *subject of action* (one who performs, executes, and directs the decision-making process). This distinction is important; for the focus of our discussion is not the question of granting legal personality to the agent, but rather the nature of the decision-making and executive role that it de facto assumes within the legal order.[153]

---

that recommendation was made, what information was considered, and which sources support the outcome. Professional responsibility and client confidentiality present operational challenges. Law firms must ensure they understand the technology they employ as required by their duties to demonstrate professional competence, and must carefully evaluate whether AI-generated content and analysis meet the standard of care expected for their jurisdictions. *See* Mark McKamey, *Legal Technology: Artificial Intelligence and the Future of the Law Practice*, 22 APPEAL REV. CURRENT L. & L. REFORM 45 (2017), https://ssrn.com/abstract=3014408; Nicole Yamane, *Artificial Intelligence in the Legal Field and the Indispensable Human Element Legal Ethics Demands*, GEO. J. LEGAL ETHICS (2020), https://www.law.georgetown.edu/legal-ethics-journal/wp-content/uploads/sites/24/2020/09/GT-GJLE200038.pdf. On the other hand, hallucination and factual accuracy also represent persistent challenges in agent systems. Agents based on large language models can present plausible-appearing but incorrect information with the confidence of factual assertions, particularly in cases where training data is sparse or ambiguous. *See* Matthew Dahl et al., *Large Legal Fictions: Profiling Legal Hallucinations in Large Language Models* (Stanford RegLab, Working Paper, 2024), arXiv:2401.01301. Stanford RegLab and HAI research empirically measures the hallucination rates of many commercial legal research AI tools. The study's findings reveal that even bespoke legal AI tools specifically designed for legal domains experience hallucination at concerning rates. . For these findings, *see* Varun Magesh, Faiz Surani, Matthew Dahl, Mirac Suzgun, Christopher D. Manning, Daniel E. Ho, *Hallucination-Free? Assessing the Reliability of Leading AI Legal Research Tools*, Journal of Empirical Legal Studies (Wiley), May 2024 (arXiv preprint), arXiv:2405.20362.

[153] Here, subjectivity ceases to be a will-centered concept; it is defined by the capacity to produce meaning and embed it in normative contexts. Thus, legal subjectivity is reconceptualized as an ontological and epistemic competence. *See* Paul Formosa, Inês Hipólito & Thomas Montefiore, *Artificial Intelligence (AI) and the Relationship Between Agency, Autonomy, and Moral Patiency* (2025) (unpublished manuscript),

From the perspective of positive law, artificial intelligence systems are not subjects of rights at present. The legal order limits the status of subjecthood to natural and legal persons; artificial intelligence, on the other hand, is positioned in the category of an instrument or thing used by actors possessing legal personality.[154] For this reason, the outcomes produced by an artificial intelligence system are, as a rule, attributed to the person using it or to the organization developing the system. However, technological developments, particularly *agentic systems* that reach high levels of autonomy and can independently conduct multi-step reasoning processes, are de facto challenging this classical understanding of instrumentality.[155] Indeed, it is anticipated that in the near future, advanced artificial

---

arXiv:2504.08853. In this study, Formosa and colleagues demonstrate that goal-directedness, contextual reasoning, and adaptability are fundamental features of AI agency; they note that current AI systems lack genuine moral agency but can simulate limited forms thereof through hybrid ethical frameworks. These findings support the argument that even though an agent is not a person possessing rights and obligations within the legal order, it occupies a decision-making position in the process of producing legal meaning and its steps can be supervised.

[154] Andrea Bertolini addresses in detail the debates concerning the recognition of 'electronic personality' for AI systems; on the one hand, he rejects the electronic personality approach in this sense by arguing that accepting machines as independent entities with their own inherent rights is neither theoretically nor normatively sound; on the other hand, he notes that the idea of electronic personality can be designed as a fiction functionally equivalent to legal personality and, within this framework, can be discussed as an instrumental model for the distribution of liability. *See* Andrea Bertolini, *Artificial Intelligence and Civil Liability* 35–36 (§ 2.1.1), 38–39 (§ 2.1.2) (Eur. Parl., Pol'y Dep't for Citizens' Rights & Constitutional Affs., Study PE 621.926 (July 2020).

[155] When international practices regarding the current use of artificial intelligence in judicial systems are examined, it is understood that there is no model today in which AI replaces human judges. On the contrary, AI is currently positioned solely as a supportive tool that reduces administrative burden and enhances procedural efficiency. With respect to Slovenia, the Ljubljana Statement on AI, published in May 2025 and presented by the European Association of Judges and Administrative Judges, states that 'AI should never replace human judges but should be seen as an opportunity to enhance, rather than undermine, the administration of justice and the rule of law.' This statement represents the clearest official expression that AI is viewed as a supportive tool rather than a decision-maker. *See* Eur. Ass'n of Judges & Admin. Judges, Ljubljana Statement on AI (May 2025), https://rm.coe.int/2025-05-16-aeaj-ljubljana-statement-on-ai/1680b5f885 (last visited Mar. 1, 2026). In Estonia, due to misleading news reports about a 'robot judge' project, the Estonian Ministry of Justice and Digital Affairs issued an official statement explicitly clarifying that they are not developing a 'robot judge' to replace human judges in small claims or general court proceedings. The Ministry emphasized that current and future work focuses on alleviating administrative burdens in courts, optimizing and automating processes, and that the aim is to support judges and court officials. *See* Est. Ministry of Just. & Digital Affs., *Estonia Does Not Develop AI Judge* (official statement), https://www.justdigi.ee/en/news/estonia-does-not-develop-ai-judge (last visited Mar. 1, 2026). China possesses the world's most comprehensive 'smart court' system, where AI automatically reviews case files, locates relevant laws and precedents, prepares draft decisions for simple cases, and detects procedural errors in judges' decisions; in particular, the *Internet Courts* (Hangzhou, Beijing, Guangzhou) operate entirely digitally. However, even here, AI functions as an 'assistant judge' that reduces the judge's workload and as an automation system managing judicial bureaucracy. In the United States, AI has become widespread particularly in criminal law, where 'risk assessment' algorithms such as *COMPAS* predict defendants' likelihood of reoffending and provide recommendations to judges during bail hearings and sentencing phases; additionally, AI is used in e-discovery processes to identify relevant documents among millions. In Malaysia, courts in the states of Sabah and Sarawak employ an AI pilot project in the sentencing process for certain types of crimes, where the system analyzes past similar cases and recommends a sentencing range to the judge; this application functions as an advisory system aimed at increasing consistency in decisions.

intelligence systems will become significant subjects vis-à-vis social structure and law. It is argued that "artificial entities," particularly those reaching high levels of autonomy and cognitive capacity, will cease to be objects and will be in a position to demand personality within the legal order as entities capable of autonomous action. Although this is an ambitious perspective from the standpoint of contemporary law, according to views increasingly articulated in legal doctrine, systems possessing advanced *cognitive capacity* and *decision-making ability* are ceasing to be mere technical instruments within social functioning; they are acquiring a determinative position in the operation of legal practice.

The distinction that must be made at this point is as follows: defining the artificial intelligence agent as a new '*subject of rights*' in terms of normative status and recognizing it as a '*de facto actor*' in legal practice are different matters. The approach we adopt in this study is to focus on the agent's position of functional subjectivity in the legal domain, rather than entering into the debate on its legal personality. For a system that drafts a contract, analyzes a case file, operates a compliance procedure, extracts the evidentiary structure in an arbitration file, or identifies normative gaps is a '*de facto actor.*' The subject that accesses knowledge, weighs information, selects methods, and produces outcomes in the process of constructing legal decisions is, on the practical plane, no longer solely human.

Consequently, even if the agent is not a person bearing rights and obligations within the legal order, it is being positioned in a decision-making role in the process of producing legal meaning. This position goes beyond the status of being an '*instrument*' in the classical sense; because the agent does not merely execute given commands, but also selects methods according to context, weighs data layers, and manages the process of semantic construction. In other words, the agent's role is to *define and complete tasks by exercising active initiative* rather than passive execution, and in this context, *to conduct the decision-making process*.

Nevertheless, as agents perform active decision-making and executive functions, it is evident that updates will be necessary in the classical doctrine of representation (*principal/agent*) and in liability regimes. In legal scholarship, it is proposed that principles of *vicarious liability* be applied, similar to the way an employer is held liable for the acts of an employee, for damages arising from the acts of advanced artificial intelligence systems, or that special regulatory liability regimes be introduced, as in the case of autonomous vehicles. However, as we mentioned above, the subject of our examination here concerns not

---

Singapore utilizes AI-supported platforms in dispute resolution, having developed online mediation and negotiation systems that assist parties in reaching agreements, particularly in commercial and international disputes; AI is also used for real-time transcription of court proceedings. In the United Kingdom, AI is primarily used in legal research and case management, with judges and lawyers utilizing AI-supported platforms to rapidly locate precedents and legal texts; additionally, certain simple cases (such as divorce and monetary claims) are conducted digitally through an online court portal. In other European Union countries, AI is used particularly to automate standard processes such as enforcement proceedings, where automatic payment orders can be generated based on documents; throughout the European Union, AI is currently considered 'high-risk,' and its use is subject to strict human oversight rules. All these examples demonstrate that AI in judicial systems functions as a system that automates administrative and repetitive judicial operations, serves as a decision support system and big data analysis tool, but has not yet attained the position of final decision-maker.

liability, but rather what agents can do and how they can do it in terms of the construction and execution of legal meaning.

### *B. Active Roles of Agents in Legal Practice*

The *agentic* orchestration of the *Mecellem semantic protocol* requires treating the artificial intelligence agent not merely as a technical tool in legal practice, but as a new executive subject positioned within a multi-layered functionality. This positioning is not a contingent expansion of function; it is the natural consequence of a meaning architecture that is ontologically structured, epistemologically grounded, and methodologically orchestrated. The role of the agent in the legal world is therefore not a one-dimensional matter of "*automation*"; rather, it is a holistic matter of construction that encompasses together the questions of how knowledge is produced, within which normative framework it is positioned, and into which operational outcomes it is transformed.

Therefore, when analyzing the agent, it is necessary to follow a tripartite chain of reasoning. First, on the *epistemic plane*, the question of how the agent understands legal data, how it parses it, and through which semantic mechanisms it produces meaning must be addressed. For all normative and operational functions depend on the accuracy and contextual coherence of the knowledge produced. A system that constructs meaning incorrectly will have both its normative references and its executive outputs impaired. For this reason, the *epistemic layer* constitutes the constitutive ground of the agent's existence in legal practice.

Second, on the *normative plane*, how the knowledge produced relates to the legal order must be examined. The agent is not merely a data-processing mechanism; it assumes a normative position by performing dynamic weighting among legislation, case law, doctrine, and other *corpus* layers. Determining which source will take precedence in which context is not merely a technical choice; it is a normative positioning articulated with the internal logic of the legal order. At this stage, the agent interacts with the internal codes of law and aligns meaning with the criteria of legal validity.

Finally, on the *operational plane*, this epistemic and normative structure is transformed into concrete execution. The agent can assume an actual role in a wide range of areas, from producing contract drafts to analyzing case files, from conducting compliance processes to developing decision recommendations that require multi-step reasoning. At this point, the issue is no longer whether the agent '*knows*'; it is into which actionable outcomes the knowledge it produces is transformed and how it functions in legal practice.

Therefore, in the subsequent part of the article, examining the agent through epistemic, normative, and operational roles respectively will ensure both conceptual coherence and intellectual progression. For normative positioning becomes possible only when the correct mode of knowing is established; executive function becomes legitimate and meaningful only when normative positioning is achieved. In this framework, the epistemic level represents the constitutive layer that determines how knowledge is produced, validated, and contextually signified within the system. This tripartite structure also reveals the constitutive logic of the *Mecellem semantic protocol* in the *agentic* age. In this context, the effectiveness of agentic functions at the macro level is directly dependent on the coherence and functionality of agentic components at the micro level. Structural or functional discontinuities between micro and macro agents disrupt agent orchestration, jeopardizing the integrity of the system and preventing agents from fulfilling the roles assigned to them.

### *1. Knowledge and Meaning Production at the Epistemic Level*

The epistemic layer refers to the agent's capacity to access knowledge, structure knowledge, signify it, and ultimately produce new meaning. On this plane, the agent is positioned not merely as a data processor, but as a cognitive subject that navigates the world of legal meaning, grasps conceptual relations, and is capable of context-sensitive inference. Thanks to the infrastructure built upon *Mecellem*'s taxonomy–ontology–semantics triangle, the artificial intelligence agent processes legal texts not merely as linguistic sequences, but as units of meaning positioned within a particular ontological order and normative arrangement. This enables the agent not to be content with parsing words, but to evaluate concepts within a contextual and systematic totality.

This capacity is not confined to superficial matching based on textual similarities. The agent first parses legal documents according to their *functional taxonomies*; then structures units of meaning through *ontological mapping* and *semantic chunking* processes; finally establishes relations among these units via *knowledge graphs* and *vectorial representations*. For instance, an agent capable of scanning thousands of pages of case files or hundreds of judicial decisions in a short time can parse relevant events, legal qualifications, party statuses, and normative references, organizing them into conceptual clusters. This process encompasses not only data extraction, but also the capacity to classify according to legal topics, establish relational connections, and generate multi-hop reasoning. Although current legal technologies have begun to automate legislation and precedent search processes, *Mecellem*'s ontological and epistemic architecture aims to elevate this automation to a more advanced level by targeting context-sensitive and ontologically grounded meaning production from complex legal document sets. This situation reveals a transformation that brings not only speed, but also structural precision to legal research or case preparation processes.

What is truly determinative in the agent's epistemic role is its ability to establish the chain extending from raw data to knowledge, from knowledge to comprehension, and from comprehension to reasoned conclusion. For instance, an agent examining a criminal case file does not content itself with identifying party names, dates, or evidence lists; it reconstructs the narrative of events, matches the elements of the offense with ontological categories, and maps the network of relations among pieces of evidence. In a statement transcript where the word '*weapon*' appears, because it knows in the ontological model that this concept is associated with categories such as '*instrument of crime*', '*evidence*', '*element of offense*', it interprets the context within the framework of criminal law. In contrast, in a dispute file arising from a contract, it can interpret the word '*weapon*' in the expression '*my only weapon was to put this document into process*' in a figurative sense as '*only way*' or '*only option*'. The fact that the same word can acquire two different legal meanings in two different contexts is the result of the contextual awareness the agent gains through the semantic layer.

This *contextual awareness* is directly related to the accuracy of the ontological model and the consistency of semantic fluidity. Ontology determines the place of concepts within the ontological order; the semantic layer, in turn, determines how these concepts will be positioned in each concrete query. The agent evaluates the concept of '*weapon*' not only by its dictionary meaning, but together with the type of file it is in, the legal field (criminal law, contract law, etc.), party relations, and normative references. Thus, knowledge ceases to be an isolated data point; it transforms into a '*comprehension*' that acquires meaning within a particular legal universe.

Another critical function of the agent at the epistemic level is its ability to make dynamic transitions between knowledge layers. By weighting among legislation, case law, doctrine, and other relevant *corpora*, it can determine through which knowledge source a particular problem will be more accurately understood. This shows that the agent becomes not merely a knowledge gatherer, but a '*cognitive agent*' that selects, prioritizes, and hierarchically organizes knowledge. Epistemic competence is measured here not only by the answer to the question '*what do I know?*', but by the answer to the question '*what do I need to know in this question and from which source should I know it?*'.

In conclusion, the epistemic role of agents has a constitutive function in the production of legal meaning. Thanks to the processes of taxonomic parsing, ontological anchoring, semantic restructuring, and dynamic weighting, the agent transforms into a '*cognitive subject*' capable of processing legal texts with a human-like comprehension. This subject may not yet possess legal personality; however, it is de facto positioned as an epistemic subject that produces knowledge, constructs meaning, and builds chains of reasoning in legal practice. As will be seen in the next stage, this epistemic depth constitutes the necessary precondition for the relationship the agent will establish with the normative order and the roles it will assume at the operational level.

### *2. Interaction with Legal Rules at the Normative Level*

The normative role of the agent is concretized in the interaction it establishes with the rule, principle, and value structure of the legal order. For a system that produces meaning at the epistemic level to be effective in legal practice, it must not only understand the text; it must also grasp the normative language and act in accordance with the claim of bindingness of this language. Law is not a descriptive field of knowledge; it is a normative order that determines 'what ought to be'. Therefore, the agent cannot content itself with identifying facts; it must also be able to determine which normative category these facts fall into, with which rule they will be associated, and to which legal consequence they will be linked.

*Mecellem*'s knowledge architecture makes normative elements a constitutive component of the system. Laws, judicial decisions, customs, and even general principles of law are categorically represented within the ontological model; the hierarchical and referential relationships among them are explicitly mapped on the knowledge graph. Thus, the normative order appears before the agent not as a collection of scattered texts, but as a structure with internal coherence and established inter-source priority relationships. Thanks to this structure, the agent can evaluate the legal nature of an action, identify the relevant norm, and determine which source is determinative within the normative hierarchy.

For instance, in an enforcement file, it can grasp that the ‘service of payment order’ is not merely a technical procedure, but the *normative threshold* of the debt collection process. It knows through procedural ontology the preclusive nature of the objection period that begins from the date of service, that if no objection is made within this period the enforcement proceedings will become final, and that subsequent compulsory execution steps will be activated. This knowledge is not a simple rule-response matching; it is the capacity to grasp the temporal and structural integrity of the normative chain. Thus, the agent produces an answer not only to the question ‘which rule exists’, but also to the question ‘at which stage, with which consequence, and with which normative bindingness is this rule applied?’

However, it would be incomplete to reduce the normative role of the agent to an automaton that merely processes coded rules. An agent operating within an advanced normative architecture can also possess the capacity to make principled evaluations, going beyond rigid rule application. For instance, when drafting a contract, it can not only incorporate the relevant legislative provisions into the text, but also analyze the conformity of the provisions with equity, the rule of good faith, or the limits of freedom of contract. It can foresee that a provision containing an excessive penalty clause may be subject to reduction by the judge in the future; it can flag ambiguous expressions with high interpretive potential in terms of semantic uncertainty. Such evaluations require, beyond knowing the letter of the norm, grasping the purpose of the norm and its function within the system.

The foundation of this understanding is again the ontological and semantic infrastructure. When the values and principles expressed as the 'spirit' of law are represented as abstract categories in the ontology, the agent can analyze the tension between norm and principle. For example, when a mandatory norm conflicts with the principle of freedom of contract, it can determine which order of priority will be applied within the framework of normative hierarchy rules. Thus, the agent transforms into an actor that not only applies the normative text, but also makes sense within the normative order.

This role at the normative level is at the center of the relationship the agent establishes with the legal order. The meaning produced at the epistemic level gains bindingness at the normative level; at the operational level, it transforms into execution. Therefore, the normative layer is a necessary condition for the agent to act legitimately and consistently in legal practice. As will be seen in the next stage, this structure that produces meaning epistemically and is positioned normatively prepares the ground for the agent's participation in actual decision-making and execution processes at the operational level.

### *3. Execution and Implementation Capacity at the Operational Level*

The operational role of the agent refers to its capacity to transform the meaning it produces epistemically and positions normatively into actual action. At this level, the issue is no longer the understanding of legal knowledge or its placement within the normative framework; it is the initiation, execution, and conclusion of a specific legal transaction. In other words, the agent's transformation of a legal evaluation it has analyzed and structured into concrete execution takes place at the operational layer. While artificial intelligence applications in the past were mostly designed as decision support tools, thanks to *Mecellem*’s *agentic* architecture, agents have evolved into structures that directly participate in the execution of decisions, manage processes, and produce outcomes.

*Mecellem* agents are not merely tools that produce output from a large language model. These new-generation agents can plan a business process from start to finish; define subtasks and execute them sequentially or in parallel; interact with external systems to receive and send data; and produce the final output to conclude the process. When this structure is built upon a *neurosymbolic* and *ontological* infrastructure, the agent's executive capabilities are guided not only technically, but also contextually and normatively. Thus, the operation becomes not random automation, but a legally meaningful and framed execution.

A simple example from corporate applications concretizes this capacity: Suppose a company has adopted a corporate policy decision not to accept liability for 'indirect damages' in any contract. This policy has been ontologically processed into the company's knowledge base within *Mecellem*'s *dynamic knowledge network*; it is represented in the system together with risk categories and normative priorities. An agent built upon this architecture can semantically scan every new contract entering the system to detect statements regarding the acceptance of 'indirect damages'; mark the relevant clause and send an automatic alert to a predefined email address. Today's technology makes it possible to execute such instructions in a fully automated manner. Indeed, in the work we conduct, matters deemed critical from the company's perspective are monitored daily through agents structured around nineteen separate contractual risk parameters; in certain risk categories, even the '*human-in-the-loop*' mechanism has been disabled[156], establishing an operational structure where the agent's findings are directly considered reliable.

Three fundamental characteristics underlie the operational success of artificial intelligence agents: the capacity to perform work, the ability to use tools, and the possession of autonomous decision loops. First, the agent can execute a specific task from start to finish toward a defined goal. This task can be a limited assignment such as making a change to a single clause of a contract, or it can be a multi-step and complex operation such as the process of preparing a lawsuit petition. What is important is that the agent can carry out this process without requiring continuous human intervention.

Second, agents use various tools to interact with the external world. A legal agent can retrieve current judicial decisions by running a search engine in the background; convert a *PDF* document into text via *OCR*; activate different retrieval strategies through *elastic search* or *vector database*; produce an output conforming to the company's letterhead format; access the calendar to record the hearing date or send an automatic notification email. This use of tools enables the agent to cease being merely a system that performs cognitive operations and transform into an actor that intervenes in the practical world.

---

[156] The statement here regarding the possibility of disabling the '*human-in-the-loop*' mechanism does not refer to the complete removal of humans from the system; rather, it refers to the optimization of human intervention at the operational stage in certain processes where ontological modeling, epistemic validation, and *agentic* decision rules have been defined and tested at a sufficient level. In other words, the principle of human oversight is maintained; however, when the necessary technical infrastructure (verified knowledge graph, explicit risk ontology, traceable inference chain) and theoretical infrastructure (normative prioritization, responsibility attribution, error tolerance threshold) are provided, in some routine and low-uncertainty operations, the agent's findings can be considered reliable and automatic workflows can be operated. Final action and institutional responsibility, at the current stage, still rest on human decision.

The third and most distinctive element is the degree of autonomy. Advanced agents can adapt their own workflows according to new situations that arise during task execution. For example, when it cannot find the legal basis it is searching for in a database, it can decide to turn to an alternative source; when it detects that a report it has prepared is contrary to internal policies, it can revise it with its own internal control modules before submitting it for human approval. Such *adaptive decision loops* enable the agent to function not merely as a tool that executes commands, but as *an organizational layer that manages processes*.

The importance of the operational role of *Mecellem* agents in legal practice manifests itself particularly in the areas of speed, efficiency, and scalability. Multi-step and repetitive operations are carried out with a lower error rate and in an uninterrupted manner compared to human intervention. For example, an agent built on the *Mecellem* architecture can monitor legislative changes and secondary regulations published by regulatory authorities in real time; it can automatically detect potential areas of non-compliance by comparing these changes with the company's existing compliance policies and internal procedures. The system does not merely notify of the change; it produces concrete revision proposals by matching the relevant normative provision with the internal policy clause and automatically prepares the necessary compliance reports in corporate format. Such an operational flow clearly demonstrates that *Mecellem* agents can function not as a tool that passively follows legal regulation, but as an executive subject that integrates normative change into corporate practice.

Another operational advantage of agents is that they reduce the error rate and can manage multiple operations simultaneously. In *Mecellem*, checklists, cross-data validations, form filling, and risk screening processes that are laborious for humans have been turned into routine subroutines for the agent. Thus, hundreds of similar contracts can be updated simultaneously; critical elements of thousands of case files can be indexed and summarized within minutes. This situation provides both speed and predictability in the application of law; it also offers a significant contribution in terms of the manifestation of justice within a reasonable time.

In conclusion, the operational layer is the practical-world counterpart of the structure built upon epistemic depth and normative consistency. Here, the agent transforms into a subject that not only knows and interprets, but also plans, executes, and produces results. It is as important to analyze the practical reflections through concrete cases as it is to draw the theoretical framework of this transformation. For this new executive capacity has the potential to fundamentally change the way law is applied and enables us to see more clearly the position of new subjects in legal practice.

## *V. MULTI-LAYERED APPLICATION OF THE MECELLEM SEMANTIC PROTOCOL IN AN ARBITRATION FILE: ARBINEXT*

The *Mecellem semantic protocol* assumes a transformative function across diverse domains of legal practice. The protocol supports fundamental functions of the legal profession across an extensive spectrum, ranging from *transactional legal services* such as contract drafting and negotiation preparation, mergers and acquisitions, financing and credit agreements, real estate transactions, and *franchise* and licensing relationships; to *dispute resolution mechanisms* including litigation management and mediation processes, enforcement and bankruptcy proceedings, and provisional remedies; to legal advisory activities encompassing regulatory compliance audits, *due diligence* examinations, personal data protection, tax planning, and intellectual property management; and extending to *strategic decision-support processes* such as corporate governance structuring, risk analysis, and prevention strategies. Nevertheless, within the framework of this article, demonstrating the added value provided by the *Mecellem semantic protocol* and making visible the practical implications of the theoretical framework assume particular significance. Arbitration proceedings, by virtue of constituting one of the dispute resolution domains wherein the conceptual, technical, procedural, and strategic dimensions of legal issues intersect in their most intensive and sophisticated form, offer a conducive ground for analyzing the multi-layered epistemic and operational contributions of the *semantic protocol*. Within this framework, in the final section of the article, the multi-layered application of the *Mecellem semantic protocol* shall be examined through the example of the *ArbiNext* project, designed as a knowledge architecture that applies *Mecellem*'s ontological, epistemological, and methodological principles to the arbitration domain; the epistemic depth and operational efficacy of the protocol will be evaluated together.

At this juncture, it becomes necessary to examine closely the structural character of the arbitration file. In contemporary arbitration practice, the file no longer consists merely of a collection of documents to which legal norms are applied; rather, it has been transformed into a dynamic knowledge network that generates continuous interaction among parties, claims, defenses, material facts, evidence, and normative references. The intensification of international commercial relations and the increasing technicality of contractual structures, in particular, have rendered arbitration files into high-volume and multi-layered datasets. The establishment of meaningful connections among contracts running to hundreds of pages, thousands of electronic communications, technical reports, expert opinions, hearing transcripts, and arbitral awards; the systematic tracing of the *evidence–fact–norm* relationship; and the reconstruction of the chain of legal reasoning have increasingly departed from sustainability through traditional document-centric methods.

This difficulty is not merely a matter of quantitative data density. The essential issue concerns how the knowledge contained within the arbitration file is structured ontologically, validated epistemologically, and operationalized methodologically. Each unit of information present in an arbitration file is not an isolated data point; rather, it is a relational entity that acquires meaning in connection with a particular legal concept, a particular event, a particular party, and a particular normative framework. For instance, the expression '*force majeure*' appearing in a contractual provision is not merely a linguistic term; it is a multidimensional legal concept that must be evaluated in conjunction with the questions of what legal elements it possesses, by what norms it is regulated, by what evidence it can be proven, and what claims it supports or refutes. Any analysis produced without grasping this multidimensionality will remain superficial and detached from context.

In traditional arbitration practice, this complexity is managed largely through the experience, intuition, and interpretive capacity of the human legal practitioner. However, in the age of artificial intelligence, particularly with the commencement of the integration of *agentic* artificial intelligence systems into arbitration processes, the question of how this intuitive and experiential knowledge is to be structured, represented, and operationalized has become unavoidable.[157] For the meaningful analysis of an arbitration file by an artificial intelligence agent depends not merely upon its ability to read texts, but upon its capacity to grasp the ontological structure of the arbitration universe, to distinguish epistemological layers, and to select methodologically correct instruments.[158] At this juncture, the *semantic protocol* developed by *Mecellem* in the legal domain offers a constitutive framework for arbitration practice as well.

The *ArbiNext* project has been designed precisely in response to this need, as a knowledge architecture that applies *Mecellem*'s *ontological*, *epistemological*, and *methodological* principles to the field of arbitration. *ArbiNext* is constructed upon a '*minimum viable ontology*' (*MVO*)[159] that restructures the arbitration file not as a random accumulation of documents, but through ontologically defined entities, epistemologically differentiated knowledge layers, and methodologically orchestrated retrieval mechanisms. This ontology establishes a common ground of meaning for both the human legal practitioner and the artificial intelligence agent by explicitly defining the fundamental building blocks of the arbitration universe, the relationships among these building blocks, and the constraints under which these relationships operate.

In this article, we shall examine in depth the ontological architecture of *ArbiNext* within the framework of *Mecellem*'s *semantic protocol*. First, we shall address how the arbitration file ought to be modeled ontologically, what entity categories need to be defined, and how the relationships among these categories are to be structured. Subsequently, we shall discuss how this ontological structure integrates with epistemological layers, how knowledge sources are differentiated, and how accuracy is secured. Finally, we shall demonstrate through concrete examples how this ontological and epistemological infrastructure is operationalized at the methodological level, how *agentic* artificial intelligence systems

---

[157] Frank Schilder, Thomson Reuters Labs, *How Agentic AI Systems Think, Learn, and Collaborate with Legal Professionals* (2025). (https://legal.thomsonreuters.com/blog/how-agentic-ai-systems-think-learn-and-collaborate-with-legal-professionals/) (*last visited Mar. 1, 2026)*

[158] Zhang *et al.*, *supra* note 133.

[159] For the intellectual development of the *minimum viable ontology* concept and the synthesis of lean startup principles with knowledge management, *see* Thomas R. Gruber, *A Translation Approach to Portable Ontology Specifications*, Knowledge Acquisition (Academic Press), Vol 5 No 2 (June 1993) https://doi.org/10.1006/knac.1993.1008; Mike Uschold, Michael Gruninger, *Ontologies*: *Principles, Methods and Applications*, The Knowledge Engineering Review, 1996;11(2):93-136. doi:10.1017/S0269888900007797; Natalya F. Noy, Deborah L. McGuinness, *Ontology Development 101: A Guide to Creating Your First Ontology*, Stanford Knowledge Systems Laboratory Technical Report No: KSL-01-05 and SMI-2001-0880, March 2001, https://protege.stanford.edu/publications/ontology_development/ontology101-noy-mcguinness.html; Steve Blank, *Why the Lean Start-Up Changes Everything*, Harvard Business Review Vol 91 No 5, May 2013, https://hbr.org/2013/05/why-the-lean-start-up-changes-everything

function upon this architecture, and what concrete outcomes are produced in arbitration practice.

### *A. Ontological Architecture: From the Order of Being to Digital Assets*

The '*minimum viable ontology*' (*MVO*) approach that forms the foundation of *ArbiNext* constitutes a domain-specific application of *Mecellem*'s principle of *ontological commitment* to the field of arbitration. This approach aims to define the smallest ontological framework sufficient for the system's core functions, rather than comprehensively modeling all possible entities and relationships within the arbitration universe. This preference is not a contingent design decision but rather a natural consequence of *Mecellem*'s philosophy of *dynamic knowledge networks*.[160] For the more complex and comprehensive an ontology becomes, the more *contextual flexibility* diminishes; the system becomes incapable of transcending predefined categories. By contrast, a minimal yet sufficient ontology both ensures conceptual stability and permits the incorporation of new entity types and relationships into the system.[161]

The theoretical foundation of the *MVO* approach rests upon the idea that ontological order constitutes a prerequisite for epistemological order. However, this order is not a comprehensive encyclopedic classification encompassing everything in the universe, but rather a hierarchical arrangement structured according to the functional needs of a particular domain of knowledge. *ArbiNext*'s *MVO* operates according to the same principle: it defines the fundamental entity categories necessary for understanding, analyzing, and managing the arbitration file; yet it construes these categories not as a fixed and closed schema, but as an extensible and context-sensitive framework.

This ontological framework organizes the arbitration universe under three fundamental entity categories: *structural entities*, *argumentation entities*, and *factual entities*. This tripartite distinction is not contingent; it reflects three fundamental functional planes inherent in the nature of the arbitration file. *Structural entities* constitute the institutional skeleton of the arbitration process: the case, parties, contracts, arbitral awards, and procedural rules fall within this category. These entities determine who is involved, when the proceedings occur, within what contractual framework they are conducted, and within what procedural order they unfold. *Argumentation entities*, in turn, represent the legal reasoning dimension of arbitration: claims, defenses, counterclaims, legal concepts, legal elements, and legal norms are situated within this layer. These entities indicate what the parties seek, on what legal grounds they seek it, and upon what normative foundation these grounds rest.

---

[160] For current literature on how ontologies, as machine-readable domain knowledge representations, enable semantic exchange, interoperability, and reasoning, *see* Elahe Parsanasab, Alihasan Ahmadipour, Esmaeil Mehraeen, *Utilization of Ontology to Develop Artificial Intelligence Systems in the Healthcare Industry*, Healthcare Informatics Research (October 31, 2025) https://doi.org/10.4258/hir.2025.31.4.320. The study addresses the challenges of developing ontologies specific to complex domains such as the healthcare sector and demonstrates how minimal yet sufficient ontological frameworks preserve contextual flexibility. This finding directly supports the theoretical justification for *ArbiNext*'s *MVO* preference.

[161] Zhang *et al. supra* note 133.

*Factual entities* express the concrete reality dimension of arbitration: evidence, events, witnesses, expert testimonies, and hearing transcripts are positioned within this category. These entities reveal upon what factual foundation legal assertions rest, what concrete events have transpired, and by what evidence these events are proven.

This tripartite ontological structure constitutes not merely a static classification of the arbitration file, but also the foundation for a dynamic process of meaning production. For in arbitration, legal meaning emerges through the interaction of these three layers. A claim acquires meaning not solely as an entity within the argumentation layer, but when it is related within a specific structural context (which case, which party, which contract), to specific legal concepts (force majeure, lesion, default, etc.), and to specific factual evidence (contract clause, email, expert report, etc.). *ArbiNext*'s *MVO*, by explicitly defining this relational integrity at the ontological level, enables the arbitration file to transcend being a fragmented collection of documents and become a holistic knowledge network.[162]

One of the most distinctive aspects of *ArbiNext*'s ontological architecture is its application of the *T-Box* (*Terminological Box*) and *A-Box* (*Assertional Box*) distinction, in accordance with classical *description logic* terminology, to the arbitration domain. This distinction is not merely a technical modeling choice; it is a concrete manifestation of *Mecellem*'s principles of *ontological commitment* and *epistemic consistency*.[163] The *T-Box* defines the conceptual schema of the arbitration universe: what entity types exist, what relationships are possible among these entities, what constraints govern these relationships, and what properties of these entities are mandatory. The *T-Box* makes no reference to any concrete case data; it answers only the question of how the arbitration world is structured. In this sense, the *T-Box* corresponds to al-Fârâbî's concept of *first intelligibles*: universal, abstract, and immutable conceptual constants.

The *A-Box*, on the other hand, is the concrete instantiation of the abstract schema defined in the *T-Box*, populated with a specific arbitration case. Each time a new case is loaded, the *T-Box* remains unchanged; only the *A-Box* expands with new individuals and relationships. In this sense, the *A-Box* corresponds to al-Fârâbî's concept of *second intelligibles*: concrete, contextual, and variable factual assertions. For instance, while the `ArbitrationCase` class is defined in the *T-Box*, a concrete case individual such as `ARB-CASE-2025-001` appears in the *A-Box*. While the `has_party` relationship is defined in the T-Box, a concrete relationship instance such as `ARB-CASE-2025-001 has_party` ABC Construction Inc. is found in the *A-Box*.

---

[162] For the theoretical framework on how semantic technologies enhance reasoning and decision-making processes in organizational knowledge management and how knowledge graphs transform into a representational order that establishes meaning within relational patterns rather than merely storing data, *see* Charlie Jones, *Semantic Technologies in Knowledge Management*. European Journal of Information and Knowledge Management. 3. 10.47941/ejikm.1750 (2024).

[163] For the theoretical foundation of the *T-Box* and *A-Box* distinction within *description logic* systems, *see* De Giacomo *et al.*, *supra* note 78. The study demonstrates how the differentiation of terminological and assertional knowledge layers constitutes a necessary modeling choice for ontological consistency checking. For a comprehensive evaluation of the application of this distinction in legal ontologies, *see* also Breuker *et al.*, *supra* note 70.

The epistemological significance of this distinction is critical. The *T-Box* guarantees the conceptual consistency of the system by preserving ontological constants. Constraints such as the requirement that an evidence node must necessarily be linked to a source (the `extracted_from` relationship is mandatory), or that a case must have at least two parties (`has_party` cardinality minimum 2), are defined in the *T-Box*, and every *A-Box* datum is validated against these constraints. Thus, the system does not accept ontologically inconsistent data; every new piece of information is positioned in conformity with the existing ontological framework. This is the arbitration-domain counterpart of *Mecellem*'s *'no source - no claim*' principle: evidence without a source is epistemically invalid; ontological constraints prevent this invalidity at the system level.

However, the ontological stability provided by the *T-Box* is balanced by the dynamic expansion of the *A-Box*. Each new arbitration case brings new parties, new claims, new evidence, and new events; the *A-Box* accommodates this novelty within the ontological framework. Thus, the system remains faithful to conceptual constants while remaining open to factual diversity. The *T-Box* answers the question what arbitration is; the *A-Box* answers the question what this case like is. This dual structure is the concrete expression in the arbitration domain of the balance *Mecellem* establishes between ontological commitment and semantic fluidity.

Another strength of *ArbiNext*'s ontological architecture is that it explicitly models the legal reasoning process in arbitration[164] on an ontological plane. This modeling does not merely define entities; it also structures the logical and normative relationships between these entities. Legal reasoning in arbitration proceeds according to a specific template: a legal concept (for example, force majeure) requires certain legal elements (an unforeseeable event, an unavoidable nature, impossibility of performance, and a causal link); these elements are governed by specific legal norms (Turkish Code of Obligations Art. 136, Art. 138); each element is proven by specific evidence (a contract clause, an email, an expert report); evidence supports or refutes specific claims (a request for time extension, a request for additional costs).

This chain is explicitly defined in the *ArbiNext* ontology: the `LegalConcept` node is connected to `LegalElement` nodes through the `requires` relation; the `LegalConcept` node is connected to `LegalNorm` nodes through the `regulated_by` relation; the `LegalElement` node is connected to `Evidence` nodes through the `proven_by` relation; the `Evidence` node is connected to `Claim` nodes through the `supports` or `contradicts` relations. This ontological chain makes visible how legal reasoning operates in arbitration. When an arbitrator evaluates a claim, they follow this chain: what legal concept does the claim rest upon? What elements does this concept require? What norms govern these elements? What evidence has been used to prove these elements? Does this evidence support or refute the claim?

[164] For a comprehensive literature review on the capacity of large language models to generate complex reasoning chains in law, medicine, and industrial domains, *see* Md Mahadi Hasan Nahid & Davood Rafiei, *Prism: Agentic Retrieval with LLMs for Multi-Hop Question Answering* (https://doi.org/10.48550/arXiv.2510.14278); Aske Plaat, Annie Wong, Suzan Verberne, Joost Broekens, Niki van Stein, Thomas Bäck, *Multi-Step Reasoning with Large Language Models: A Survey*, arXiv:2407.11511v3 [cs.AI] (2025). These studies demonstrate that multi-step reasoning provides epistemic depth and methodological consistency in artificial intelligence systems.

*ArbiNext*'s ontological model does not merely define these questions at an abstract level; it also shows how this chain is instantiated in each concrete arbitration file. For example, while the concept of '*force majeure*' is defined as an abstract `LegalConcept` in the *T-Box*, it appears in the *A-Box* as a concrete node such as `Concept-fm` in a specific case. This node is connected through the `requires` relation to concrete element nodes such as `Elem-fm-01`: Unforeseeable event, `Elem-fm-02`: Unavoidable nature, and `Elem-fm-03`: Impossibility of performance and causal link. Each element node is connected through the `proven_by` relation to concrete evidence nodes such as `Evi-000023`: Contract clause snippet and `Evi-000045`: Amendment text. Each evidence node is connected through the `supports` relation to concrete claim nodes such as `CLM-001`: Request for time extension.

This ontological chain constitutes not only a static structure of the arbitration file but also a map of a dynamic reasoning process. By navigating through this chain, questions such as what legal basis a claim rests upon, what elements support this basis, what evidence proves these elements, and from what sources this evidence is derived can be answered systematically. This forms the ontological foundation of *Mecellem's* multi-step reasoning capacity. The system can answer complex legal questions not by examining only a single node but by following chains of relationships between nodes.

Moreover, this ontological chain is used not only for information access but also for epistemic validation. The constraints defined in the *T-Box* monitor the consistency of *A-Box* data at *runtime*. For example, if the *T-Box* defines the constraint that 'every `LegalElement` must be connected to at least one `Evidence` through the `proven_by` relation' the system can automatically check whether an element node remains without evidence. If an element is left without evidence, this situation is reported as a deficiency in the burden of proof. Thus, the ontology not only defines entities but also monitors their epistemic validity.

The power of the *ArbiNext* ontology lies not only in defining entity types but in explicitly specifying the relationships between these entities and the constraints to which these relationships are subject. Ontological relationships express the structural rules of the arbitration universe. For example, the `has_party` relation indicates which parties an arbitration case has; however, this relation is not arbitrary. The cardinality constraint defined in the *T-Box* specifies that every case must have at least two parties. This constraint derives from the nature of arbitration: arbitration is the resolution of a dispute between at least two parties; unilateral arbitration is ontologically inconsistent.

Similarly, the `has_contract` relation indicates which contracts an arbitration case is based upon; the cardinality constraint specifies that every case must be based on at least one contract. This guarantees the contractual foundation of arbitration at the ontological level. The `requires` relation indicates which elements a legal concept necessitates; the cardinality constraint specifies that every concept must have at least one element. This ensures at the ontological level that legal concepts cannot be empty and devoid of content. The `regulated_by` relation indicates which norms regulate a legal concept; the cardinality constraint specifies that every concept must be based on at least one norm. This is the ontological counterpart of the principle that there is no concept without norms.

One of the most critical relation types is the `extracted_from` relation. This relation indicates from which document evidence is extracted; the cardinality constraint specifies that every evidence must be based on exactly one source. This is the concrete expression in the arbitration ontology of *Mecellem*'s '*no source - no claim*' principle. An evidence without a source is epistemically invalid; the ontological constraint prevents this invalidity at the system level. When the system attempts to create an evidence node without a specified source, it rejects this operation due to ontological inconsistency. Thus, epistemic discipline is structurally guaranteed at the ontological level.

Other critical relations are the `supports` and `contradicts` relations. These relations indicate which claims or legal elements evidence supports or refutes. However, these relations are also not arbitrary. The disjoint class constraint defined in the *T-Box* specifies that the same evidence cannot both support and refute the same element. This is the ontological counterpart of logical consistency: evidence, with respect to a particular element, can be either favorable, unfavorable, or neutral; however, it cannot be both favorable and unfavorable at the same time. This constraint prevents contradictory evidence evaluations in the arbitration file at the ontological level.

Temporal relations also occupy an important place in the ontological model. The `occurred_before` relation indicates the chronological ordering between events; the `triggered_by` relation indicates the causal connection between events; the `resulted_in` relation indicates which consequences an event led to. These relations represent the temporal dimension of the arbitration file at the ontological level. For in arbitration, the sequence of events and causal connections are determinative for legal evaluation. For example, the fact that the site access obstruction event occurred before the delay notification event strengthens the contractor's force majeure claim. The *ArbiNext* ontology, by explicitly modeling these temporal relations, enables the systematic analysis of chronological and causal connections between events.

Finally, source traceability relations ensure the epistemic reliability of the ontological model. While the `extracted_from` relation indicates from which document evidence is extracted, the `cited_in` relation indicates in which documents an evidence or norm is cited; the `mentioned_in` relation indicates in which documents an entity is mentioned. These relations enable tracing back to the source of every claim in the arbitration file. *Mecellem*'s *epistemic consistency* principle requires not only that knowledge be correct, but also that it be obtained from the correct source through the correct method. *Provenance* relations structurally guarantee this epistemic discipline at the ontological level.

### ***B. Epistemological Layers: Differentiation and Verification of Knowledge Sources***

One of the fundamental principles of *Mecellem*'s epistemological regime is that legal knowledge does not consist of a single homogeneous source, but rather of multiple knowledge layers with different epistemic statuses. In the field of arbitration, this principle is embodied in the distinction between *normative corpus* and *enterprise corpus*. The *normative corpus* contains public and binding legal sources: legislation, supreme court precedents, arbitration rules, and generally accepted legal principles. This layer constitutes the normative framework of arbitration; it determines which rules are valid, which principles will be applied, and which procedures will be followed. The *enterprise corpus*, on the other hand, contains institution-specific practices, contracts, correspondence, factual cases, and inter-party relations. This layer constitutes the concrete context of arbitration; it shows what the parties want, which events occurred, and which evidence is available.

These two layers have epistemically different statuses. The *normative corpus* is a knowledge source with general validity, supported by public authority, and resistant to change. The *enterprise corpus*, on the other hand, is a specific, contextual, and variable knowledge source. However, this difference does not imply a hierarchy. *Mecellem*'s *epistemic consistency* principle requires that knowledge sources be weighted not according to a fixed order of priority, but according to the nature of the query. The same principle applies in arbitration: If the question concerns the definition of a general legal concept, the normative corpus gains weight. If the question concerns how the parties understood a specific contractual provision, the enterprise corpus gains weight.

The *ArbiNext* ontology structurally supports this epistemic distinction. Documents are classified by the `doc_type` property: types such as `contract_master`, `contract_annex`, `amendment_signed`, `email`, `notice`, `expert_report`, `hearing_transcript` indicate which epistemic layer the document belongs to. For example, the `contract_master` type represents the main contract text; this document is the most authoritative expression of the parties' will and has the highest epistemic weight within the enterprise corpus. In contrast, the email type represents correspondence between parties; these documents provide contextual information about how the contract was interpreted but are not as binding as the contract text itself. By recognizing these epistemic differences through ontological types, the system assigns appropriate weight to each document type according to context.

The ontological differentiation of these epistemic layers is critically important for *agentic* artificial intelligence systems. This is because an agent cannot produce consistent decisions without knowing which knowledge source is more authoritative in which context. If the system evaluates the contract text and inter-party email correspondence on the same epistemic plane, it produces incorrect results. The *ArbiNext* ontology structurally guarantees this epistemic hierarchy through document types, relationship weights, and verification mechanisms. Thus, the agent not only collects data but also draws information from the epistemically correct source with the correct weight.

*ArbiNext*'s evidence verification mechanism is the concrete application of *Mecellem*'s *epistemic consistency* principle in the field of arbitration. In arbitration, evidence constitutes the fundamental building blocks of legal reasoning; however, the epistemic value of evidence depends not only on its content but also on the verifiability of its source. In the *ArbiNext* ontology, each `Evidence` node has a *verified* property. This property carries a *boolean* value: *true* or *false*. When an evidence node is first created, the *verified* value is assigned as *false*. Evidence is elevated to *verified=true* status only after source verification is performed by the *Verifier Agent*.

This verification process operates through ontological relationships. The *Verifier Agent* follows the `extracted_from` relationship of the `Evidence` node to reach the source document; re-extracts the relevant *snippet* from the original document; compares the extracted *snippet* with the snippet stored in the `Evidence` node; if there is a match and the `source_checksum` is verified, it updates the verified value to true. If there is no match or if a change is detected in the source document, the *verified* value remains *false* and the related `supports/contradicts` relationships are deleted. Thus, the system works only with verified evidence; unverifiable evidence is considered epistemically invalid and is not included in the reasoning chain.

The ontological foundation of this verification mechanism is the cardinality constraint of the `extracted_from` relationship: each piece of evidence must be based on exactly one source. This constraint explicitly determines the epistemic origin of the evidence. Evidence cannot be linked to multiple sources because this situation creates ambiguity about which document the evidence was extracted from. Evidence cannot be linked to no source because this situation eliminates the epistemic foundation of the evidence. Therefore, the `extracted_from` relationship is not merely a data connection; it is the ontological guarantee of epistemic reliability.

The verification mechanism also takes into account the temporal dimension. Documents, especially contracts, can change over time: amendments can be made, annexes can be updated, new versions can be created. The *ArbiNext* ontology tracks temporal versions of documents with properties such as `signed_date` and `version`. When verifying evidence, the *Verifier Agent* checks not only whether the *snippet* exists but also whether the document version from which the evidence was extracted is still valid. If the document has been updated and the relevant clause has changed, the evidence automatically drops to *verified=false* status. This temporal verification is the counterpart of *Mecellem*'s *temporal validity* principle in the field of arbitration: information must not only be correct but also current.

In arbitration, legal questions often cannot be answered with a single node; they require following multiple relationship chains. For example, the question 'what are the provisions for the contractor's exemption from delay liability?' requires multi-step reasoning: Find the relevant contract documents (`has_contract` relationship), identify evidence related to liability and delay in these documents (`extracted_from` relationship), determine which legal concepts this evidence supports (`supports` relationship), show which elements these concepts require (`requires` relationship), check which evidence proves these elements (`proven_by` relationship).

This multi-step chain is realized through *graph traversal* methods over ontological relationships. The system starts from an initial node (for example `ArbitrationCase`) and reaches target nodes by following relevant relationships. This traversal is not random; the relationship types and cardinality constraints defined in the *T-Box* determine which paths are valid. For example, there is no direct transition from a `Claim` node to an `Evidence` node; however, the `Claim` → `LegalElement` → `Evidence` chain is ontologically valid. By navigating in accordance with these ontological constraints, the system produces epistemically consistent results.

The *graph traversal* method also offers contextual filtering capabilities. For example, filters such as retrieve only evidence with *verified=true*, show only evidence with *stance=pro*, sort only evidence with `relevance_score > 0.8` can be applied during traversal. These filters provide epistemic quality control at the ontological level. By working only with verified, favorable, and high-relevance-score evidence, the system reduces epistemic noise and increases reasoning quality.

The multi-step reasoning capability is the fundamental element that distinguishes *ArbiNext* from classical information retrieval systems.[165] Classical systems can only answer single-step queries such as 'which parties are involved in this case?', 'which articles are in this contract?'. *ArbiNext*, however, can answer multi-step queries such as 'what is the proof status of the force majeure claim?', 'which evidence supports which claims and from which sources was this evidence extracted?'. This capability is made possible by the explicit definition of ontological relationships and the systematic application of *graph traversal* methods. Thus, the arbitration file ceases to be a static document collection; it becomes a navigable, queryable, and restructurable knowledge network.

### *C. Methodological Orchestration: Concrete Application of Agentic Structure in Arbitration Practice*

The *agentic* architecture of *ArbiNext* divides the arbitration process into ontologically defined tasks and assigns each task to a specialized agent.[166] This task

[165] For how hybrid graph traversal algorithms (combining *depth-first* and *breadth-first* strategies) facilitate efficient information discovery and multi-step reasoning in complex networks, *see* Rowanda Ahmed, Belaynesh Chekol, Mahmoud Alsaleh, *HDBMS: A Context-Aware Hybrid Graph Traversal Algorithm for Efficient Information Discovery in Social Networks*, arXiv:2508.14092v1 [cs.SI] (Aug 14, 2025). For how the agentic deep graph reasoning approach enables self-organizing knowledge networks that dynamically generate and refine relationships, *see* also Markus J. Buehler, *Agentic Deep Graph Reasoning Yields Self-Organizing Knowledge Networks*, J. Mater. Res. 40, 2204–2242 (2025). These findings validate the thesis that explicit definition of ontological relationships and systematic graph traversal enable multi-step query capacity. For a comprehensive literature review on the capacity of large language models to generate complex reasoning chains in law, medicine, and industrial domains, *see* Aske Plaat, Annie Wong, Suzan Verberne, Joost Broekens, Niki van Stein, Thomas Bäck, *Multi-Step Reasoning with Large Language Models, a Survey*, arXiv:2407.11511v3 [cs.AI] (Nov 02, 2025).

[166] The *agentic* structure referred to here is not identical to the commonly used '*AI Agents*' concept. While *AI Agents* are characterized as modular systems directed by large language models and designed for task-specific automation—that is, singular agents completing independent tasks—'*agentic AI*' architecture in the *Mecellem* context represents a qualitative paradigm shift in which multiple specialized agents collaborate through coordinated orchestration to achieve complex, higher-level objectives. This distinction between '*AI Agents*' and '*agentic AI*' is not merely a semantic difference; it reflects fundamental differences in how systems are architecturally organized, how autonomy is distributed, how goals are structured and pursued, and how learning and adaptation occur. Multi-agent systems and orchestration mechanisms represent the most distinctive feature of *agentic AI*. While single-agent systems are valuable for specific tasks, complex problems generally benefit from architectural approaches in which multiple specialized agents with different capabilities, tools, and domains of expertise collaborate toward common goals. Orchestration enables agents to communicate, share context, and collaborate effectively to complete complex tasks or workflows. Different orchestration patterns serve different purposes: centralized orchestration provides tight supervision and consistent execution, hierarchical orchestration enhances scalability, adaptive orchestration dynamically

distribution is not coincidental; it reflects the natural workflow and ontological structure of the arbitration file. The *Router Agent* determines the scope of the arbitration file: which contracts are relevant, which parties are active, which document types are available. This agent works on `ArbitrationCase` and `Contract` nodes; it delineates the structural boundaries of the file using `has_contract` and `has_party` relationships. The *Planner Agent* plans which node categories will be scanned and how many agents will work in parallel. This agent performs *fan-out* calculations using the node counts in the ontological model; thus, the processing load is distributed in a balanced manner.

*Evidence Miner Agents* are the most critical agents of the arbitration file. These agents specialize in different document types: *Contract Miner* searches for evidence in contract texts, *Email Miner* in email correspondence, *Minutes Miner* in hearing minutes, *Notice Miner* in notices and notifications, *Expert Miner* in expert reports. Each *miner* uses *semantic chunking* and *embedding* strategies specific to its own document type.[167] For example, while *Contract Miner* parses the contract text at the article, clause, and paragraph level; *Email Miner* divides email chains into chronological and topic-based clusters. This specialization is the *agentic* counterpart of *Mecellem*'s document-type-specific semantic parsing principle.

Each *miner* adds the evidence it finds to the *A-Box* as an *Evidence* node; links it to the source document with the `extracted_from` relationship; determines whether the evidence is favorable, unfavorable, or neutral through the `stance` property; calculates the degree of relevance of the evidence with the `relevance_score` property; explains why the evidence was classified in this way with the `rationale_short` property. These properties represent not only the content of the evidence but also its epistemic status at the ontological level. Thus, each piece of evidence enters the system with its own epistemic identity.

The *Verifier Agent* validates the evidence produced by the *miners*. This agent follows the `extracted_from` relationship to return to the source document; retrieves the original *snippet* again; verifies the `source_checksum`; if verification is successful, it marks it as *verified=true*. The `supports/contradicts` relationships of evidence that fails verification are deleted. This process makes epistemic quality control automatic and systematic.[168] The *Writer Agent*, on the other hand, works only with evidence marked as

---

reassigns responsibilities according to changing conditions, while emergent orchestration allows agents to self-organize with coordination patterns emerging from local interactions rather than central direction.

[167] For current studies on the capacity of legal artificial intelligence systems for autonomous legal reasoning, evidence mining, and dispute resolution, *see* Sushrita Rakshit, James Hale, Kushal Chawla, Jeanne M. Brett, Jonathan Gratch, *Emotionally-Aware Agents for Dispute Resolution* (2025) https://arxiv.org/pdf/2509.04465. For how *NLP* techniques such as Legal-BERT and RoBERTa are used for entity recognition, article extraction, and multi-label classification, *see* Srinivasa Kalyan Vangibhurathachhi, *Advanced Natural Language Processing For Legal Document Analysis*, (2025). 10.13140/RG.2.2.35504.06409.

[168] For how continuous testing, formal verification, error monitors, and voting systems in artificial intelligence systems ensure that outputs are authentic, relevant, and reliable, *see* Lalli Myllyaho, Mikko Raatikainen, Tomi Männistö, Tommi Mikkonen, Jukka K. Nurminen, *Systematic Literature Review Of Validation Methods For AI Systems*, Journal of Systems and Software, Volume 181 (2021) https://doi.org/10.1016/j.jss.2021.111050.

*verified=true*; lists favorable evidence using `supports` relationships; lists unfavorable evidence using `contradicts` relationships; produces the final report.

The *Claim Analysis Agent* breaks down claims into legal concepts. This agent analyzes `Claim` nodes; identifies which legal concepts are at issue; creates `LegalConcept` nodes; adds `LegalElement` nodes with the `requires` relationship; connects `LegalNorm` nodes with the `regulated_by` relationship. Thus, claims become not merely party statements but structured arguments linked to legal concepts, elements, and norms. The *Timeline Agent* establishes chronological and causal relationships between events. This agent creates `Event` nodes; establishes chronology with `occurred_before` relationships; shows causal connections with `triggered_by` relationships; links outcomes with `resulted_in` relationships. The *Burden of Proof Agent* performs burden of proof analysis. This agent updates the proven property of `LegalElement` nodes; checks `proven_by` relationships; reports which elements have been `proven` and which remain incomplete.

These agent roles represent the *agentic* reconstruction of the arbitration process. In traditional arbitration practice, these functions are carried out intuitively and holistically by human legal practitioners. *ArbiNext*, however, ontologically decomposes these functions, assigns each function to a specialized agent, and reconstructs the arbitration file through the orchestration of these agents. This approach not only increases processing speed but also ensures that each step is traceable, auditable, and reproducible. Thus, the arbitration process ceases to be an intuitive art and becomes an ontologically structured, epistemically validated, and methodologically orchestrated knowledge production process.[169]

---

These mechanisms are critically important especially in high-risk environments such as law and strengthen the theoretical foundations of the process of returning to the source document by following the extracted_from relationship and verifying the *source checksum*. For how ontological structuring in the context of explainable artificial intelligence (*XAI*) enhances reliability and regulatory compliance, *see* also T. Raheem, G. Hossain, *Agentic AI Systems: Opportunities, Challenges, and Trustworthiness*, 2025 IEEE International Conference on Electro Information Technology (eIT), Valparaiso, IN, USA, 2025, 10.1109/eIT64391.2025.11103638.

[169] This approach is strongly supported in current literature. A comprehensive study published in 2026 reveals that multi-agent systems are based on layered communication protocols. The study emphasizes that protocols such as *Model Context Protocol* (*MCP*) are critically important for external data access, while the *Agent-to-Agent* (*A2A*) protocol is essential for inter-agent coordination, negotiation, and task delegation. The research demonstrates how the complex orchestration of specialized agents (*retrieval, reasoning, validation, monitoring*) is applied in domains such as scientific research, autonomous driving, and news gathering. Architectural models including centralized, distributed, hybrid, and fully decentralized structures are defined, and the different advantages and disadvantages of each in terms of scalability, coordination complexity, and governance are analyzed. For more detailed information, *see* Apoorva Adimulam, Rajesh Gupta, Sumit Kumar, The Orchestration of Multi-Agent Systems: Architectures, Protocols, and Enterprise Adoption, arXiv:2601.13671v1 [cs.MA] 20 Jan 2026. On the other hand, Anthropic's industry report published in 2025 also validates the practical effectiveness of the specialized agent approach proposed in the article. The report demonstrates that a ninety-two percent performance increase was achieved through the orchestration of specialized sub-agents working in parallel. This system integrates capabilities such as iterative planning, tool use, and memory management, supporting the functional differentiation we propose in the article for Router Agent, Planner Agent, and Evidence Miner Agents. *See* Anthropic, Engineering at Anthropic, How We Built

One of *ArbiNext*'s most distinctive contributions is its ability to determine the *histology* of the arbitration file at the ontological level. *Histology* is the branch of biology that examines the structure of tissues; here, it refers to the internal structure of the arbitration file, its layers, and the relationships between these layers. The histology of an arbitration file is determined not only by which documents are present but also by which ontological categories these documents belong to, which epistemic layers they occupy, and within which relational network they are positioned. In other words, the *histology* of the file is the concrete manifestation of its *ontological* structure.

This *histological determination* is a prerequisite for the correct selection of methodology. For the method by which an arbitration file is analyzed depends on the internal structure of that file. If the file consists predominantly of contract texts and there is intensive correspondence between the parties, the *Contract Miner* and *Email Miner* work predominantly; knowledge graph and ontological inference mechanisms come to the fore. If the file consists predominantly of technical reports and expert opinions, the *Expert Miner* works predominantly; semantic proximity and embedding mechanisms gain higher weight. If the file consists predominantly of hearing minutes and oral statements, the *Minutes Miner* and *Hearing* nodes come to the fore; temporal relationships and chronological analysis become more important.

Therefore, in *ArbiNext*, methodology is determined according to the *histology* of the file. This is the arbitration-domain counterpart of *Mecellem*'s principle of methodological pluralism: a single method is not valid for all files; each file's unique ontological structure requires its own unique methodological approach. The system analyzes the histology of the file at the ontological level and dynamically determines which agents will work with what weight, which retrieval strategies will be used, and which epistemic layers will be prioritized. This dynamic determination is executed by the *agentic* orchestration function.

The practical consequence of this approach is as follows: *ArbiNext* does not process each arbitration file with a uniform template; it recognizes each file's unique ontological structure and creates a methodological configuration appropriate to that structure. Thus, the system both maintains ontological stability and ensures contextual flexibility. Ontology determines what the file is; histology determines how the file will be processed; methodology determines with which tools the file will be analyzed. This triple determination transforms *ArbiNext* from being a static schema into a knowledge architecture that is file-specific, dynamic, and context-sensitive.

On the other hand, the capacity of agents in *ArbiNext* to pose correct queries is not merely a matter of technical optimization; it is the methodological counterpart of epistemic sensitivity. For the quality of the result produced by an agent depends directly on the quality of the question that agent asks. An incorrectly formulated query can produce incorrect results even in an ontologically correct model. Therefore, agents in *ArbiNext* not

---

Our Multi-Agent Research System, Jun 13, 2025, https://www.anthropic.com/engineering/multi-agent-research-system (*last visited Mar. 1, 2026*) only use the ontological model; they are also structured to formulate queries in a manner appropriate to the ontological structure.

This structuring is based on two fundamental principles. The first is the correct determination of the ontological type of the query. A query follows different ontological paths depending on whether it is structural (which parties exist), argumentative (which claims have been made), factual (which events have occurred), or relational (which evidence supports which claims). The agent determines which node categories will be scanned and which relationship types will be used by identifying the ontological type of the query. For example, while a structural query works on `ArbitrationCase`, `Party`, and `Contract nodes`, an argumentative query works on `Claim`, `Defense`, `LegalConcept`, and `LegalElement` nodes.

The second is the correct determination of the contextual parameters of the query. A query must be enriched not only with ontological type but also with contextual filters such as time interval, party status, document type, and epistemic layer. For example, the question "what are the claimant's claims?" should retrieve not only `Claim` nodes but also `Claim` nodes connected through the `has_claim` relationship with the `Party` node where *role=claimant*. The question "what is the verified supporting evidence" should apply not only `Evidence` nodes but also *verified=true* and *stance=pro* filters. This contextual filtering ensures epistemic quality control at the query level.

In *ArbiNext*, agents pose correct queries by systematically applying these two principles. This capacity demonstrates that agents not only use the ontological model but also query it in an epistemically meaningful way. The ability to pose correct queries combines the agents' speed advantage with epistemic sensitivity. The system produces not only fast but also accurate results. This accuracy stems from the structural power of the ontological model and the agents' capacity to use this model in an epistemically correct manner.

In conclusion, the agents' ability to pose correct queries is the fundamental determinant of *ArbiNext*'s operational success. This capacity emerges from the integration of three elements: the clear and consistent definition of the ontological model, the agents' capacity to recognize ontological types and relationships, and the enrichment of queries with contextual parameters. This triple integration elevates *ArbiNext* beyond being merely a document management system and transforms it into a knowledge architecture that provides epistemic sensitivity and methodological precision in arbitration practice.

### *D. Holistic Meaning Construction in Arbitration*

The *ArbiNext* project is a concrete application of *Mecellem*'s ontology-epistemology-methodology triangle in the field of arbitration. This application is not merely a technical adaptation but a reconceptualization of the epistemic and methodological challenges inherent in arbitration practice within the *Mecellem* framework. The complexity of an arbitration file manifests itself on three fundamental planes: On the ontological plane, the arbitration universe is highly heterogeneous. Numerous entity types such as parties, contracts, claims, defenses, evidence, events, legal concepts, legal elements, legal norms, witnesses, experts, hearings, and arbitral awards coexist within the same file. Each of these entities possesses different ontological statuses: some are structural (case, party), some are argumentative (claim, defense), and some are factual (evidence, event). This ontological diversity necessitates the correct placement of entities into categories and the explicit definition of relationships between them.

On the epistemological plane, the sources of information in an arbitration file possess different epistemic densities. The main contract text is the most authoritative expression of the parties' will; amendment texts demonstrate the evolution of the contract; email correspondence reflects the parties' actual conduct; expert reports provide expert opinion on technical matters; hearing transcripts record the parties' statements. Each of these sources has different epistemic values and must be evaluated with different weights depending on the context. For instance, when contract interpretation is at issue, the main contract text carries the highest epistemic weight. However, when the question concerns how the parties applied the contract, email correspondence and actual conduct may acquire higher epistemic weight.

On the methodological plane, the analysis of an arbitration file requires different methodological tools. Structural and deterministic questions (which parties exist, which contracts are relevant) can be answered through knowledge graphs and ontological inference mechanisms. Analogical and interpretive questions (does this event constitute force majeure, how should this provision be interpreted) require *semantic proximity*, *vectorial representation*, and *embedding* mechanisms. Evidence discovery and contextual association require *graph traversal* and multi-step reasoning capacity. *ArbiNext*'s *agentic* architecture orchestrates this methodological diversity: each agent employs the method appropriate to its task, and inter-agent coordination is ensured through the ontological model.

This threefold integration distinguishes *ArbiNext* from a simple document management system. The system does not merely store documents; it structures them ontologically. It does not merely provide access to information; it distinguishes epistemic layers. It does not merely answer queries; it selects methodologically correct tools. Thus, the arbitration file ceases to be a static archive and becomes a dynamic, queryable, and reconfigurable knowledge universe. This universe offers a common ground of meaning for both human lawyers and artificial intelligence agents.

One of the most critical problems artificial intelligence systems face in the legal domain is the risk of *hallucination*. The tendency of large language models to generate data can lead to obtaining incorrect results when making legal assessments. This situation can result in legal evaluation errors that may have serious consequences in arbitration practice. *ArbiNext*'s ontological and epistemic architecture is designed to structurally minimize this risk.[170]

The fundamental cause of hallucination risk is the system operating without ontological commitment. When an artificial intelligence model produces output without knowing which entities actually exist, which relationships are ontologically valid, and which claims are epistemically verifiable, hallucination becomes inevitable. *ArbiNext*, however, validates every output through the ontological model. When an agent makes an evidence claim, this claim must have a counterpart in the *A-Box*; the `Evidence` node in the *A-Box* must be linked to a source document through the `extracted_from` relationship; the source document must have a *verified=true* status. This triple validation chain prevents hallucination at the ontological level.

---

[170] For how the principle of explainability in ensuring the transparency and reliability of artificial intelligence systems encompasses both comprehensibility in the epistemological sense and accountability in the ethical sense, *see* Floridi *et al.*, *supra* note 93. This principle demonstrates how ontological structuring enhances reliability and regulatory compliance by grounding artificial intelligence outputs on explicit knowledge representations and constitutes the normative justification for *ArbiNext*'s mechanism of validating every output through the ontological model.

Moreover, *ArbiNext*'s *Verifier Agent* mechanism continuously monitors the authenticity of evidence at runtime. When an evidence node is created, the *Verifier Agent* returns to the source document, retrieves the *snippet* again, and checks for a match. If the relevant *snippet* cannot be found in the source document, the evidence node is marked as *verified=false* and is not included in the reasoning chain. This mechanism guarantees that the system works only with verifiable evidence.[171] Thus, the hallucination risk is structurally controlled through the epistemic validation process.

This approach addresses the most critical requirement for the use of artificial intelligence in arbitration practice: reliability. An arbitrator or lawyer can only trust the output produced by an artificial intelligence system when they know that the output is epistemically verifiable. *ArbiNext*'s ontological and epistemic architecture structurally ensures this reliability. Every claim has a counterpart in the ontological model; every piece of evidence is based on a verified source; every relationship complies with ontological constraints. Thus, the system ceases to be a prediction mechanism that produces hallucinations; it becomes an epistemically reliable, ontologically consistent, and methodologically auditable knowledge production system.

As we have demonstrated in this section, arbitration practice has entered a profound epistemic and methodological transformation process in the age of artificial intelligence. This transformation cannot be explained merely by the addition of technological tools to arbitration processes. The real transformation lies in the necessity of a new semantic protocol regarding how the arbitration file is to be interpreted, structured, and operated. The *ArbiNext* project presents a concrete example of this new semantic protocol by applying *Mecellem*'s ontological, epistemological, and methodological principles to the field of arbitration.

*ArbiNext*'s *minimum viable ontology* approach reconstructs the arbitration universe through ontologically defined entities, explicitly determined relationships, and structural constraints. The *T-Box* and *A-Box* distinction establishes the balance between ontological constancy and factual dynamism. The ontological modeling of the legal reasoning chain

---

[171] For an industry report on the German example of how legal artificial intelligence validation systems incorporate continuous validation, bias reduction, and compliance with legal standards, *see* Dirk Hartung, Lauritz Gerlach, *The Market for Digitalization of Legal Services in Germany*, Legal Tech Monitor https://legaltechverband.de/wp-content/uploads/2025/03/Legal-Tech-Monitor-2025_EN.pdf (2025). The report reveals that these systems transform legal workflows by supporting eDiscovery, contract analysis, compliance monitoring, and litigation analytics. For market analysis predicting that the legal artificial intelligence software market will nearly double by 2029, *also see Legal AI Software - Company Evaluation Report* (2025) https://www.researchandmarkets.com/reports/6096975/legal-ai-software-company-evaluation-report (*last visited Mar. 1, 2026*)

makes visible the reasoning process extending from concept to evidence, from evidence to claim. The differentiation of epistemic layers ensures that knowledge sources are evaluated with correct weights. The evidence verification mechanism guarantees epistemic reliability at the ontological level. Multi-step reasoning and *graph traversal* capacity enable the systematic answering of complex legal questions.

*Agentic* orchestration, in turn, operationalizes this ontological and epistemological infrastructure at the operational level. Specialized agents work on different dimensions of the arbitration process; each agent processes a specific layer of the ontological model; inter-agent coordination is ensured through ontological relationships. The ontological determination of file histology enables the methodology to be determined according to context. The agents' accurate query execution guarantees epistemic precision at the methodological level. Thus, *ArbiNext* analyzes the arbitration file not only quickly but also in an epistemically accurate, ontologically consistent, and methodologically traceable manner.

In conclusion, the *ArbiNext* project concretely demonstrates what kind of transformation the *Mecellem semantic protocol* can create in the field of arbitration. This transformation is not merely a technical efficiency increase; it is the reconstruction of the epistemic foundations of arbitration practice. This new *semantic protocol*, built upon the principles of ontological commitment, epistemic consistency, and methodological orchestration, removes arbitration from being a random pile of documents; it transforms it into a structured, comprehensible, and operable knowledge architecture. This architecture enables the meaningful participation of artificial intelligence agents in arbitration processes. Agents are no longer merely tools that produce speed; they become cognitive actors capable of producing epistemically accurate, ontologically consistent, and methodologically disciplined decisions. This is the new phase of arbitration practice in the age of artificial intelligence: a holistic transformation extending from meaning to ontology, from ontology to epistemology, from epistemology to methodology, and from methodology to *agentic* orchestration.

## *VI. CONCLUSION*

Law has historically been a discipline positioned between certainty and flexibility, stability and change, norm and context. This tension is inherent in the ontological fabric of law and has never been completely resolved in any period; rather, it has been reformulated in each era. The call raised by Lee Loevinger approximately seventy years ago pointed precisely to a new phase of this ancient tension. That call was not an isolated 'initial spark'; it was an attempt to reorganize the conflict between law's claim to logical consistency and life's experiential flow through scientific method. Holmes's warning that the life of the law has not been logic but experience had shaken the *axiomatic* security of formalism. Loevinger, in turn, transformed this line into a methodological program, arguing that legal research must be carried to a measurable, comparable, and verifiable ground.

However, Loevinger's most significant limitation manifested here as much as his most important contribution: Science provides data, analyzes, makes trends visible; but it does not produce normative will. Statistical accuracy does not inherently generate normative legitimacy. Today, it is not possible to say that this boundary has completely disappeared; but the nature of the transformation has fundamentally changed. The age of artificial intelligence is carrying this ancient tension to a new plane. The issue is no longer merely how norms should be interpreted; it is within which knowledge regime legal meaning is produced, on which order of being it is based, and how this production is to be legitimized. Legitimacy is increasingly tied not to the correctness of the outcome, but to the ontological and epistemic traceability of the chain of reasoning that leads to that outcome.

At this point, the distinction between operational decision and normative legitimacy becomes decisive. Artificial intelligence systems can propose consistent outcomes under certain data and rules; they can even produce reasoned draft decisions, risk analyses, and norm-evidence mappings. However, this production does not inherently generate binding force without an ontological transfer of authority. In law, legitimacy arises not only from the outcome; it emanates from the order of authority, the chain of reasoning, and the explicit representation of normative hierarchy. This representation is necessary not only to show how the decision was produced, but also to ensure that the decision is questionable; for law is a mode of existence built upon contestability. The legitimacy of a norm arises not only from its correctness, but from the fact that this correctness is publicly questionable and opposable.

A human can only 'object' to what they can comprehend; to what they cannot comprehend, they are merely 'subjected.' This observation positions epistemic justice as a precondition of legal justice. Epistemic justice is not only equality of access to knowledge, but also the capacity to participate in the production processes of knowledge and to question these processes. The concepts of 'testimonial injustice' and 'hermeneutical injustice' conceptualized by Fricker[172] acquire new meaning in this context. When an algorithmic system ontologically fails to recognize the experiences of certain social groups or does not provide the conceptual tools to interpret these experiences, it produces not merely a technical deficiency, but an epistemic injustice. Therefore, ontological completeness is not a standard of technical perfection, but a constitutive element of epistemic justice. Contestability, in turn, is the procedural manifestation of this justice; because objection must be directed not only at the outcome, but also at the epistemic regime that produced that outcome.

*Proof* (*evidence*, *argument* or *'burhan'*) is the sole instrument that liberates the human from this passive state of 'subjection' and transforms them back into an active legal subject who reasons. For this reason, transparency is not an epistemic virtue, but a normative necessity. Where objection is not possible, there is no reasoning; where there is no reasoning, there is no legitimacy; and where there is no legitimacy, law remains nothing but force. Law remains law by virtue of the possibility of objection that can be directed against it. Human oversight is therefore not a simple ethical reflex, but a constitutive element of the architecture of legal legitimacy; however, for this oversight to be genuine oversight and to enable contestability, it depends on ensuring epistemic auditability, that is, the traceability of the reasoning, the data source, and the chain of inference. For human oversight to be genuine oversight, the overseer must possess epistemic authority. Epistemic authority is not merely possessing knowledge, but being able to evaluate the validity of that knowledge and, when necessary, to question it. However, as digital systems become increasingly complex, an 'automation bias' emerges: humans may tend to accept algorithmic recommendations uncritically rather than evaluating them critically. This means the transfer of epistemic authority to the algorithm and reduces human oversight to a nominal approval mechanism. Therefore, for human oversight to be meaningful, the system must make visible not only the outcome, but also the reasoning process.

---

[172] *See* Fricker, *supra* note 39.

At this juncture, the connection established with classical legal thought is not coincidental. Law has never been constructed by observing events with the naked eye; the courtroom is not the venue where the incident occurred, and the judge is not a direct witness. The legal order is founded not upon sensory experience, but upon evidence. The Mecelle's maxim *'that which is established by burhan is as if established by direct observation'*[173] presents this reality not merely as a procedural technique, but as the epistemological architecture of law. In al-Fârâbî's logical system, *burhan* denotes necessary and consistent reasoning that definitively demarcates the boundary between *ẓann* and *yaqīn*; *burhan* produces not probability, but necessity. The Mecelle translates this logical certainty into legal certainty: if a ruling has been established through a *burhanic* chain of reasoning, it must now be accepted as if it were actually witnessed. It is precisely at this point that the *jurimetrics* approach pioneered by Loevinger is completed by al-Fârâbî's epistemic certainty. Measured data acquires legal value only when it passes through the filter of *burhanic* reasoning. In the age of artificial intelligence, the demands for explainability, traceability, and contestability find their classical counterpart precisely here.

However, a critical distinction must be made here: while al-Fârâbî's *burhan* produces deductive certainty, legal reasoning often also includes inductive and *inferential* forms. The judge may perform an induction that ascends from the concrete case to the general principle, just as they may make an inference that arrives at the best explanation from the available evidence. The ability of digital systems to ontologically distinguish these three types of inference and to make transparent the epistemic weight of each is a precondition for the reconstruction of *burhan* in the digital environment.

In Islamic legal methodology, *qiyas* (analogical reasoning) is the fundamental instrument for producing rulings in new matters where the *nass* (scriptural text) does not provide a direct answer. *Qiyas* is the transfer of a known ruling to an unknown matter through *illa* (effective cause). This reasoning by analogy bears a striking parallel with machine learning's generalization of patterns to new situations. However, the distinguishing feature of *qiyas* is that it necessitates the explicit definition and justification of the *illa*. Algorithmic similarity can transform into legal *qiyas* only when the *illa* is ontologically defined and epistemologically grounded; otherwise, it merely produces correlation.

H.L.A. Hart's rule of recognition [174], locates the ultimate criterion of legal validity in social facts. In the information age, however, algorithmic systems require meta-rules that determine which norms are to be considered 'authoritative'. These meta-rules become part of the ontological design. Thus, ontological design itself transforms into a normative act; the infrastructure that determines what is to be recognized as law ceases to be a technical choice and becomes a legal responsibility.

---

[173] Mecelle-i Ahkâm-ı Adliyye Article 75: 'That which is established by definitive proof is accepted as clear and certain'. *See* Ali Haydar Efendi, *supra* note 1.

[174] Hart, *supra* note 81.

In classical definitions of law, the regulated subject has most often been depicted as subject–subject relations or the domain of a subject's disposition over objects. In this approach, the legal order is a network of relations centered on human will. However, in the digital environment, the mode of production of legal meaning is changing. Normative outcomes now arise not only from direct interactions between individuals, but from the relationships among datasets, model representations, ontological categories, and chains of algorithmic inference. It can therefore be argued that law is evolving into a discipline that increasingly regulates the relational structure established among digital entities themselves.

In the digital world, a 'relation' is not merely a normative bond between two human beings, but a structural configuration among entity categories and data nodes defined within the system. If this configuration is not established in an ontologically explicit manner, the claim that the outcomes produced by the system are just becomes weakened. For in the age of artificial intelligence, legal decisions often arise not from a direct encounter between two wills, but from the processing of connections among digital representations.

At this point, the relation among *being–body–conscience* discussed at the beginning of the article acquires a decisive significance. Failing to recognize an entity ontologically, excluding it from the model, is not merely a technical deficiency; the unrepresented entity is effectively rendered nonexistent within the system. The negation of the body, in turn, implies—in terms of conceptual origin—a violation of conscience. Thus, ontological deficiency is not only an epistemic error; it is a normative rupture. Algorithmic injustice often arises not from erroneous judgment, but from entities that are misrepresented or not represented at all.

For this reason, ontological commitment is not merely a conceptual preference; it is a precondition for algorithmic justice. If an entity type is not defined, if a relational model is not established, or if the epistemic status of a knowledge source is not determined, the claim that the judgments produced by the system are just loses its theoretical foundation. Without ontological explicitness, epistemic consistency cannot be ensured; without epistemic consistency, methodological trust cannot be formed; and without methodological trust, systemic justice sometimes weakens, sometimes entirely loses its institutional grounding.

In this context, the *semantic protocol* offered by *Mecellem* is not an attempt to add a technology to law. The real issue is the restructuring of the legal mode of thinking under the ontological and epistemic conditions of the age of artificial intelligence. *Mecellem* conceives legal knowledge not as an object of static classification, but as a dynamic, context-sensitive, and temporally evolving knowledge network. On the ontological plane, it explicitly defines entity types; on the epistemological plane, it hierarchically calibrates knowledge sources; and on the methodological plane, it orchestrates reasoning.

Epistemic transparency makes visible which source is activated with what weight. This visibility is not merely an ethical expectation; it is a minimum normative standard in terms of judicial review and institutional accountability. The documentation of the justification process enables the reproducibility and contestability of the decision.

Contestability encompasses not only access to the outcome, but also the accessibility of the ontological schema and weighting parameters used.

Methodological orchestration serves to preserve the distinction between statistical prediction and normative judgment. In an environment where prediction and decision are not differentiated, the system merely produces probability; whereas law must produce normative judgment. Preserving this distinction appears critical for law to maintain its essence in the age of artificial intelligence.

While static normative structures carry the risk of obsolescence within accelerating temporality, they can exist not between rigid codification and uncontrolled flow, but within an ontologically grounded dynamism. The temporal dimension of law is one of the most critical areas of ontological modeling. The entry into force, suspension, or repeal of a norm is not merely a technical data update, but the temporal modality of normative validity. To the extent that the ontological architecture can represent this temporal dimension, it can capture the dynamism of law without reducing it to rigid codification. This representational capacity also necessitates conceiving law not as a merely technically updated body of norms, but as a semantic field that temporally reconstitutes itself.

This transformation requires conceiving law not as a fixed normative structure, but as a continuously reconstituted field of meaning and relationality. Law is not a frozen order; it is a dynamic process that, while preserving past experience, orients itself toward the uncertainty of the future, processing the tension between stability and change within its own internal logic. The acceleration and pluralization of modern social time necessitate thinking of law not as a static '*being*', but as a field of '*becoming*' whose layers are in constant interaction. Therefore, the validity of law is grounded not only in its normative coherence, but in its capacity to generate its own movement.

This *ontodynamic* and temporal character of law brings with it not only the functioning of the existing normative structure, but also the expansion of the domain of being with which law comes into contact. Within this framework, it is possible to make a prediction: In the near future, law is likely to cease being merely a normative system regulating relations among humans and become a meta-discipline that also regulates the mutual representations of digital entities, data relations, and algorithmic inference processes. This transformation will not eliminate the human-centered nature of law; however, it will necessitate that legal order also be established through digital ontology.

In the final analysis, justice in the age of artificial intelligence will be measured not only by reaching the correct outcome, but by explicitly demonstrating which entities it recognizes, which relations it accepts, and within which epistemic regime it renders judgment. Ontological commitment, in this sense, constitutes the theoretical core of algorithmic justice.

*Mecellem*'s contribution lies in offering a framework that enables systematic thinking about this ontological responsibility of law in the age of artificial intelligence. While it does not claim to be a final and closed model, it strongly asserts that legal meaning-production can no longer be sustained without taking into account ontological grounding and epistemic transparency. This amplifies the step that Loevinger called the '*next step forward*' by reinforcing it with legal philosophy; yet it also keeps alive his most valuable warning: Statistical accuracy alone is not legitimacy.

Within this framework, *Mecellem* is not merely a technological solution; it is a holistic epistemic approach that enables legal reasoning to be executed in a more transparent, traceable, cautious, contestable, and contingency-free manner through new tools. Holmes's emphasis on experience, Loevinger's *jurimetrics* line, and al-Fârâbî's *burhan* epistemology are not disconnected references in this approach; they are components of a single intellectual trajectory aimed at making legal inference more robust, more justified, and more auditable. The system does not replace reasoning; it elevates the quality of reasoning and increases the structural transparency of legal inference. What will determine the future of law is not solely the computational power of large models, but an explainable ontological architecture upon which these models are built—one that is auditable, contestable, and carries the pursuit of *burhanic* certainty at every stage.